\documentclass[prd,nofootinbib,preprintnumbers,floatfix]{revtex4}  
\usepackage[plainpages=false, colorlinks=true, anchorcolor=blue, linkcolor=blue, citecolor=blue, bookmarks=false]{hyperref}
\newcommand{\rthis}[1]{\textcolor{black}{#1}}
\usepackage{graphicx}
\usepackage[export]{adjustbox}
\usepackage{listings}
\usepackage{dcolumn}
\usepackage{bm}
\usepackage{longtable}
\usepackage{multirow}
\usepackage{hyperref}
\usepackage{url}

\begin{document}
\renewcommand{\arraystretch}{2.2}


\title{Search for Cosmological time dilation from Gamma-Ray Bursts - A 2021 status update}

\author{Amitesh  \surname{Singh}$^{1}$}%
 \altaffiliation{asingh10@go.olemiss.edu}

\author{Shantanu \surname{Desai}$^{2}$ }%
 \altaffiliation{shntn05@gmail.com}

\affiliation{$^{1}$ University of Mississippi, University, MS, USA-38677}

\affiliation{$^{2}$ Indian Institute of Technology, Hyderabad, Kandi, Telangana, India}

\date{\today}

\date{\today}

\begin{abstract}
We carry out a search for signatures of  cosmological time dilation in the light curves  of Gamma Ray Bursts (GRBs), detected by the  Neil Gehrels Swift Observatory. For this purpose, we calculate two different durations ($T_{50}$ and $T_{90}$) for a sample of 
247 GRBs in the  fixed rest frame energy interval of 140-350 keV, similar to a previous work~\cite{zhang2013cosmological}. We then carry out a power law-based regression analysis   between the durations and redshifts. This search  is done using both the unbinned  as well as the  binned data, where both the weighted mean and the geometric mean was used.
For each analysis, we also  calculate the intrinsic scatter to determine the tightness  of the relation. We  find that the weighted mean-based  binned data for long GRBs and the geometric mean-based binned data  is consistent with the cosmological time dilation signature, whereas  the analyses using unbinned durations show a very large scatter. We  also make our analysis codes and the procedure for obtaining the light curves and estimation of $T_{50}$/$T_{90}$ publicly available.
\end{abstract}

\maketitle


\section{Introduction}

Gamma-ray bursts (GRBs) are  short-duration single-shot transient events detected in keV-MeV energy range,  and constitute one of the biggest and brightest explosions in the universe, with isotropic-equivalent energies  between $10^{48}$ and $10^{55}$ ergs~\cite{Kumar}. GRBs have been traditionally bifurcated into two categories based on $T_{90}$ and $T_{50}$, which represent the durations over which 90\% and 50\% of fluence are detected, respectively~\cite{Kouveliotou93}. GRBs with $T_{90}> 2$ secs are known as long bursts and associated with core-collapse supernovae, whereas those with $T_{90}<2$ seconds are known as short bursts and usually associated with compact object mergers~\cite{Levan,Piran,Nakar,Berger}. There are, however, exceptions to this broad dichotomy, and additional sub-classes of GRBs have also been identified~\cite{Horvath98,Horvath02,Horvath06,Horvath08,Zhang09,Horvath10,Bromberg,Kulkarni,Tarno19,Horvath18}. The smoking gun evidence for the association of short GRBs with compact binary object mergers (involving neutron stars) was, however, provided by the simultaneous detection of gravitational waves and gamma rays from GW170817/GRB170817A~\cite{LIGO}, which also enabled tests of a whole slew of modified gravity theories~\cite{Woodard}.

Although GRBs were first detected in the 1970s~\cite{firstgrb} there were still lingering doubts as to whether GRBs are local or cosmological until the 1990s (see the contrasting viewpoints on this issue by
~\cite{Paczynski} versus ~\cite{Lamb}, as of 1995.) This issue was unequivocally resolved in favour of a cosmological origin,  after the first precise localization and redshift determination in 1997 from the Beppo-Sax satellite~\cite{vanPara}. After the launch of the Neil Gehrels Swift observatory (Swift, hereafter), a large number of GRBs with confirmed redshifts have been detected~\cite{SWIFT}.

Despite this, no smoking gun signature of cosmological time dilation has emerged in the GRB light curves ever since the earliest proposed attempts~\cite{Piran92,Paczynski92}, unlike the corresponding signature found in Type
Ia supernovae~\cite{Goldhaber}. We note, however, that such a time dilation has also not been seen in quasar light curves, although the reasons have been attributed to other astrophysical effects which mask the cosmological time dilation signal~\cite{Hawkins}. 
Before the launch of the Swift satellite, there were a few indirect pieces of evidence for cosmological time dilation in GRB light curves~\cite{Norris,Che97,Deng98,Lee2000,Chang01}.  The main difficulty in coming to a robust conclusion in these early studies stemmed from the  uncertainty in the intrinsic GRB duration and very little statistics of GRBs with well-measured redshifts. However, even after the launch of Swift and FERMI gamma-ray space telescope, there were  no smoking gun signatures of cosmological time dilation in the GRB light curves~\cite{Sakamoto,Petrosian13,Gruber,Lien,Crawford} or via indirect methods such as through the skewness of $T_{90}$ distributions~\cite{Tarno}.

The first novel attempt at  measuring  a robust signature of cosmological time dilation using a large sample of Swift GRBs was carried out by \citet{zhang2013cosmological} (Z13, hereafter). Z13 pointed out that all previous studies of cosmological time dilation prior to their work were carried out from the durations in the observer frame using a fixed energy interval, usually corresponding to the energy senstivity of the detector, which was the main reason for not seeing  a smoking-gun  signature.
To circumvent this, they calculated the durations for a sample of 139 Swift GRBs using a fixed energy interval between 140-350 keV in the GRB rest frame, followed by calculating the observed durations using $E_{obs}= E_{rest}/(1+z)$, where $E_{obs}$ and $E_{rest}$ denote the photon energy in the observer and rest frame, respectively. Z13 showed after binning the data as a function of redshift that  the durations were correlated with $(1+z)$, i.e., ($T_{90} \propto (1+z)^{0.94 \pm 0.26}$). This analysis was then extended by ~\cite{Butler} to 237 Swift GRBs and 57 Fermi-GBM GRBs using three independent measures of duration: $T_{90}$, $T_{50}$, and $T_{R45}$. They showed that the binned data for the  Swift GRB sample is broadly consistent with cosmological time dilation.

We independently carry out the same procedure as Z13 by  using the latest Swift GRB database. We then carry out a power-law  based regression analysis between the durations and redshifts  using both the unbinned well as binned data. We also thoroughly document the procedure for obtaining the Swift light curves for every GRB and obtain the $T_{90}$, which would benefit readers for carrying out any analysis using the GRB light curves. 
This manuscript is organized as follows.  Our data  analysis and results are presented in Sect.~\ref{sec:results}. We conclude in Sect.~\ref{sec:conclusions}. Detailed documentation of the procedure to get the GRB light curves and calculate $T_{50}$/$T_{90}$ can be found in Appendix~\ref{app}.

\section{Data  Analysis and Results}

\label{sec:results}
We construct light curves for all the GRBs detected by Swift with confirmed redshifts in the fixed  rest frame energy interval between 140-350 keV (similar to Z13), and thereby determine $T_{50}$ and $T_{90}$. The full step-by step procedure to download these light curves and obtain $T_{50}$/$T_{90}$ is documented in Appendix~\ref{app}. We have also uploaded these codes on a {\tt github} link, which can be found at the end of the manuscript. Out of all the GRBs containing redshift data present in the Swift database (around 400), we were able to determine the durations for 247 of them.  A tabulated summary of these durations along with 1$\sigma$ rs in the fixed rest frame energy interval (140-350 keV) as well as the full energy range in the observer frame for all GRBs can be found in Table~\ref{Table1:Swift_GRBs}.

We now describe the procedure for checking if the aforementioned GRB durations   are consistent with the signature of   cosmological time dilation.  For this purpose, we carry out a  power law-based regression analysis between  the observed durations and the redshift using the following  equation
\begin{equation}
   \centering
    y=A x^{B}
    \label{equation_model}
\end{equation}
where $A$ and $B$ are parameters of our model (unknown before the analysis, $x$ is an input variable with $x=(1+z)$ , where $z$ is the redshift, and $y$ denotes the burst interval (either $T_{90}$ or $T_{50}$) in seconds. For cosmological time dilation, $B$ should be equal to one.

To get the best-fit parameters we maximize the likelihood as follows:
\begin{eqnarray}
-2\ln L &=& \large{\sum_i} \ln 2\pi\sigma_i^2 + \large{\sum_i} \frac{[y_i-Ax_i^B]^2}{\sigma_i^2}
\label{eq:eq8}  \\
\sigma_i^2 &=& \sigma_{y_i}^2+m^2\sigma_{x_i}^2+\sigma_{int}^2\nonumber
\end{eqnarray}
where $x_i$ is $(1+z)$ data for every GRB. Similar to our works on galaxy clusters~\cite{Gopika_2020,Pradyumna,Bora_feb} and also~\cite{Tian}, we have added in quadrature, an unknown intrinsic scatter $\sigma_{int}$ as an additional free parameter while fitting for Eq.~\ref{equation_model}. This can be used to parameterize  the tightness of the scatter in the relation. A value of $\sigma_{int} > 1$ indicates that the relation has a lot of scatter  and that any relation between the two variables cannot be discerned. On the other hand, a small value of scatter points to a deterministic relation between the two variables. Note that Z13 have investigated for putative correlations between the durations and redshifts using the Pearson correlation coefficient. However, the Pearson correlation coefficient does not take into account the errors in the observables. The magnitude of the intrinsic scatter would be a more robust diagnostic of the scatter.

We use the {\tt emcee} MCMC routine \cite{foreman2013emcee} to sample the above likelihood and obtained 68\%, 90\%, and 95\% marginalized credible intervals on each of the parameters. We use uniform priors on all the free parameters : $0<A<50$, $-1<B<4$, $-5<\ln(\sigma_{int})<5$.
In all the cases, the marginalized contours were obtained using the {\tt Corner} package, and estimates are performed over parameters of the model which we are trying to optimize.  

We now present our results.  We carried out multiple analyses for both $T_{50}$ and $T_{90}$, using both the unbinned data as well as using the binned data. The binning was done in equal-sized redshift bins and with two different ways of averaging the data in each bin: viz. weighted mean (as in Z13) and the geometric mean (as in ~\cite{Butler}). Our results are summarized in Table~\ref{tab:unbinned_analysis} and  ~\ref{tab:binned_analysis}.

\begin{itemize}

\item  {\bf Analysis using unbinned data for all GRBs}

\label{sec:comprehensive_analysis}
We first fit the $T_{50}$ and $T_{90}$ obtained for all the 247 GRBs in our sample to Eq.~\ref{equation_model}. The marginalized credible intervals for the unbinned data are given by Fig.~\ref{fig:corner_T50_unbinned_data} and Fig.~\ref{fig:corner_T90_unbinned_data} for both  $T_{50}$ and $T_{90}$, respectively. As we can see, the intrinsic scatter is greater than 100\%, indicating that it is difficult to discern a deterministic  relation between $T_{50}$ or $T_{90}$ and the redshift. Furthermore, the power-law index $B$ is not equal to the cosmological expected value of one to within 1$\sigma$. Fig.~\ref{fig:burst_intervals_with_redshifts} shows the plots for both the burst intervals, $T_{90}$ and $T_{50}$, along with the best-fit obtained from fitting the full unbinned data.


\begin{figure}
    \centering

    \includegraphics[scale=0.5]{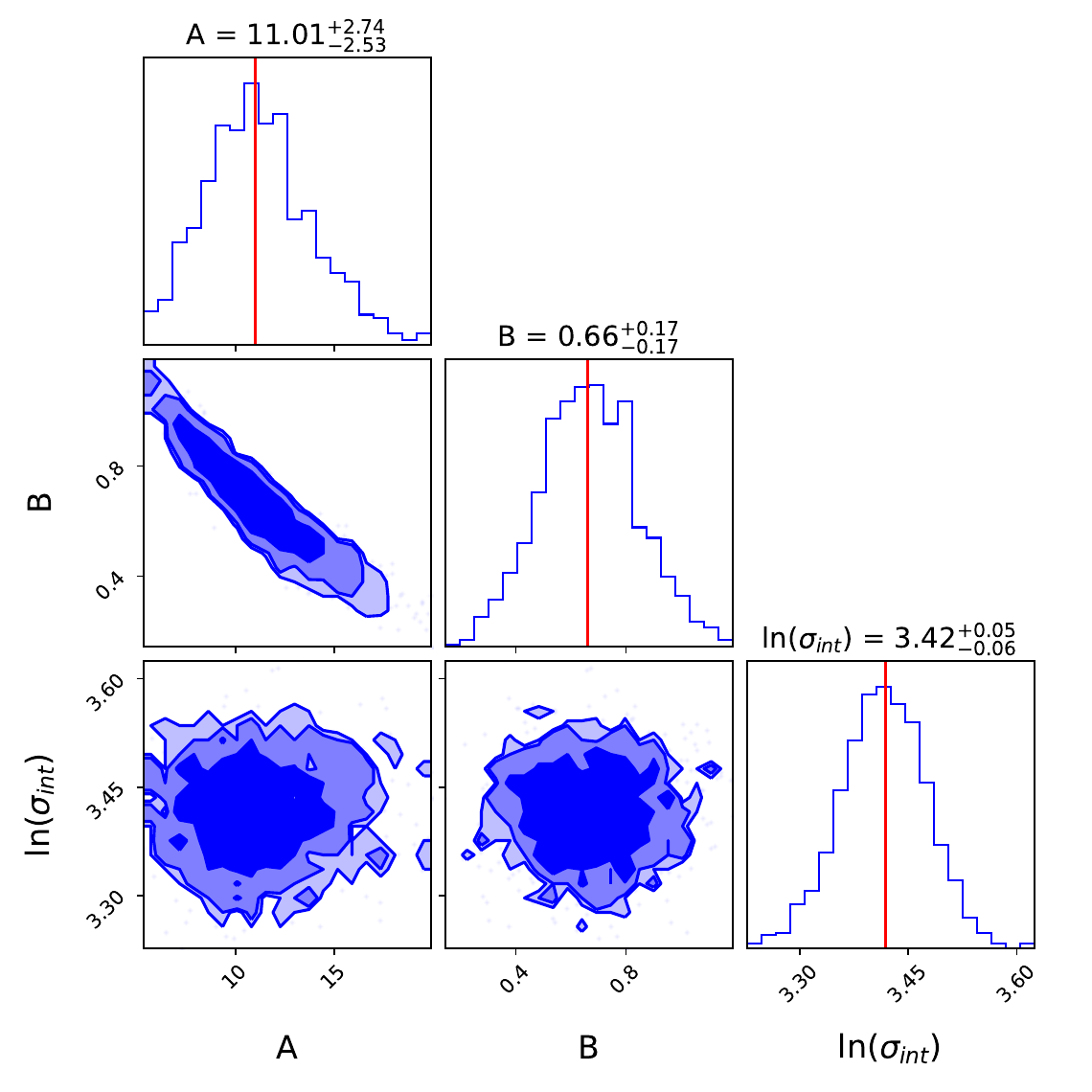}

    \caption{Figure showing the contour plot for  $T_{50}$ in the rest-frame energy range 140-350 keV, for the unbinned GRB data. The contours represent 68\%, 90\%, and 95\% credible intervals. As we can see, the intrinsic scatter is quite high ($>$ 100\%), which implies that it is hard to discern a deterministic relation between $T_{50}$ and the redshift.} 
    \label{fig:corner_T50_unbinned_data}
\end{figure}

\begin{figure}
    \centering

    \includegraphics[scale=0.5]{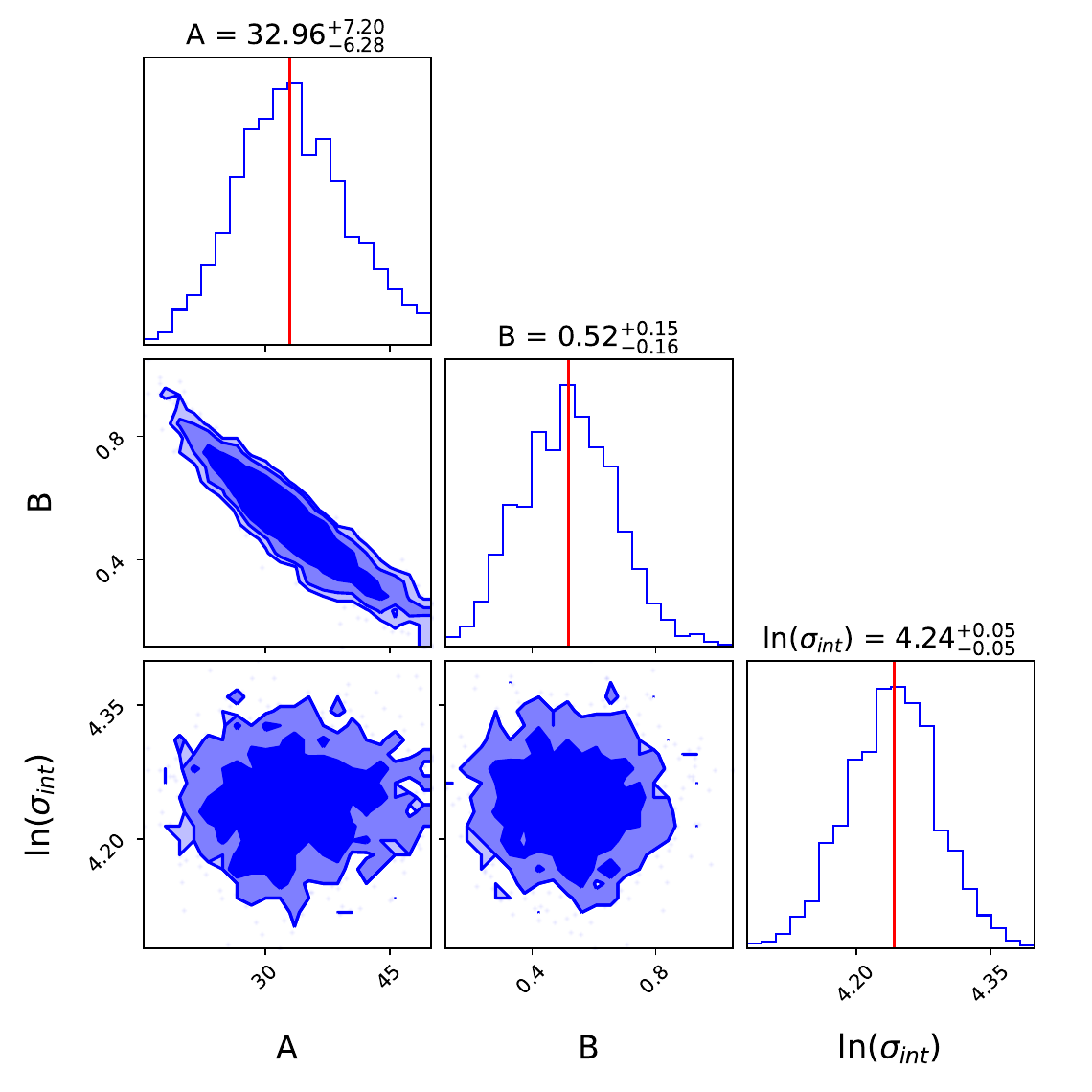}

    \caption{Figure showing the contour plot  $T_{90}$ in the rest-frame energy range 140-350 keV, for the unbinned GRB data. The contours represent 68\%, 90\%,  and 95\% credible intervals. Again, the intrinsic scatter is very high ($>$ 100\%).} 
    \label{fig:corner_T90_unbinned_data}
\end{figure}

\begin{figure}
    \centering

    \includegraphics[scale=0.15]{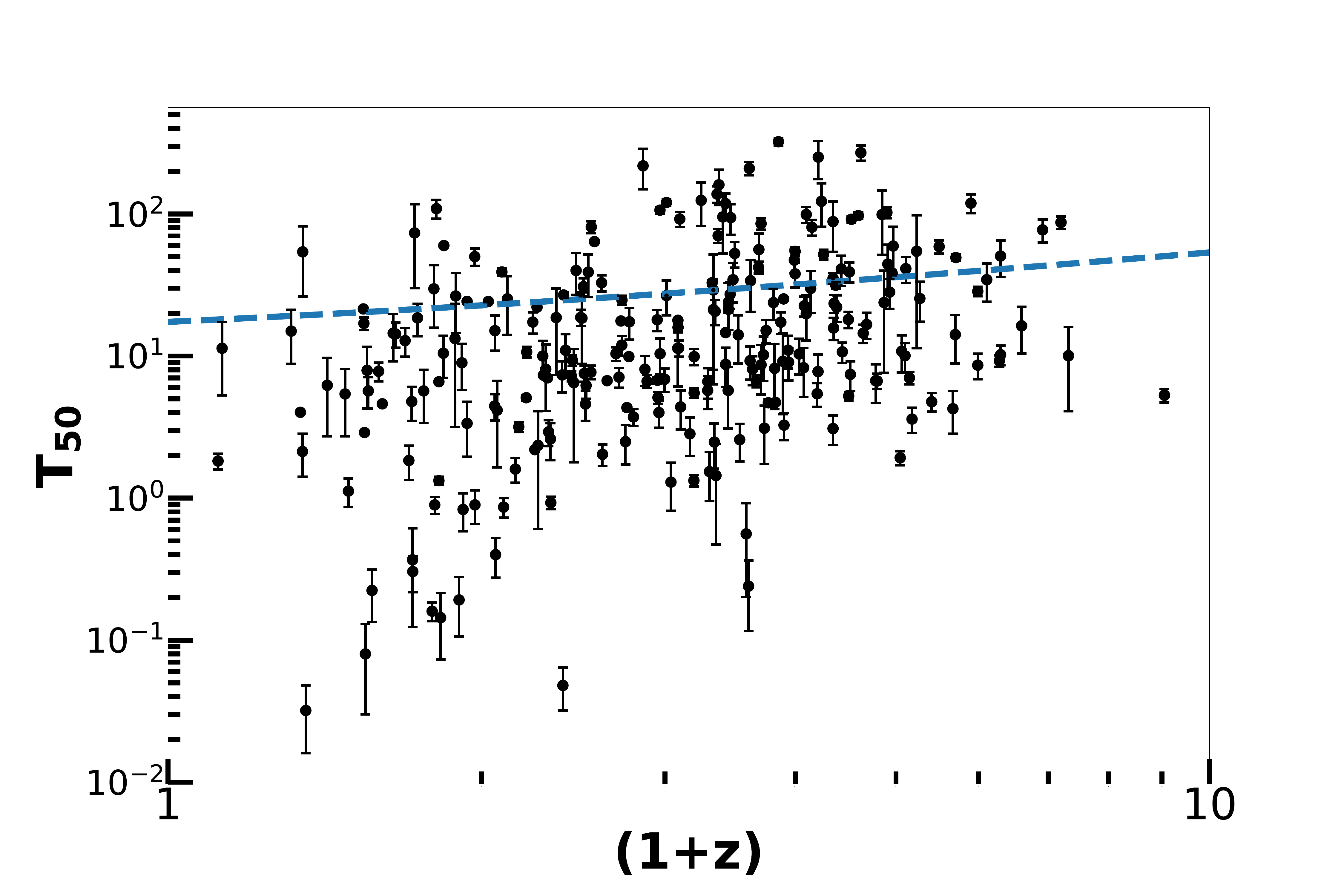}
    \includegraphics[scale=0.15]{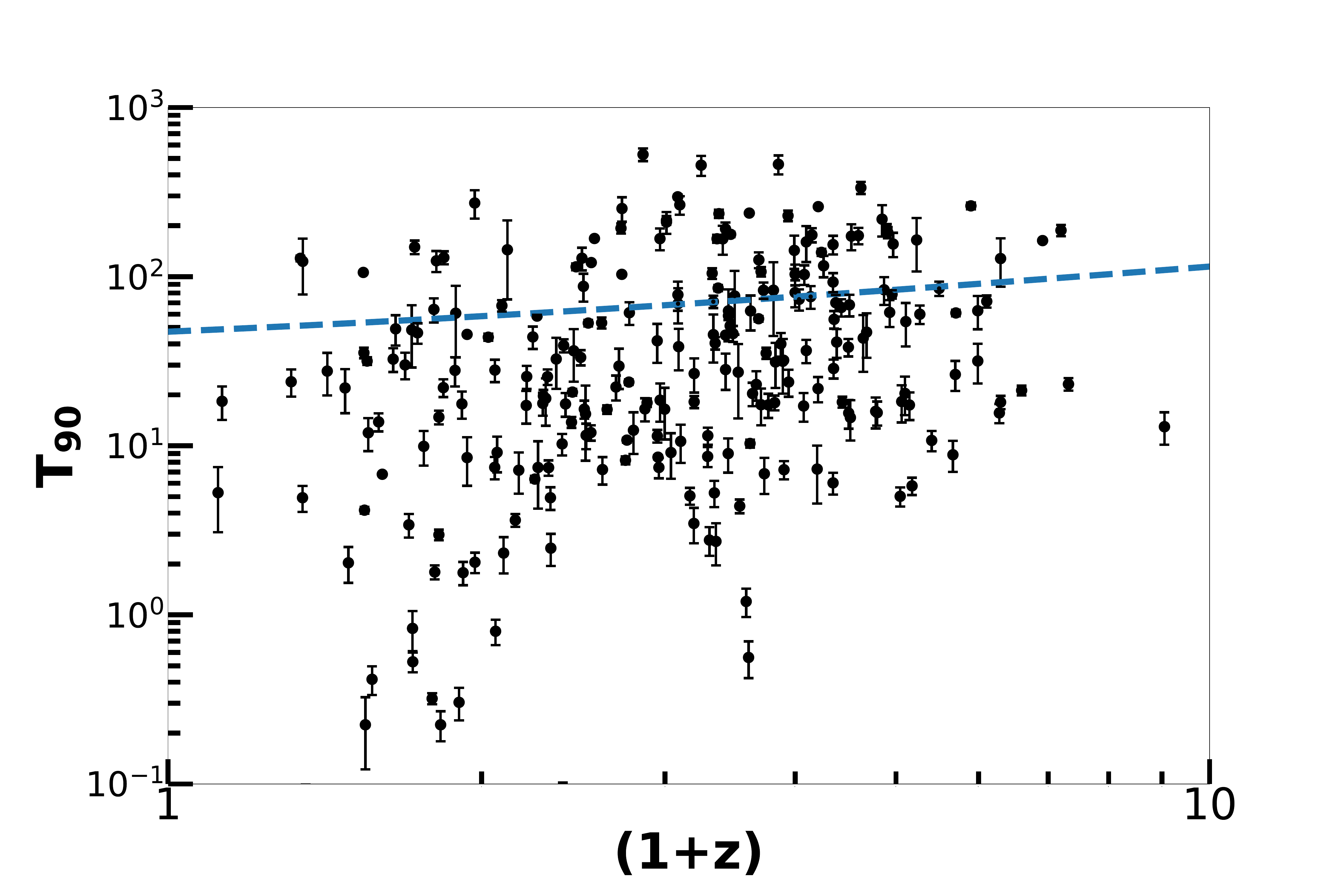}

    \caption{Figure showing the relationship between the burst intervals, $T_{90}$ and $T_{50}$  in the observer frame  energy range $E_{1}=140/(1+z)$ keV and $E_{2}=350/(1+z) $ keV (denoted by $T_{90}$, $T_{50}$), for the unbinned GRB data. The data points have been plotted with their respective error bars. The best fits are shown by the dashed blue lines and those are $T_{50}=\rthis{11.01}(1+z)^{\rthis{0.66}}$ and $T_{90}=\rthis{32.96}(1+z)^{\rthis{0.52}}$ respectively. } 
    \label{fig:burst_intervals_with_redshifts}
\end{figure}

\item  {\bf Analysis using weighted-mean binned data for all GRBs}

We now re-analyze this data  by binning the data into six bins with equal redshift intervals and using  the error-weighted means of both the burst intervals, $T_{90}$ and $T_{50}$ in each redshift bin.
The marginalized contours for $A$ and $B$ for such a   binned analysis for both $T_{50}$ and $T_{90}$ can be found  in Figs.~\ref{fig:Contour_T50_all_grbs} and \ref{fig:Contour_T90_all_grbs}, respectively. We find that $B$ is discrepant from the cosmological expected value of one at 1.2$\sigma$ ($T_{50}$)  and 2$\sigma$ ($T_{90}$).   The intrinsic scatters are approximately 75\% for both $T_{50}$ and $T_{90}$. This shows that there is still no evidence for   cosmological time dilation signature using the complete binned sample.
The binned data along with the best fit can be found in Fig.~\ref{fig:weighted_means}.

\begin{figure*}
    \centering
    \includegraphics[scale=0.5]{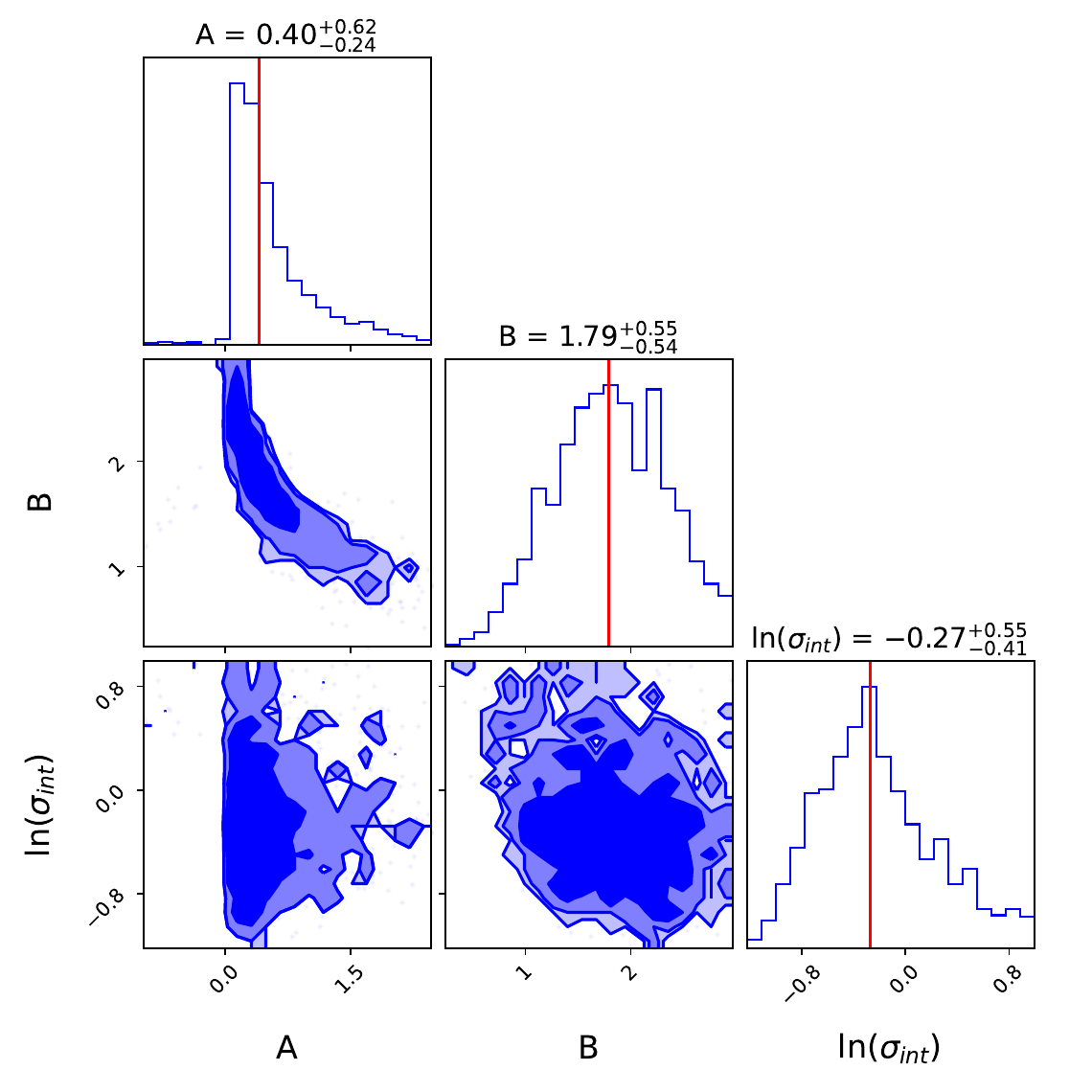}

    \caption{Contours for $A$ and $B$ for $T_{50}$ in the rest-frame energy range 140-350 keV. The input data is the weighted mean of binned data of $T_{50}$ vs the binned redshifts (1+z). This data is then fit to Eq~\ref{equation_model}. We have shown 68\%, 90\%, and 95\% credible intervals. The intrinsic scatter we obtain is about 76\%. The value of $B$ is discrepant  with the signature of cosmological time dilation at about  $1.2\sigma$. }
    \label{fig:Contour_T50_all_grbs}
\end{figure*}

\begin{figure*}
    \centering
    
    \includegraphics[scale=0.5]{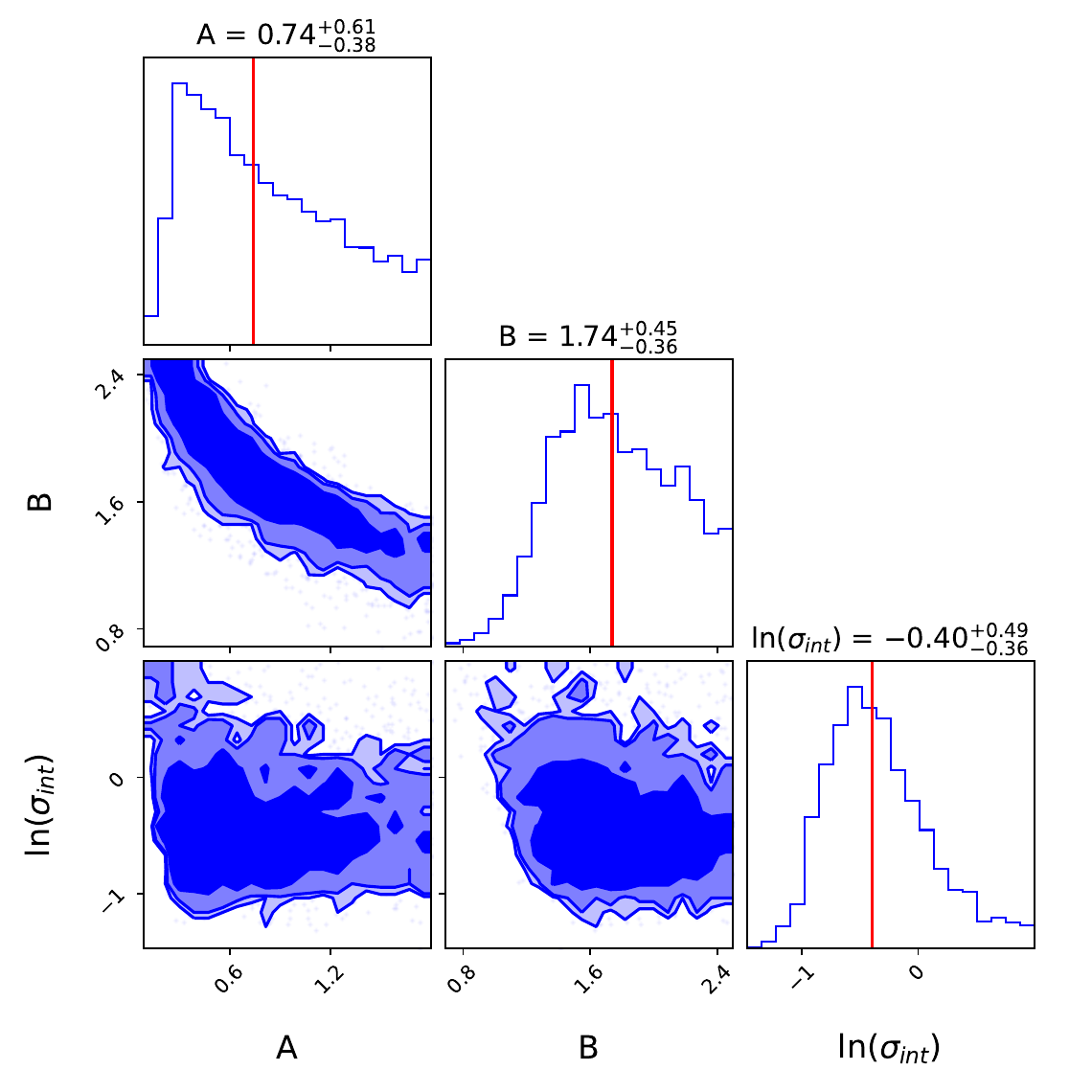}

    \caption{Contours for $A$ and $B$ using  90\% burst interval, $T_{90}$ in the rest-frame energy range 140-350 keV.  Input data is the weighted mean of binned data of the $T_{90}$ vs the binned redshifts (1+z). This data is then fit to Eq.~\ref{equation_model}. We have shown 68\%, 90\%, and 95\% credible intervals. The intrinsic scatter is equal to \rthis{76.3}\%.  The value of $B$ is discrepant  with the signature of cosmological time dilation at about  $2\sigma$.  }
    \label{fig:Contour_T90_all_grbs}
\end{figure*}

\begin{figure*}
     \centering
    \includegraphics[scale=0.15]{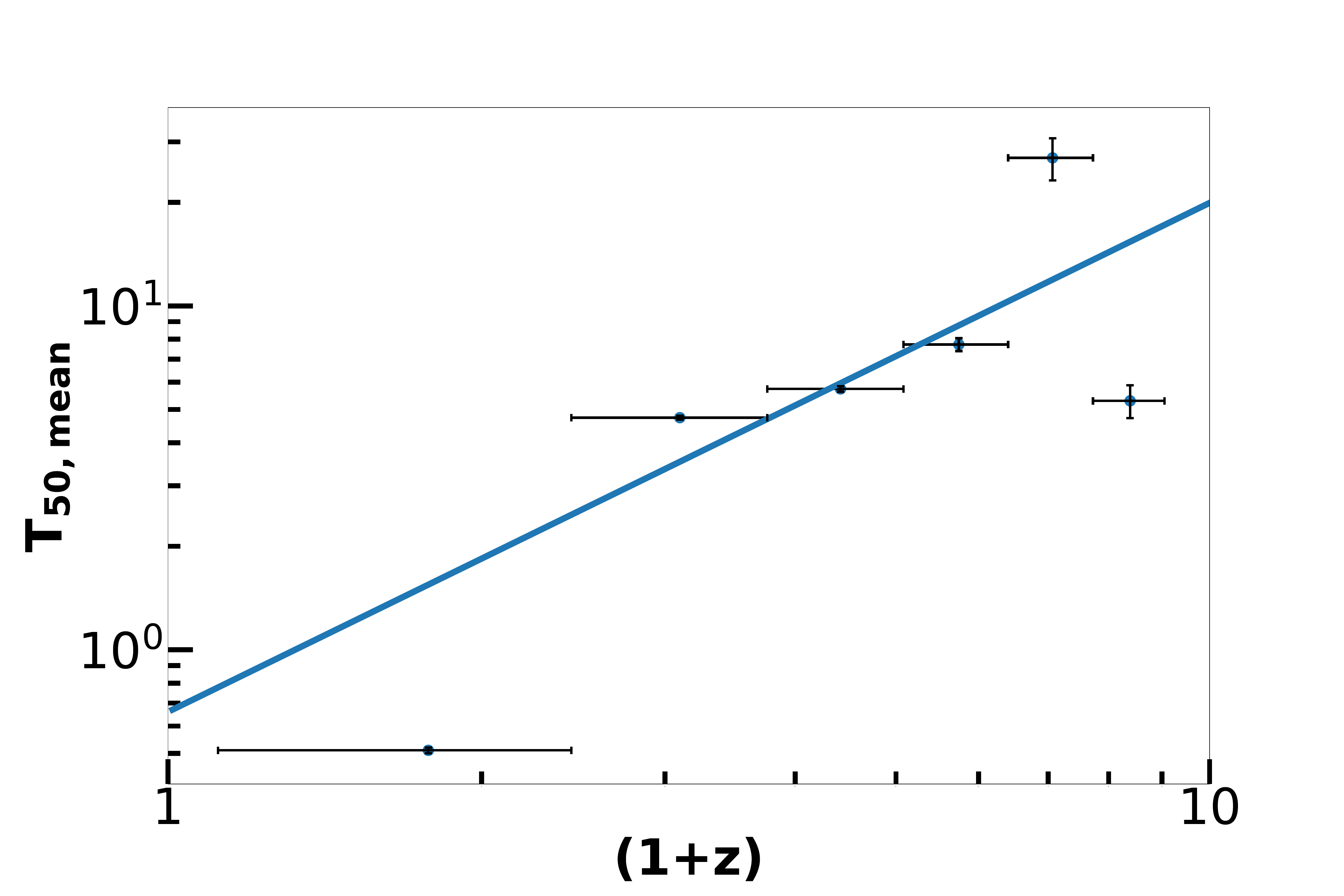}
    \includegraphics[scale=0.15]{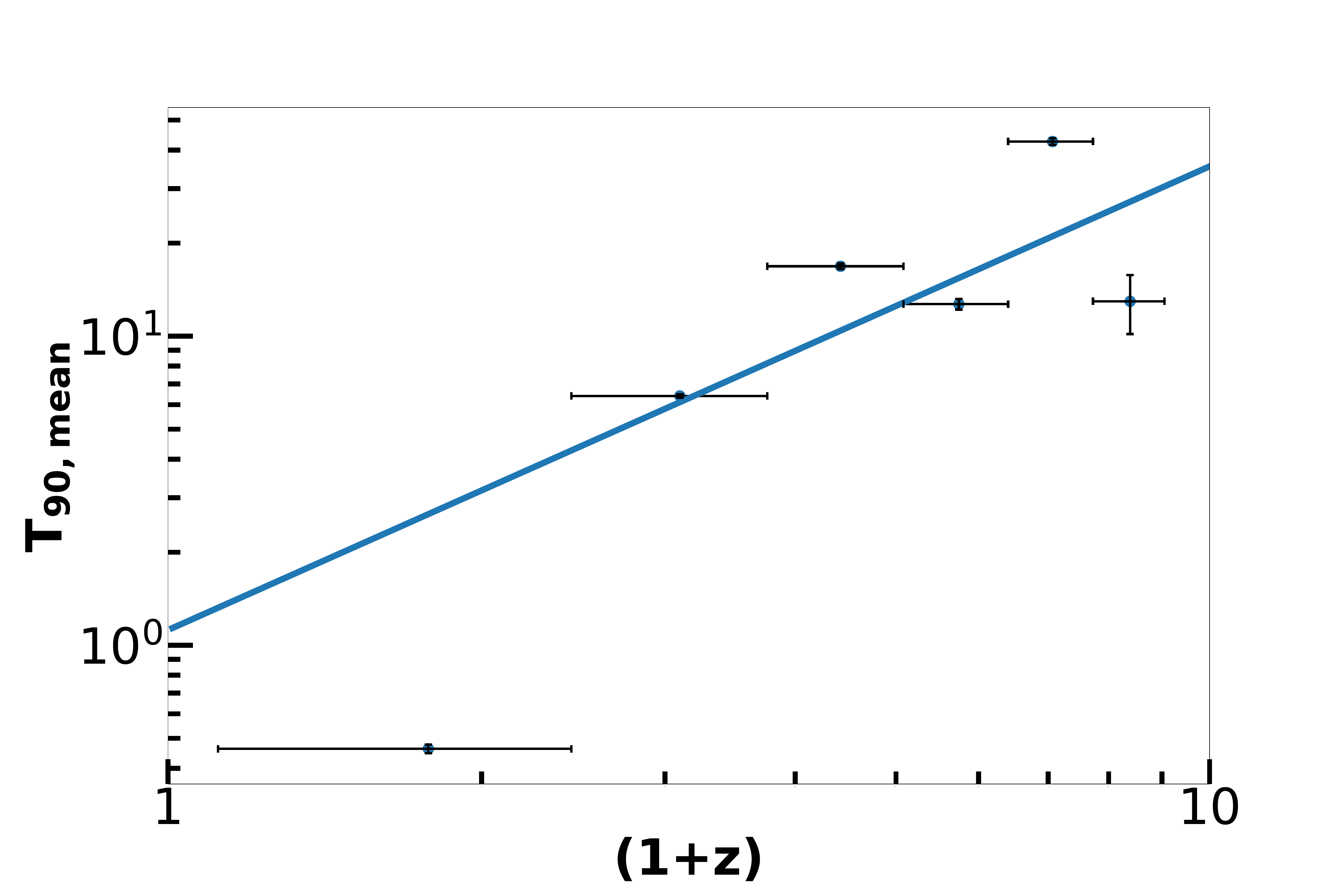}
    \caption{Plots showing the weighted means of the burst intervals with the binned redshift data (containing 6 groups equally spaced in $z$), where the horizontal errors bars denote the redshift range in each group. The blue line curve is our best fit for the data obtained through the estimation of parameters $A$ and $B$. The best-fits shown are $T_{50}=\rthis{0.40}(1+z)^{\rthis{1.79}}$ and $T_{90}=\rthis{0.74}(1+z)^{\rthis{1.74}}$.}
    \label{fig:weighted_means}
\end{figure*}

\item {\bf Analysis  using unbinned data for long GRBs} 

\label{sec:analysis_T90_raw_gt_2s}
Z13 and Ref.~\cite{Butler} did their analysis
using only long GRBs, with $T_{90} >$ 2 secs. The reason for this exclusion of short GRBs is to have a pure sample with an intrinsically similar distribution. We now present similar  results using only long  GRBs  with the cut of $T_{90,raw} > 2$ secs. We show the marginalized contours for both $A$ and $B$, for both $T_{90}$
and $T_{50}$ in Figs.~\ref{contour_T50_with_T90_raw_gt2} and ~\ref{contour_T90_with_T90_raw_gt2}. Similar to the full unbinned GRB dataset, the intrinsic scatter is $> 100\%$ making it impossible to discern any relation between the duration and redshift. Furthermore, the values of $B$ are not equal to one (at more than $1\sigma$) for both $T_{90}$
and $T_{50}$ Fig.~\ref{fig:burst_interval_T90>2sec} shows the  burst intervals as a function of  redshift for this subset along with the best fit.

\begin{figure}
    \centering
    \includegraphics[scale=0.5]{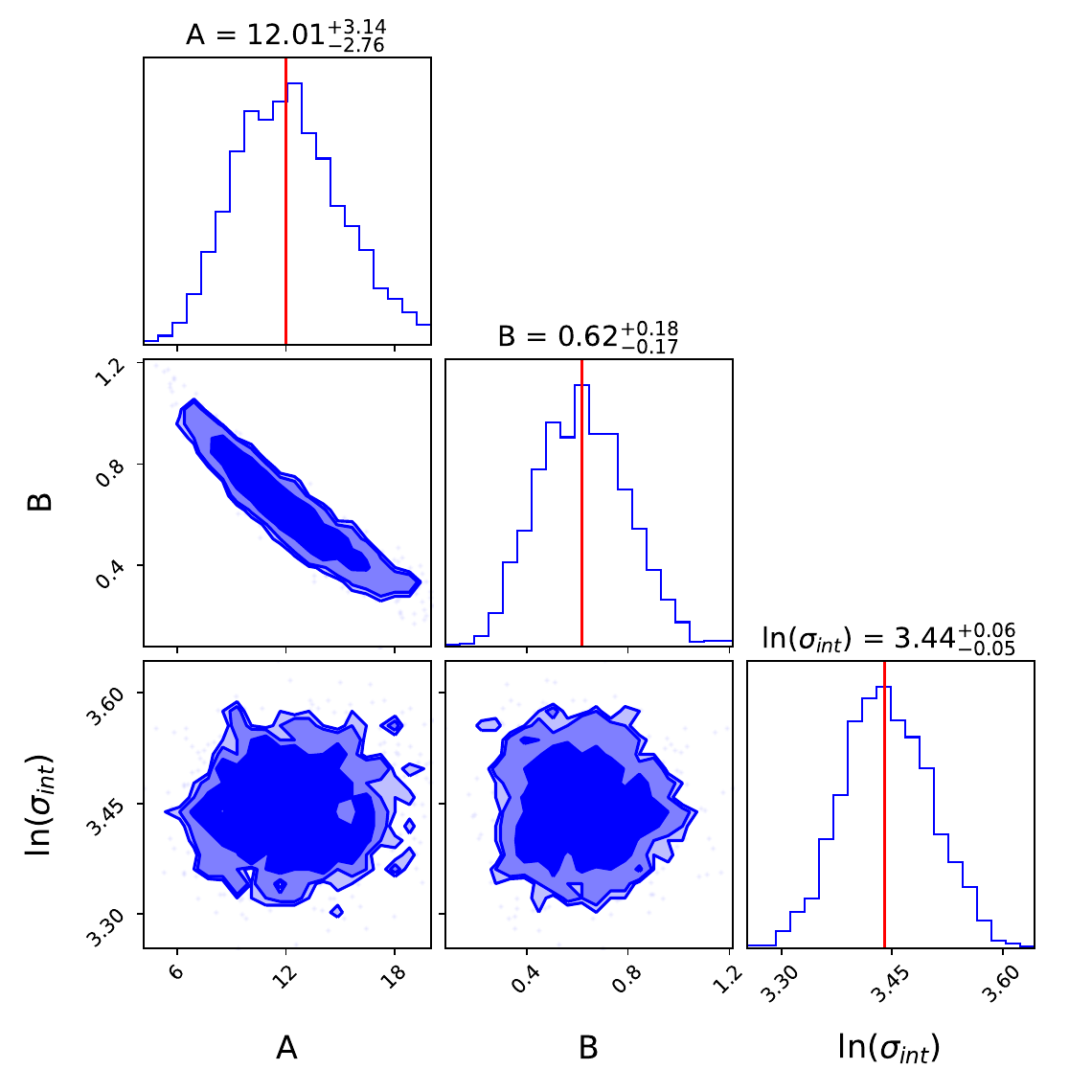}
    
 \caption{Contours for $A$ and $B$ for $T_{50}$, in the rest-frame energy range 140-350 keV  for long GRBs which have  $T_{90,raw}$ greater than 2 secs. We have shown 68\%, 90\%,  and 95\% credible intervals. The intrinsic scatter is greater than 100\%.}
 \label{contour_T50_with_T90_raw_gt2}
\end{figure}

\begin{figure}
    \centering
    \includegraphics[scale=0.5]{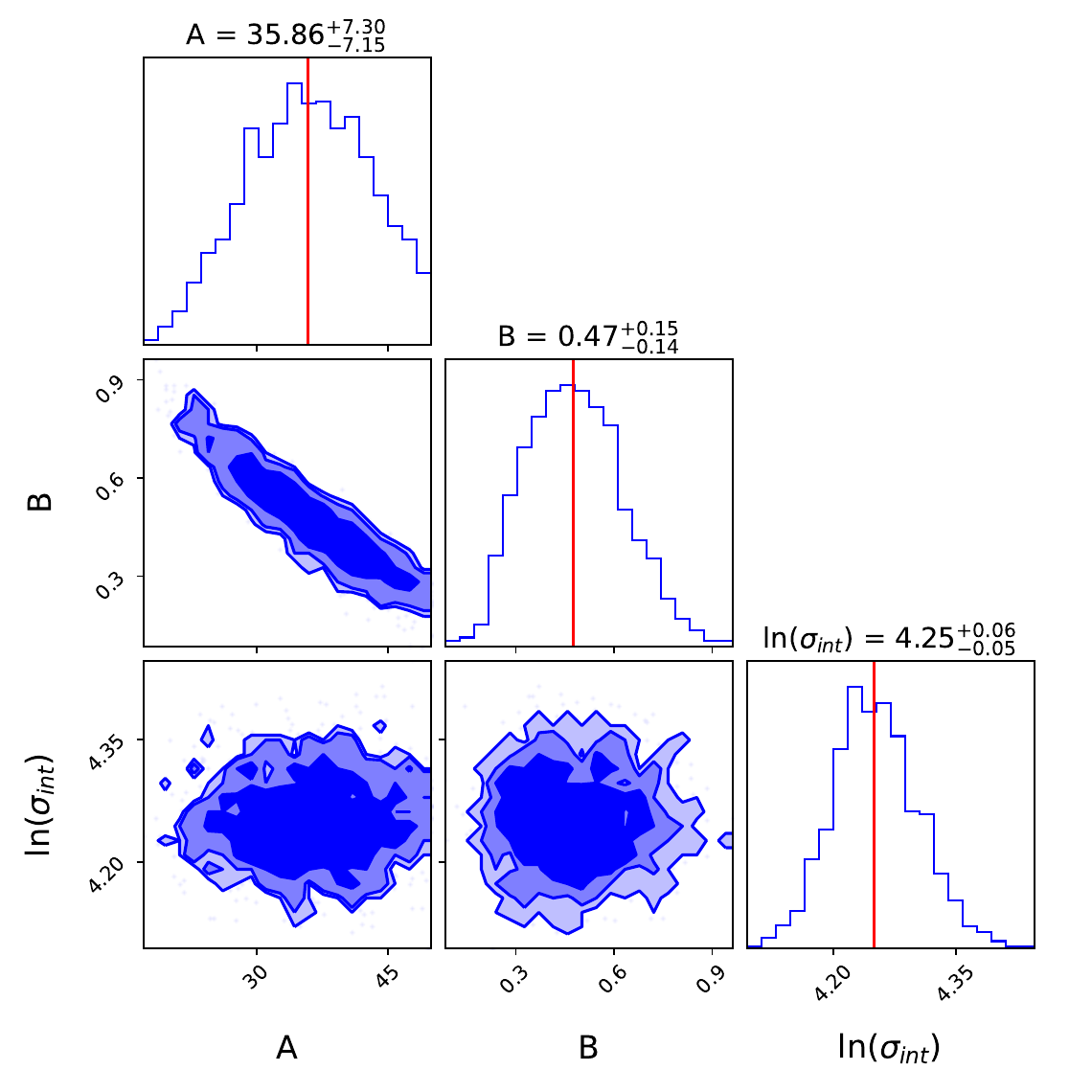}
    
 \caption{Contours for $A$ and $B$ for $T_{90}$ in the rest-frame energy range 140-350 keV, for long GRBs which had a $T_{90,raw}$ greater than 2 secs. The intrinsic scatter is greater than 100\%. }
 
 \label{contour_T90_with_T90_raw_gt2}
\end{figure}

\begin{figure*}
    \centering

 \includegraphics[scale=0.15]{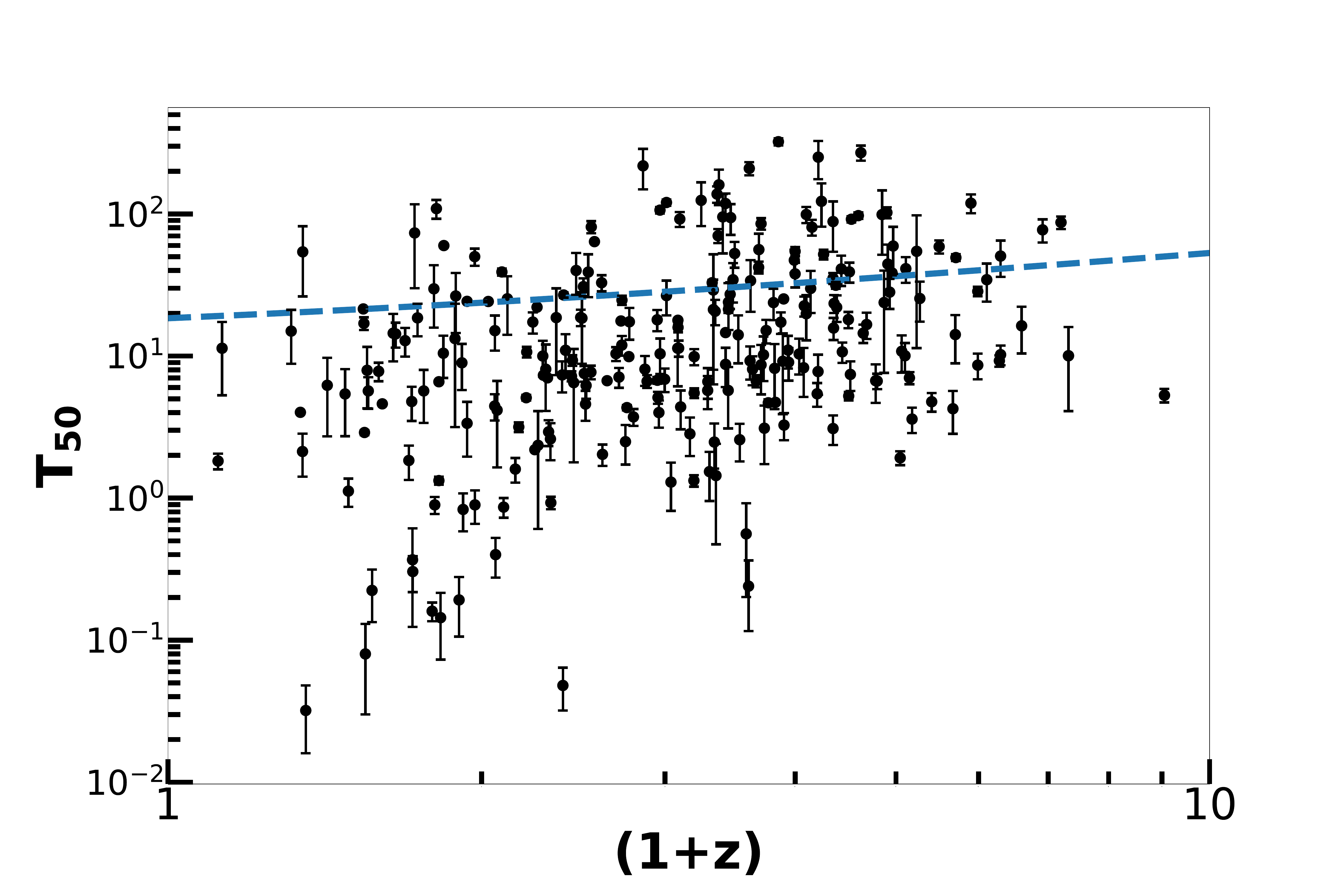}
    \includegraphics[scale=0.15]{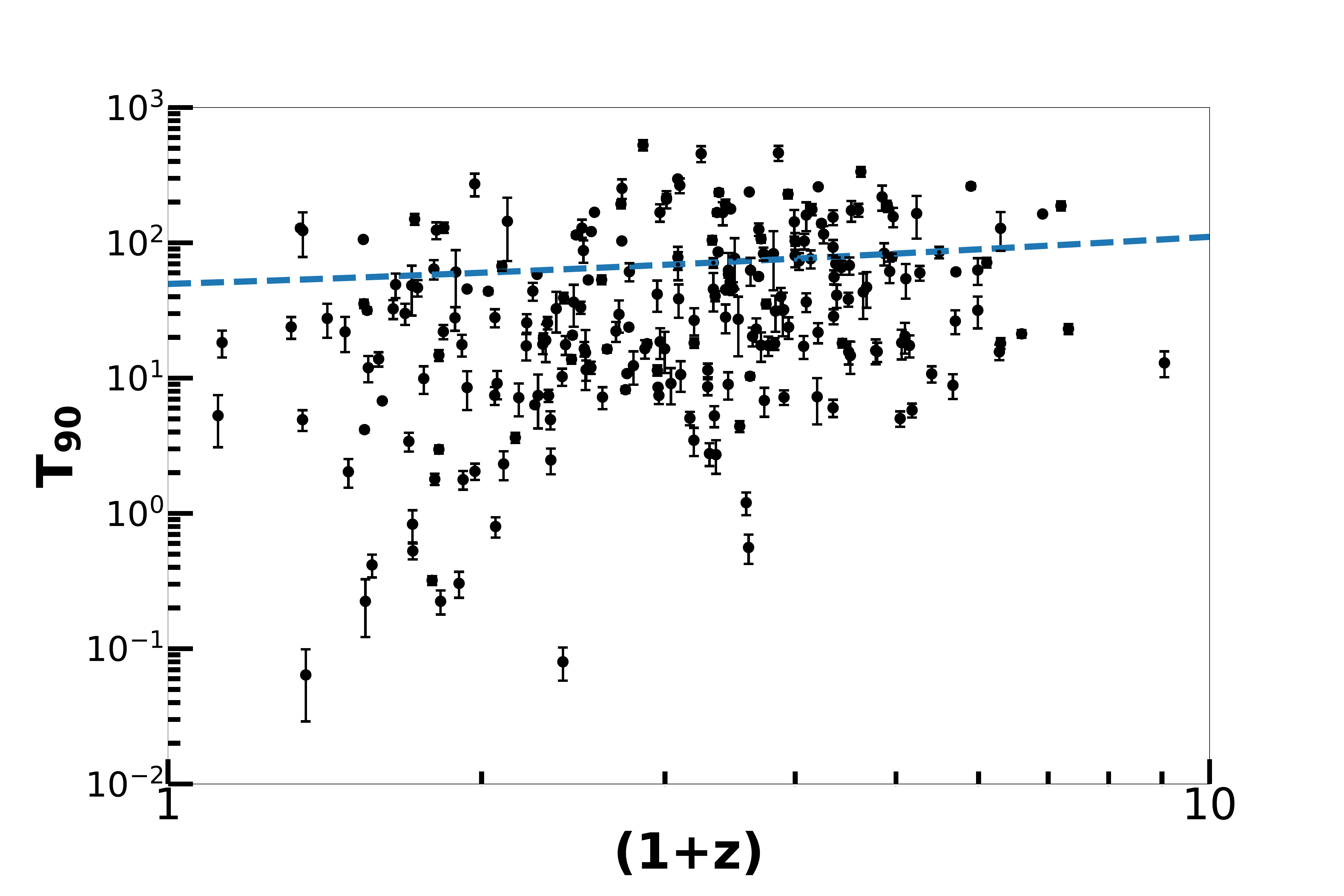}
    \caption{Figure showing the relationship between the burst intervals, $T_{90}$ and $T_{50}$ in observer energy range (denoted by $T_{90}$, $T_{50}$), for long GRBs which had $T_{90,raw}>2 $ seconds.  Through curve fitting, we obtain $T_{50}=\rthis{12.01}(1+z)^{\rthis{0.62}}$ and $T_{90}=\rthis{35.86}(1+z)^{\rthis{0.47}}$. }
    \label{fig:burst_interval_T90>2sec}
\end{figure*}

\item {\bf Analysis using binned data for long GRBs}

We now bin the data for long GRBs and use the weighted mean in each redshift bin.
The marginalized contours for both $A$ and $B$, which are obtained by fitting the model in Eq.~\ref{equation_model}, using the binned data  can be found in  Figs~\ref{contour_T50_gt2_binned} and \ref{contour_T90_gt2_binned}. We obtain intrinsic scatters of 57\% and 55\% for $T_{50}$ and $T_{90}$, respectively. The values of $B$ for both the binned durations
are consistent with the cosmological time dilation signature ($B=1$) to within 1$\sigma$.
We  show the weighted mean burst intervals for these long GRBs, with redshifts in Fig.~\ref{fig:weighted_mean_T90>2sec}, along with the best-fit.

\begin{figure*}
    \centering
    \includegraphics[scale=0.5]{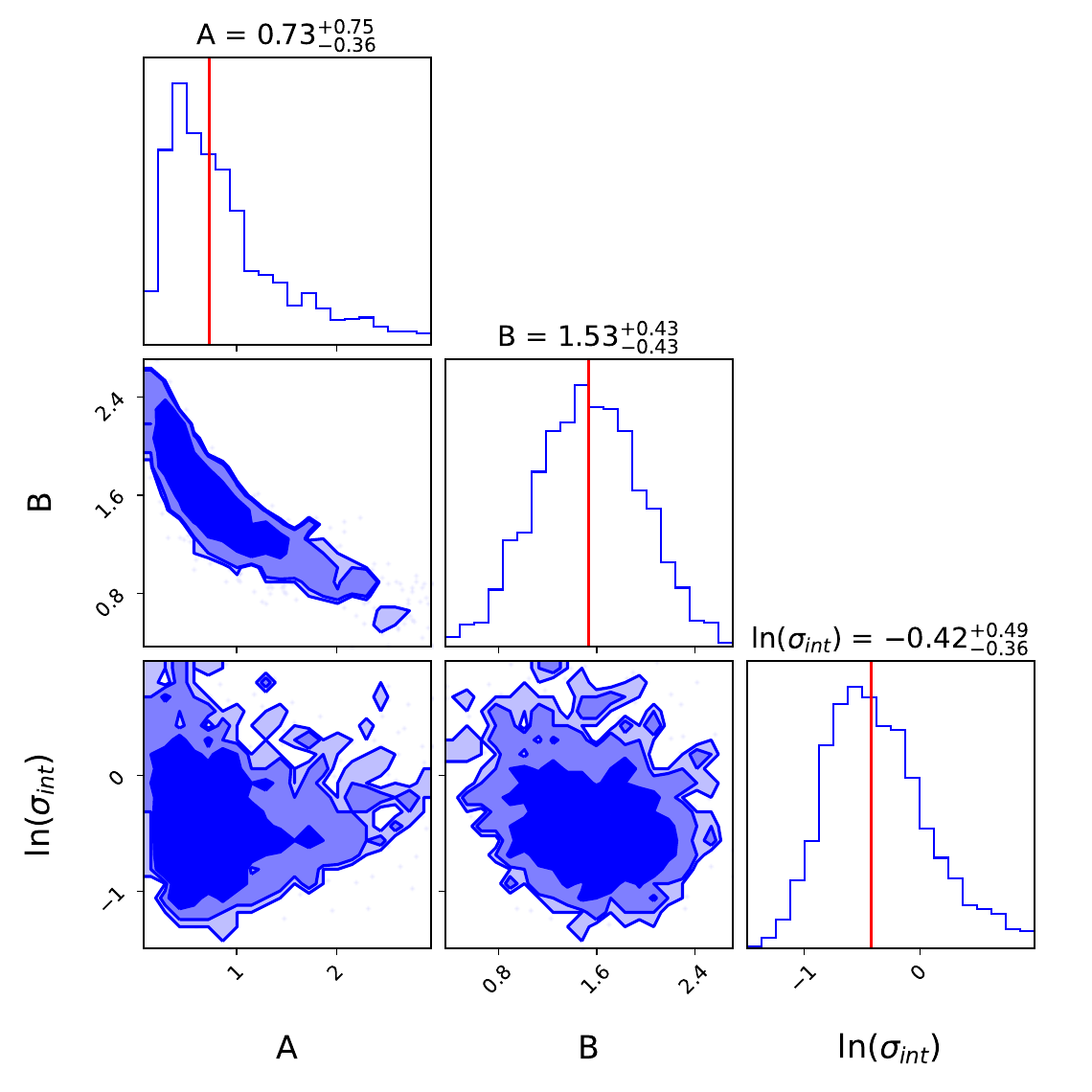}
    \caption{Plot showing 68\%, 90\%, and 95\% credible intervals  for $A$ and $B$ for $T_{50}$ in the rest frame energy range of 140-350 keV, using the binned redshift data (containing 6 groups equally spaced in $z$), for long GRBs.  The intrinsic scatter is \rthis{65.7}\%. The value of $B$ is consistent with a cosmological time dilation signature within 1$\sigma$. }
    \label{contour_T50_gt2_binned}
\end{figure*}

\begin{figure*}
    \centering
    \includegraphics[scale=0.5]{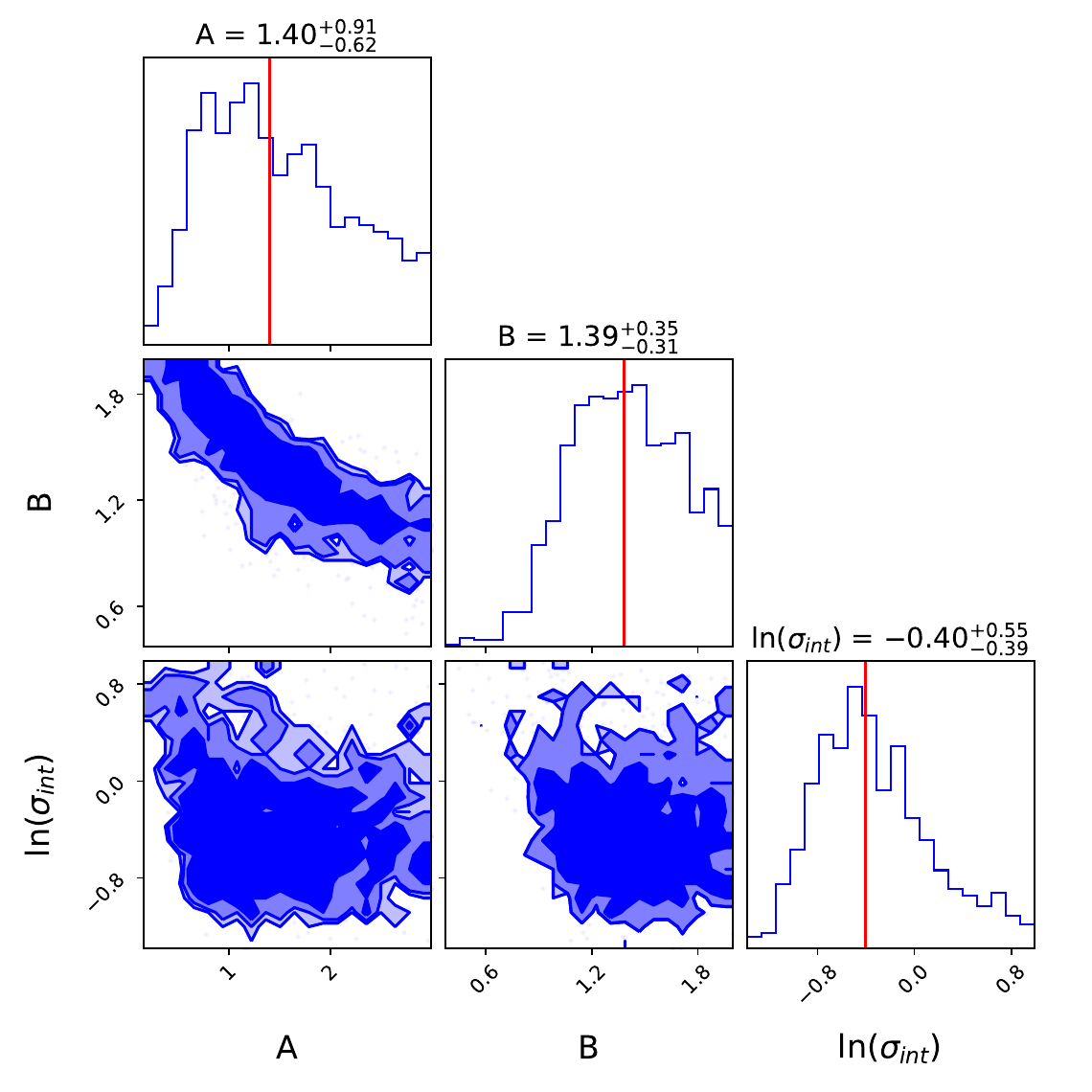}
    \caption{Plot showing 68\%, 90\%,  and 95\% credible intervals of $A$ and $B$ for $T_{90}$  with the binned redshift data (containing 6 groups equally spaced in $z$), for long GRBs. The intrinsic scatter is equal to  \rthis{67.03}\%. The value of $B$ is consistent with a cosmological time dilation signature within 1$\sigma$.}
    \label{contour_T90_gt2_binned}
\end{figure*}


 \begin{figure*}
    \centering
    \includegraphics[scale=0.13]{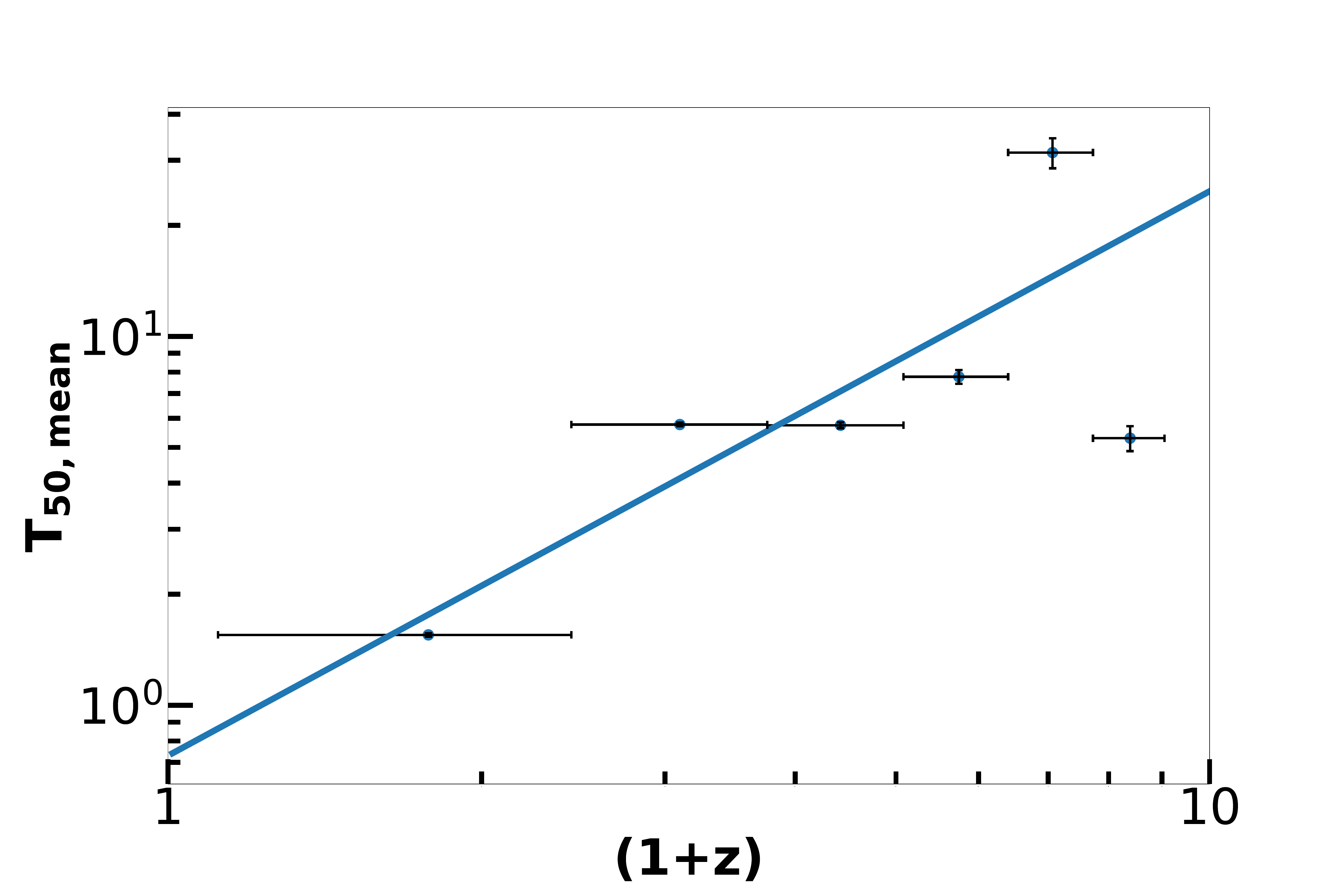}
    \includegraphics[scale=0.13]{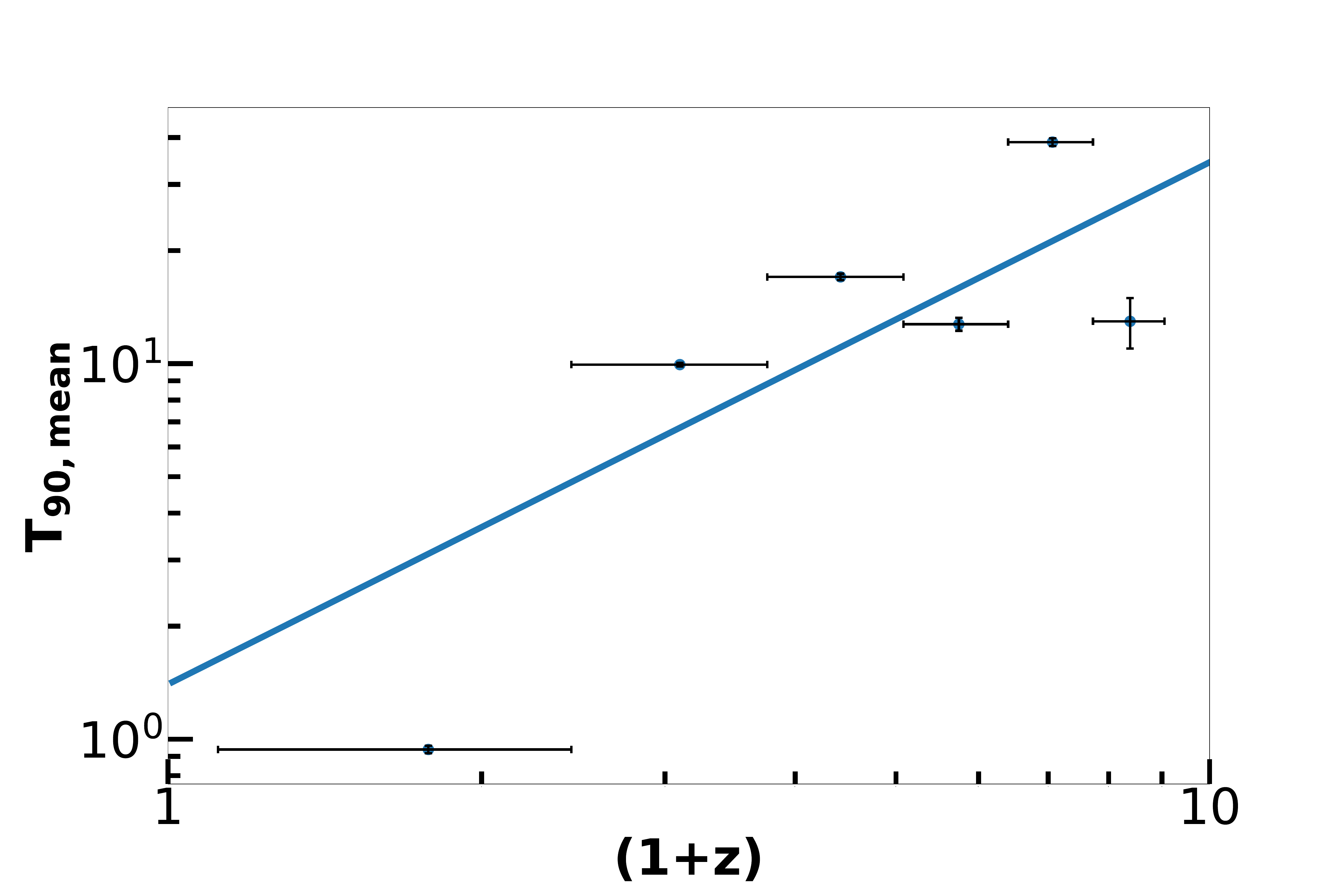}

    \caption{Plots showing weighted means of burst intervals using only the long GRB sample, with the binned redshift data (containing 6 groups equally spaced in $z$), where the horizontal error bars denote the redshift range in each group. . The blue line is the best fit curve obtained through the estimation of parameters $A$ and $B$ (cf. Eq.~\ref{equation_model}). We obtain $T_{50}=\rthis{0.73}(1+z)^{\rthis{1.53}}$ and $T_{90}=\rthis{1.40}(1+z)^{\rthis{1.39}}$.}
    \label{fig:weighted_mean_T90>2sec}
\end{figure*}

\item {\bf Analysis using geometric mean binned data for all GRBs}

\label{sec:geometric_mean_analysis}
Ref.~\cite{Butler} pointed out that any weighted mean would be affected by outliers and one should instead use more robust estimates such as geometric mean and median. For this purpose, we also redo our search for cosmological time dilation analysis by  considering the geometric mean of $T_{50}/T_{90}$ in each redshift bin. The contours for $A$ and $B$ after such a regression analysis can be found  in Fig.~\ref{fig:GM_contours_T50} and Fig.~\ref{fig:GM_contours_T90} for $T_{50}$ and  $T_{90}$ respectively. For  $T_{50}$ the intrinsic scatter we get from this fit is $<\rthis{60}\%$. For $T_{90}$,
the intrinsic scatter which get is  $\ln(\sigma_{int})=\rthis{1.84}^{+0.55}_{-1.33}$.
Although the best-fit value of the  scatter is greater than 100\%, it is is consistent with 27\% to within 68\% c.l. Therefore, we get the tightest scatter using the geometric mean for both $T_{50}$ and $T_{90}$. The power-law exponent ($B$) we get  is consistent with a cosmological signature, for  $T_{50}$ and $T_{90}$ to within 1$\sigma$. The burst intervals as a function of redshift ($1+z$) are shown  in Fig.~\ref{fig:geometric_mean_burst_intervals} along with the best-fit.
\end{itemize}

The results of all our regression analyses using  both the unbinned as well as binned data are summarized  in  Tables~\ref{tab:unbinned_analysis} and \ref{tab:binned_analysis}, respectively.
For the unbinned analyses, the intrinsic scatter is $>100$\%, which implies that we cannot draw any definitive conclusion between durations and redshifts.
Our main result is that only after using the geometric mean based binned data  (full sample)\footnote{We have not done a  geometric mean-based analysis on only long GRBs, but we would expect our conclusions to be the same as those with the full sample.} and weighed mean based binned data (long GRBs),   we find evidence for cosmological time dilation for both these GRB durations. Among the binned analyses the data obtained using geometric mean show the smallest scatter.  Therefore, we concur with Z13, that GRBs show evidence for cosmological time dilation only in a statistical sense, but not on a per-GRB basis. 
\begin{figure*}
    \centering
    \includegraphics[scale=0.5]{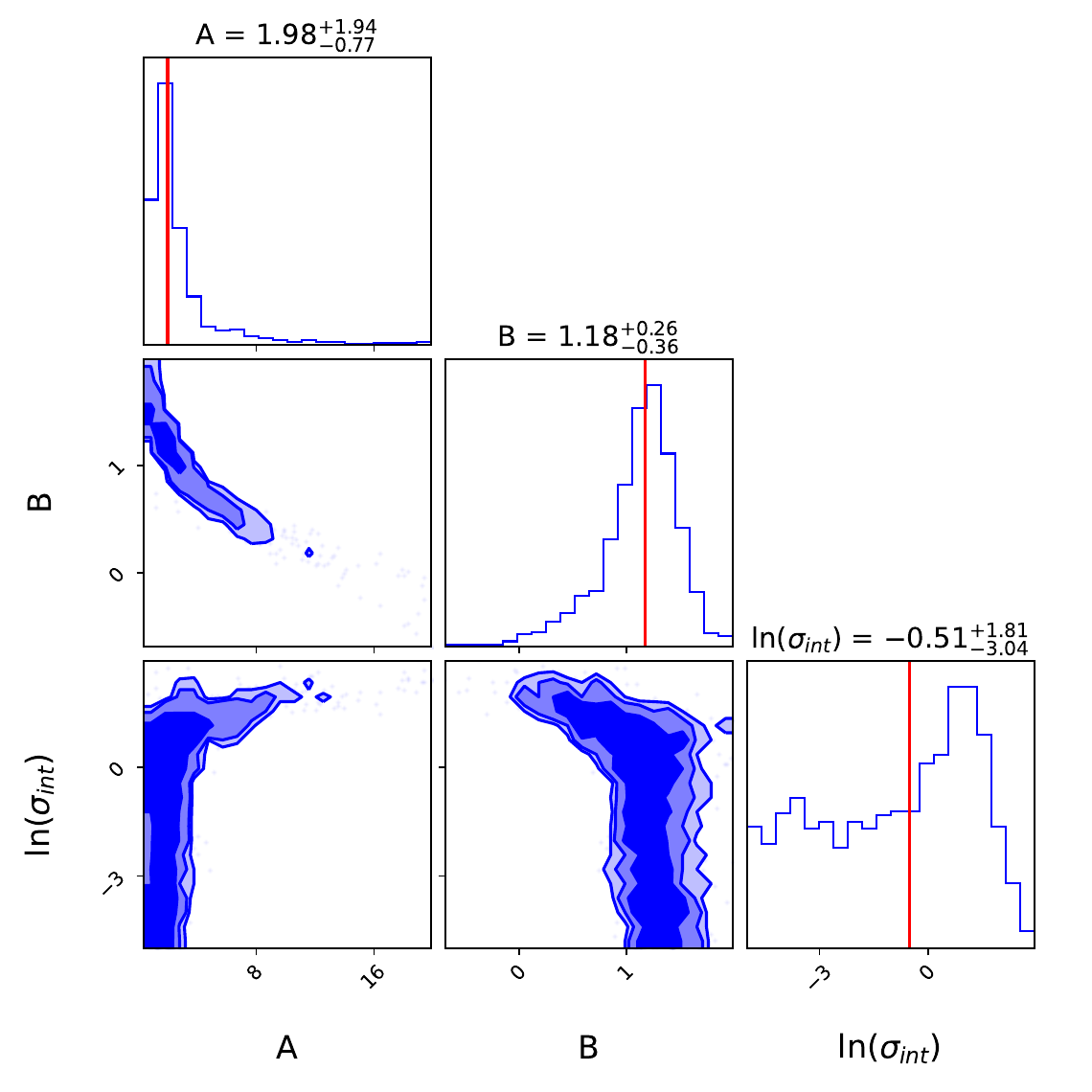}
    \caption{Contours for 68\%, 90\%, and 95\% credible intervals $T_{50}$  in the rest frame energy range of 140-350 keV, using binned data where the geometric mean of $T_{50}$ in each bin was used. The intrinsic scatter ($\sigma_{int}$) is very low,
    indicating that there is a deterministic  relationship between binned durations and the redshifts. The value of $B$ is consistent with the cosmological time dilation signature. }
    \label{fig:GM_contours_T50}
\end{figure*}

\begin{figure*}
    \centering
    
    \includegraphics[scale=0.5]{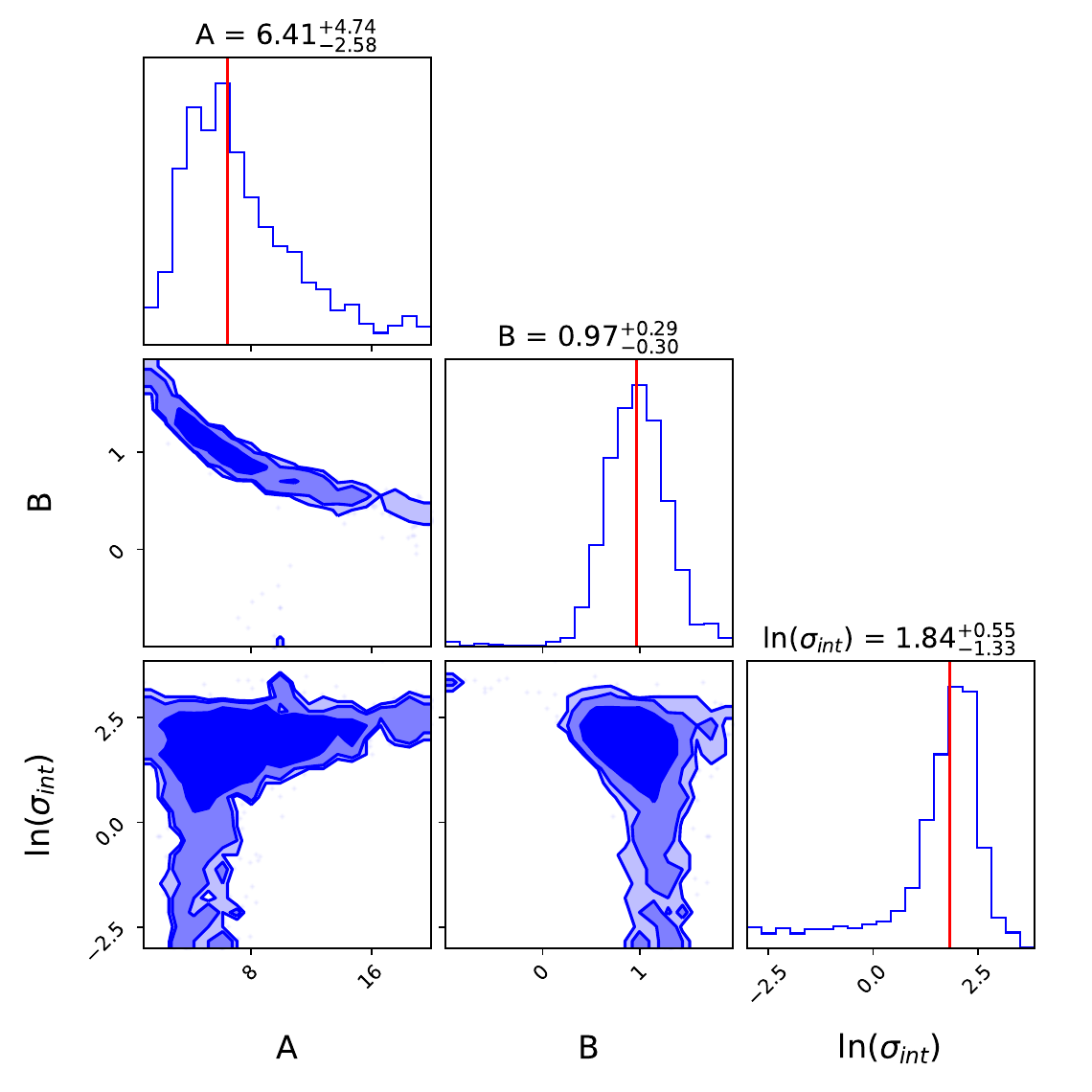}

    \caption{Contours showing 68\%, 90\%, and 95\% credible intervals for $A$ and $B$ for  $T_{90}$ in the rest frame energy range of 140-350 keV, using binned data where  geometric mean  of $T_{90}$ in each bin was used. Although the median intrinsic scatter is greater than 100\%, it is consistent 
    within 1$\sigma$. However, the value of $B$ is consistent with the cosmological time dilation signature to within 1$\sigma$. } 
   
    \label{fig:GM_contours_T90}
\end{figure*}

\begin{figure*}
    \centering
    \includegraphics[scale=0.15]{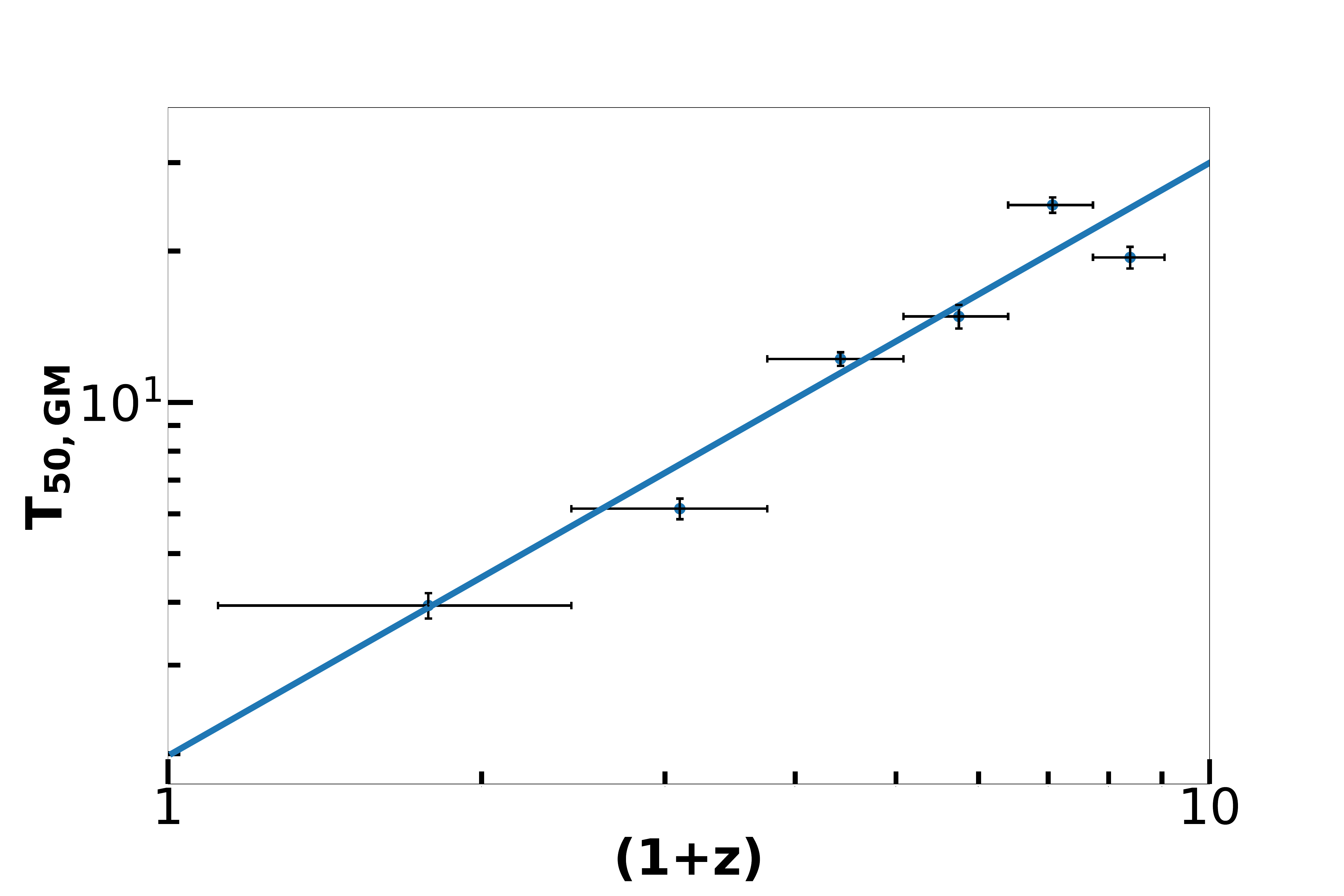}
    \includegraphics[scale=0.15]{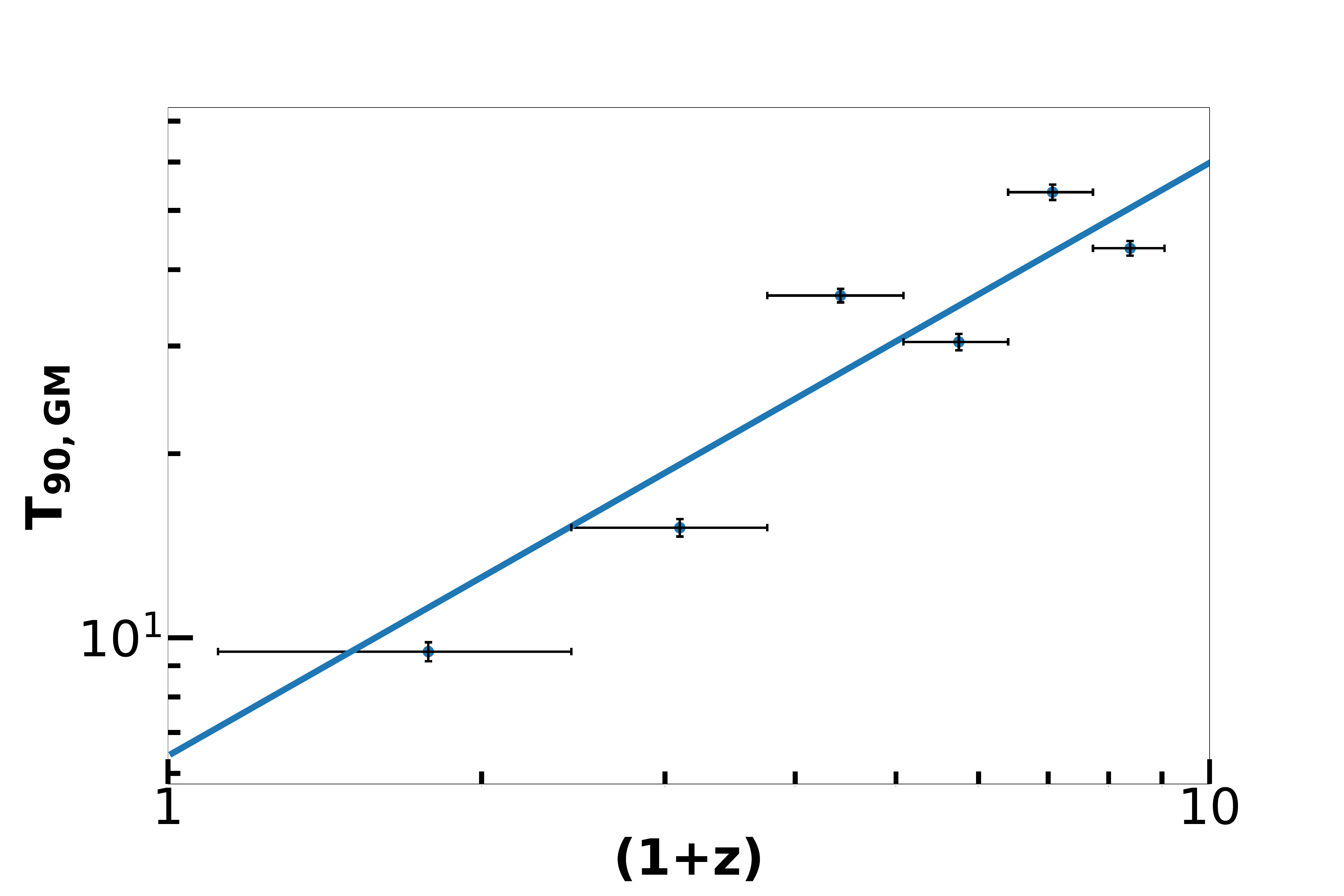}

    \caption{Plots showing geometric means of burst intervals with the binned redshift data (containing 6 groups equally spaced in z), where the horizontal error bars denote the redshift range in each group. The blue line is the best fit curve obtained through the estimation of parameters $A$ and $B$ (cf. Eq.~\ref{equation_model}). We obtain $T_{50}=\rthis{1.98}(1+z)^{\rthis{1.18}}$ and $T_{90}=\rthis{6.41}(1+z)^{\rthis{0.97}}$.}
    \label{fig:geometric_mean_burst_intervals}
\end{figure*}






\begin{table}
    \centering
    \caption{Table showing a summary of best fit values of $A$, $B$, and the logarithm scatter ($\sigma_{int}$) for different kinds of analysis for unbinned GRB datasets.}
    
    \begin{tabular}{|c|c|c|c|c|c|c|}

    \hline
    \renewcommand{\arraystretch}{5}
    Dataset & $A_{T_{50}}$ & $B_{T_{50}}$ & $\ln\sigma_{int,T_{90}}$ & $A_{T_{90}}$ & $B_{T_{90}}$ & $\ln\sigma_{int,T_{90}}$\\
    \hline
    
  Full Sample &  $\rthis{11.01}^{+2.74}_{-2.53}$ & $\rthis{0.66}^{+0.17}_{-0.17}$ & $\rthis{3.42}^{+0.5}_{-0.6}$  &  $\rthis{32.96}^{+7.82}_{-6.66}$ & $\rthis{0.52}^{+0.15}_{-0.16}$  & $\rthis{4.24}^{+0.05}_{-0.05}$   \\
    
    Long GRBs  &   $\rthis{12.01}^{+3.14}_{-2.76}$ & $\rthis{0.62}^{+0.18}_{-0.17}$ & $\rthis{3.44}^{+0.06}_{-0.05}$ & $\rthis{35.86}^{+7.30}_{-7.15}$ & $\rthis{0.47}^{+0.15}_{-0.14}$ &  $\rthis{4.25}^{+0.06}_{-0.05}$ \\
    
    \hline
    \end{tabular}

    \label{tab:unbinned_analysis}
\end{table}

\begin{table}
    \centering
    \caption{Table showing a summary of best fit values of $A$, $B$, and the logarithm of intrinsic scatter ($\sigma_{int}$) for different kinds of analysis for binned GRB datasets. Note that the first two rows use the weighted mean in each redshift bin. The values of $B$ are consistent with a cosmological time dilation signature for long GRBs and using geometric mean}
    \begin{tabular}{|c|c|c|c|c|c|c|}
    \hline
    Dataset & $A_{T_{50}}$ & $B_{T_{50}}$ & $\ln \sigma_{int,T_{50}}$ & $A_{T_{90}}$ & $B_{T_{90}}$ & $\ln\sigma_{int,T_{90}}$ \\
    \hline
    
    Full Sample &  $\rthis{0.40}^{+0.62}_{-0.24}$ & $\rthis{1.79}^{+0.55}_{-0.54}$ &  $\rthis{-0.27}^{+0.55}_{-0.41}$   &    $\rthis{0.74}^{+0.61}_{-0.38}$ & $\rthis{1.74}^{+0.45}_{-0.36}$ & $\rthis{-0.40}^{+0.49}_{-0.36}$\\
    
    Long GRBs & $\rthis{0.73}^{+0.75}_{-0.36}$ & $\rthis{1.53}^{+0.42}_{-0.44}$ &  $\rthis{-0.42}^{+0.49}_{-0.36}$  & $\rthis{1.40}^{+0.91}_{-0.62}$ & $\rthis{1.39}^{+0.35}_{-0.31}$ & 	$\rthis{-0.40}^{+0.55}_{-0.39}$ \\
    
    Geometric mean binning & $\rthis{1.98}^{+1.94}_{-0.77}$ & $\rthis{1.18}^{+0.26}_{-0.36}$ &  $<\rthis{-0.51}$   &  $\rthis{6.41}^{+4.74}_{-2.58}$ & $\rthis{0.97}^{+0.29}_{-0.30}$ & $\rthis{1.84}^{+0.55}_{-1.33}$ \\
    \hline
    \end{tabular}

    \label{tab:binned_analysis}
\end{table}

\subsection{Tests with rest frame durations}
\begin{figure}
    \centering
    \includegraphics[scale=0.15]{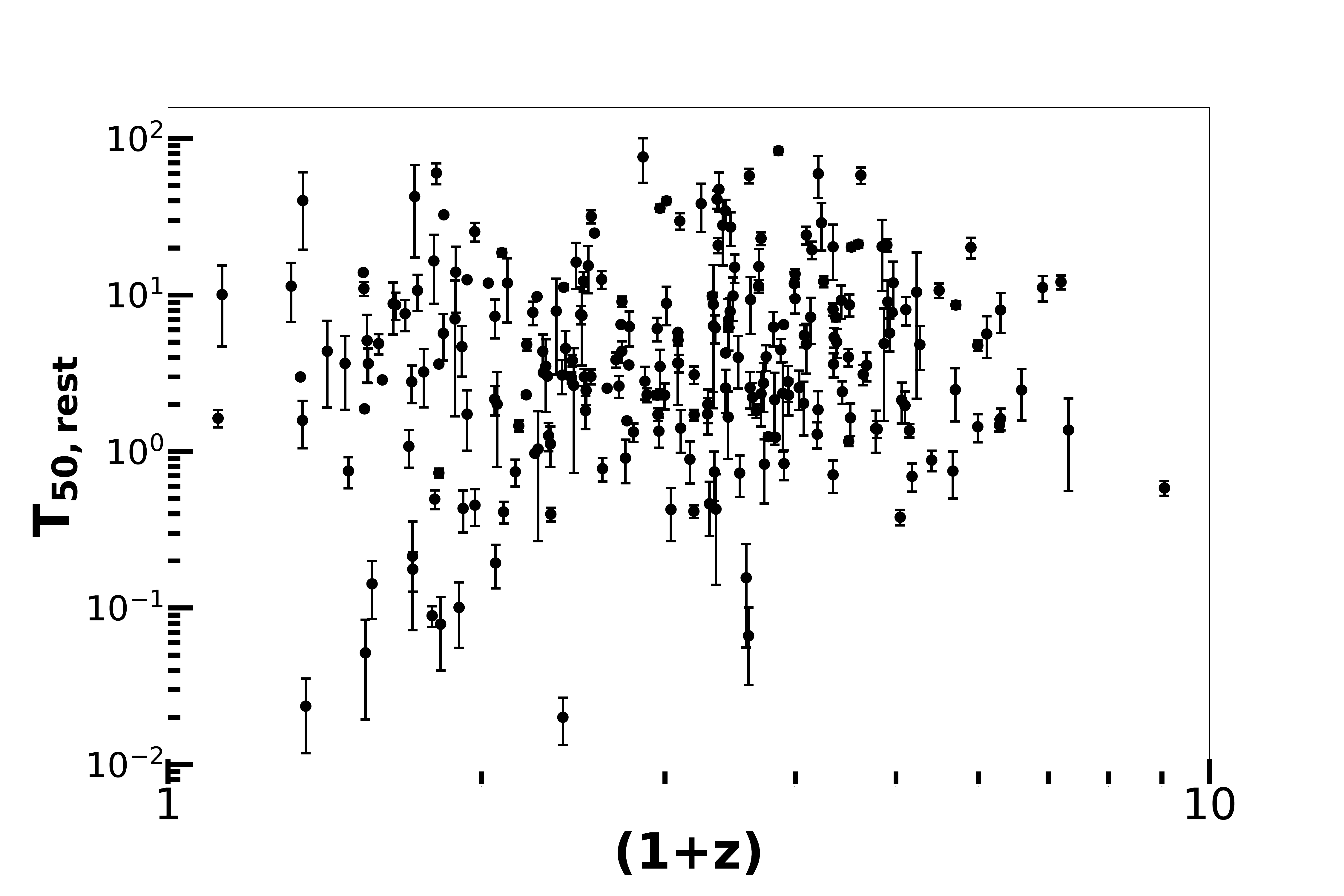}
    \includegraphics[scale=0.15]{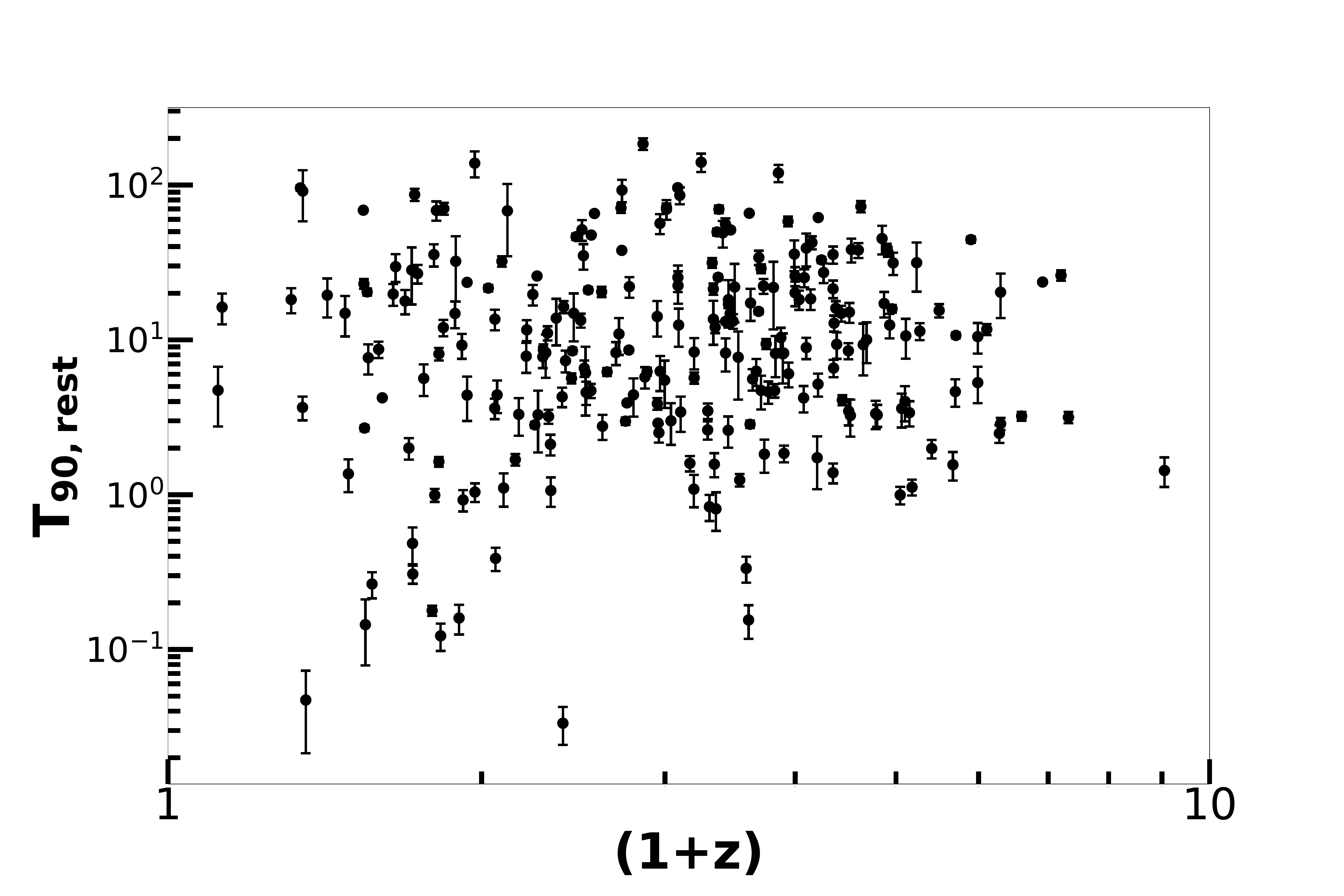}
    \caption{Plots showing the correlations between $T_{50,rest}$ and $T_{90,rest}$ in the rest frame of GRBs with the redshifts. The rest frame durations are obtained by dividing the $T_{50}/T_{90}$  tabulated in the last two columns of Table~\ref{Table1:Swift_GRBs} by ($1+z$).}
    \label{fig:rest_frame_raw_burst_intervals}
\end{figure}

Similar to Z13, we carry out some additional tests to completely ascertain the redshift evolution of the intrinsic GRB duration, as  found from some of the binned analyses. We calculate the rest frame $T_{90,rest}$ and  from the $T_{90}$ determined earlier using $T_{90,rest} = T_{90}/(1+z)$ (and same for $T_{50}$). These rest-frame durations are  shown in Fig.~\ref{fig:rest_frame_raw_burst_intervals}. Similar to Z13, we find a large dynamic range in the durations with median values of $3.8$ and $10$ seconds for $T_{50,rest}$ and $T_{90,rest}$, respectively, while the standard deviations are $\rthis{13}$ and $27$ seconds for $T_{50,rest}$ and $T_{90,rest}$ respectively. We also checked for an evolution effect of the rest frame durations with redshifts using the same procedure as done for $T_{50}$ and $T_{90}$ earlier. The best-fit values for $B$ are equal to  $0.11^{+0.20}_{-0.19}$ and $-0.11^{+0.18}_{-0.17}$ for $T_{50,rest}$ and $T_{90,rest}$ respectively. Therefore, the values of $B$ are significantly different from that expected from cosmological time dilation ($B=1$). Furthermore, the logarithm of the intrinsic scatter is given by $\ln (\sigma_{int})$ is equal to  $3.12^{+0.05}_{-0.05}$ and $2.17^{+0.06}_{-0.05}$ for $T_{50,rest}$ and $T_{90,rest}$ respectively, implying that the intrinsic scatter is greater than 100\%. Therefore,  This implies that there is no evolution of the rest frame durations, in accord with Z13's conclusions.

\begin{figure}
    \centering
    \includegraphics[scale=0.15]{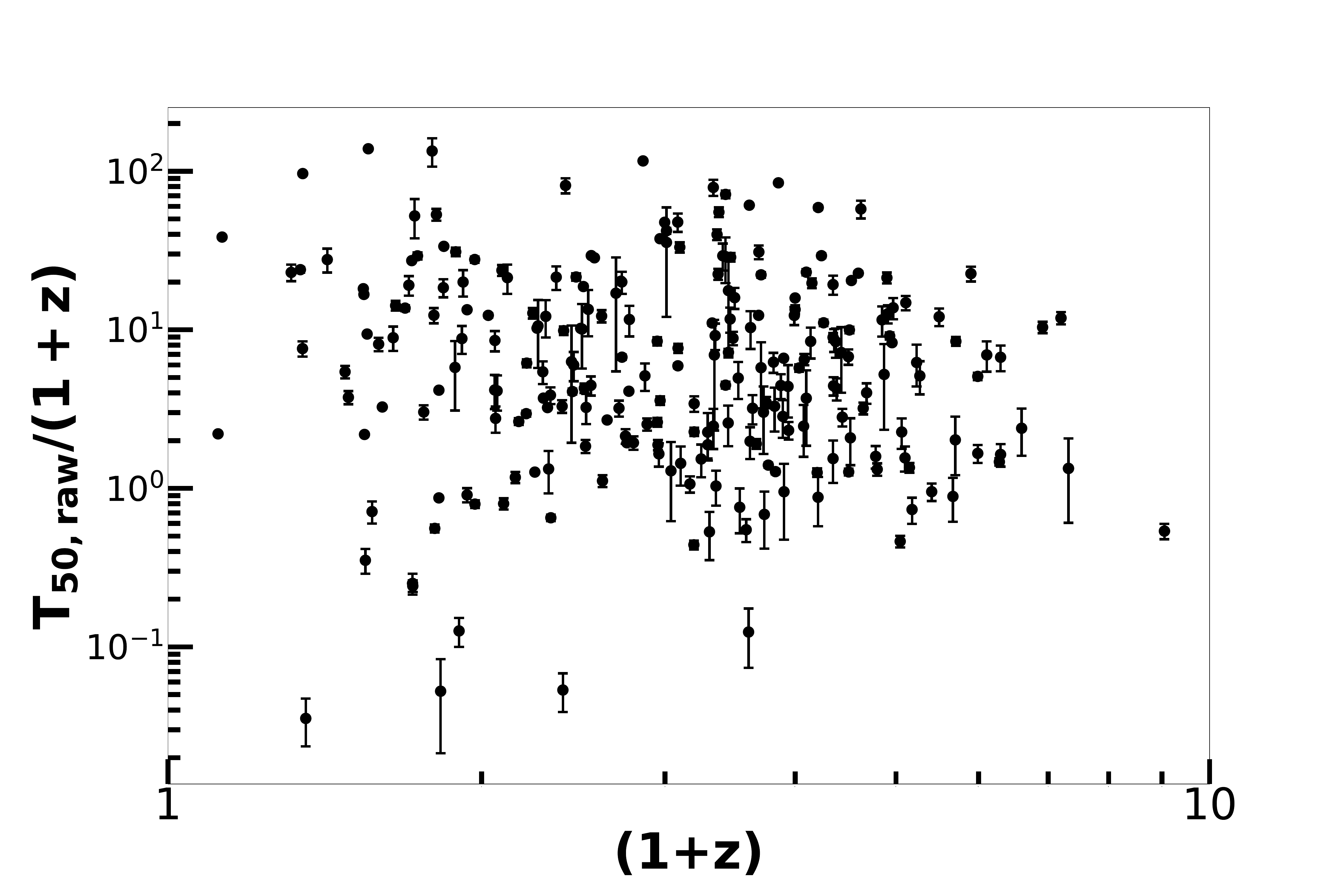}
    \includegraphics[scale=0.15]{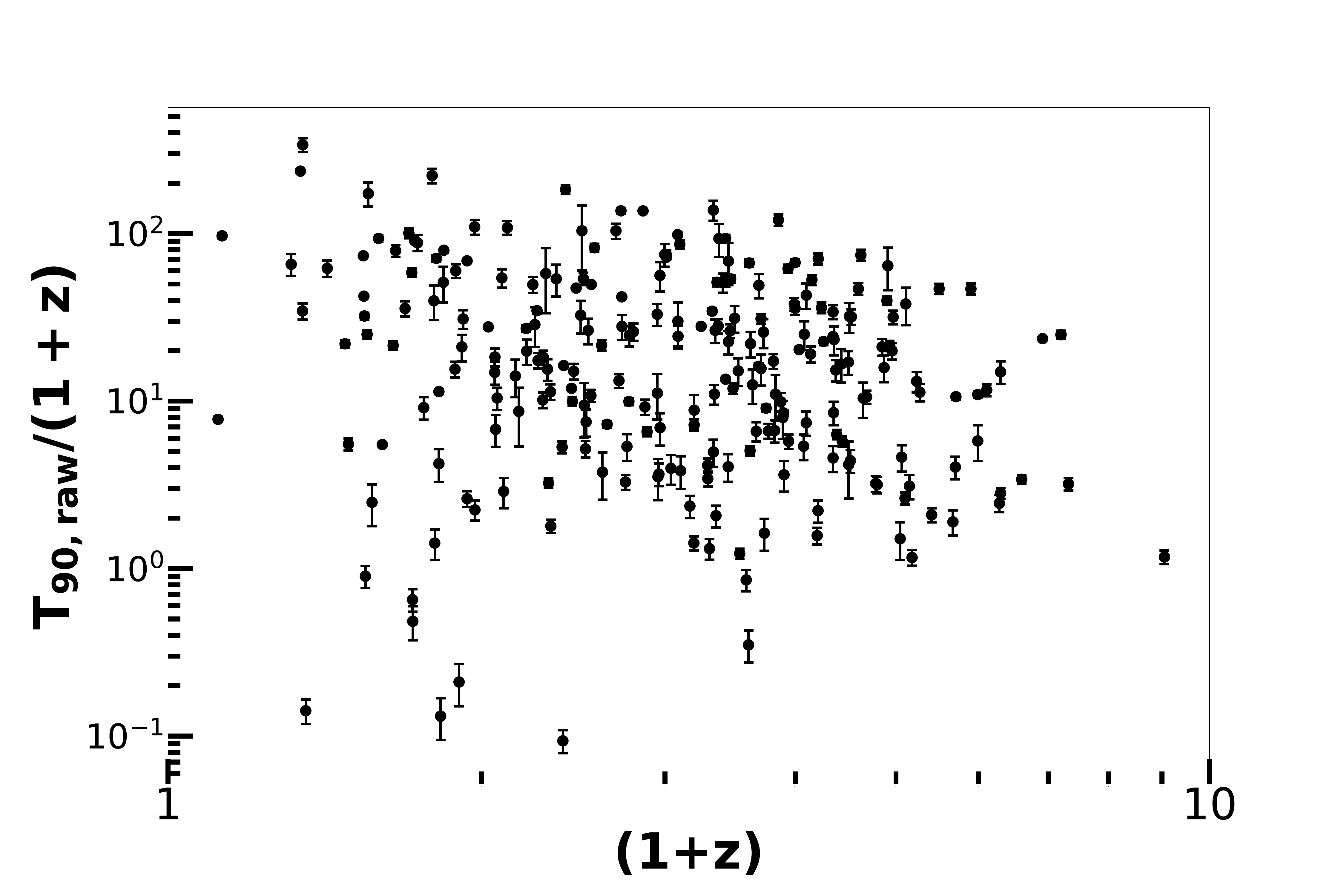}
    \caption{Plots showing the correlations between the durations in the rest frame obtained using  $T_{90,raw}/({1+z})$,  (where ``raw'' indicates data measured between 15-350 keV in the observer frame), and the redshifts. All symbols carry the usual meaning.}
    \label{fig:T_raw_upon_(1+z)_frame_raw_burst_intervals}
\end{figure}

 Finally, we  check for any possible redshift evolution of rest-frame duration estimated directly from the raw durations (i.e. $T_{50,raw}/(1+z)$) as a function to redshift to independently confirm if we see a decreasing trend as found in some previous works~\cite{Pelangeon,Petrosian13}. These $T_{50,raw}/(1+z)$ and $T_{90,raw}/(1+z)$ plots can be found in Fig.~\ref{fig:T_raw_upon_(1+z)_frame_raw_burst_intervals}. 
 We find that $B$ is equal to $-0.59^{+0.21}_{-0.21}$ and $\rthis{-0.82}^{+0.13}_{-0.10}$
for  $T_{50,raw}/(1+z)$ and $T_{90,raw}/(1+z)$, respectively. Furthermore, the logarithm of the intrinsic scatter is given by $\ln (\sigma_{int})$ which is equal to  $2.91^{+0.04}_{-0.05}$ and $3.55^{+0.05}_{-0.05}$. Therefore, we again concur with Z13, that there is no evidence that the  measure of rest frame duration analyzed in ~\cite{Pelangeon,Petrosian13} shows any evolution with redshift.

\section{Conclusions}
\label{sec:conclusions}

Even though we know for more than two decades that almost all GRBs are located at cosmological distances, the cosmological time dilation signature for GRBs has not been unequivocally demonstrated. 
Therefore, we independently carry out a study similar to Z13 (and also ~\cite{Butler}), using the latest updated GRB sample from the Swift satellite. We estimated $T_{50}$ and $T_{90}$ for all the  GRBs in the Swift catalog, for the energy band of 140-350 keV in the rest frame. This corresponds to an energy band of $140/(1+z)$ and $350/(1+z)$ in the observer frame. These $T_{50}$ and $T_{90}$ intervals calculated for all the Swift GRBs are tabulated  in Table~\ref{Table1:Swift_GRBs}. 

We then carry out a regression analysis between the duration and redshifts using $T_{90}$($T_{50}$) = $A(1+z)^{B}$ using both the unbinned and binned durations, where the binning was done using both a error weighted-mean  as well as geometric-mean based average.
We also analyzed the full sample sample as well as only the long GRB sample. For a cosmological time dilation signature, $B$ should be equal to one. 
Our results from all these myriad analyses are summarized in Table~\ref{tab:unbinned_analysis} and Table~\ref{tab:binned_analysis} for the unbinned and the binned data, respectively.  
We conclude that the burst intervals obtained from the geometric mean-based  and weighted mean-based (long GRBs) binned  analyses are consistent with the cosmological time dilation phenomenon (within $1\sigma$) for both $T_{50}$ and $T_{90}$. (cf. Table~\ref{tab:binned_analysis}). The smallest intrinsic scatters are obtained using the geometric mean based binned data. The weighted mean analysis
for the full sample is discrepant with respect to the cosmological time signature at 1.2$\sigma$ and 2$\sigma$ for $T_{50}$ and $T_{90}$, respectively. The unbinned analyses show intrinsic scatters $>$ 100\% as well as values of $B$ significantly different from $1$, implying that no definitive conclusion between the durations and the redshifts can be inferred.

\rthis{Finally, we should point out that although choosing a fixed energy frame interval may avoid the biases associated  by recording different parts of the GRB intrinsic light curves, our results could still be affected by detector related selection biases. As Z13 pointed out for GRBs at higher redshifts and low SNRs, only the brightest GRBs would be detected and some of the observations could be underestimated. Secondly, the BAT effective area is energy dependent and drops sharply at energies less than 25 keV and greater than 100 keV~\cite{Barthelmy}. Another possibility, pointed out in ~\cite{Butler} is that the measured duration of a GRB could also be affected by the gradual loss of the final fast-rise exponential decay  pulse tail due to the degraded
signal-to-noise ratio~\cite{Littlejohns}. A detailed modeling of these effects is beyond the scope of this work. However in a future work, we shall be using the deconvolved photon fluxes for estimating $T_{90}$, which could get rid of these instrumental effects~\cite{von}. }

All the codes which are utilized in this analysis, including the {\tt batbinevt} and {\tt battblocks} code generators have been made publicly available and can be found at  \href{https://github.com/Amiteshh/GRB-Analysis}{GRB Analysis Code (Github)}.

\section*{ACKNOWLEDGEMENT}
We are grateful to I-Non Chiu for useful correspondence about this work. We thank Amy Lien  for help regarding Swift GRB database and {\tt battblocks} command. We would also like to acknowledge Eleonora Troja,  for  assistance regarding GRB redshift data. We also appreciate the  Swift Help team for their prompt response to all our queries. \rthis{Finally, we are grateful to the anonymous referee for several constructive feedback on this manuscript.}

\appendix

\section{Data retrieval and light curve analysis}
\label{app}
\begin{figure*}
\centering
\includegraphics[scale=0.5]{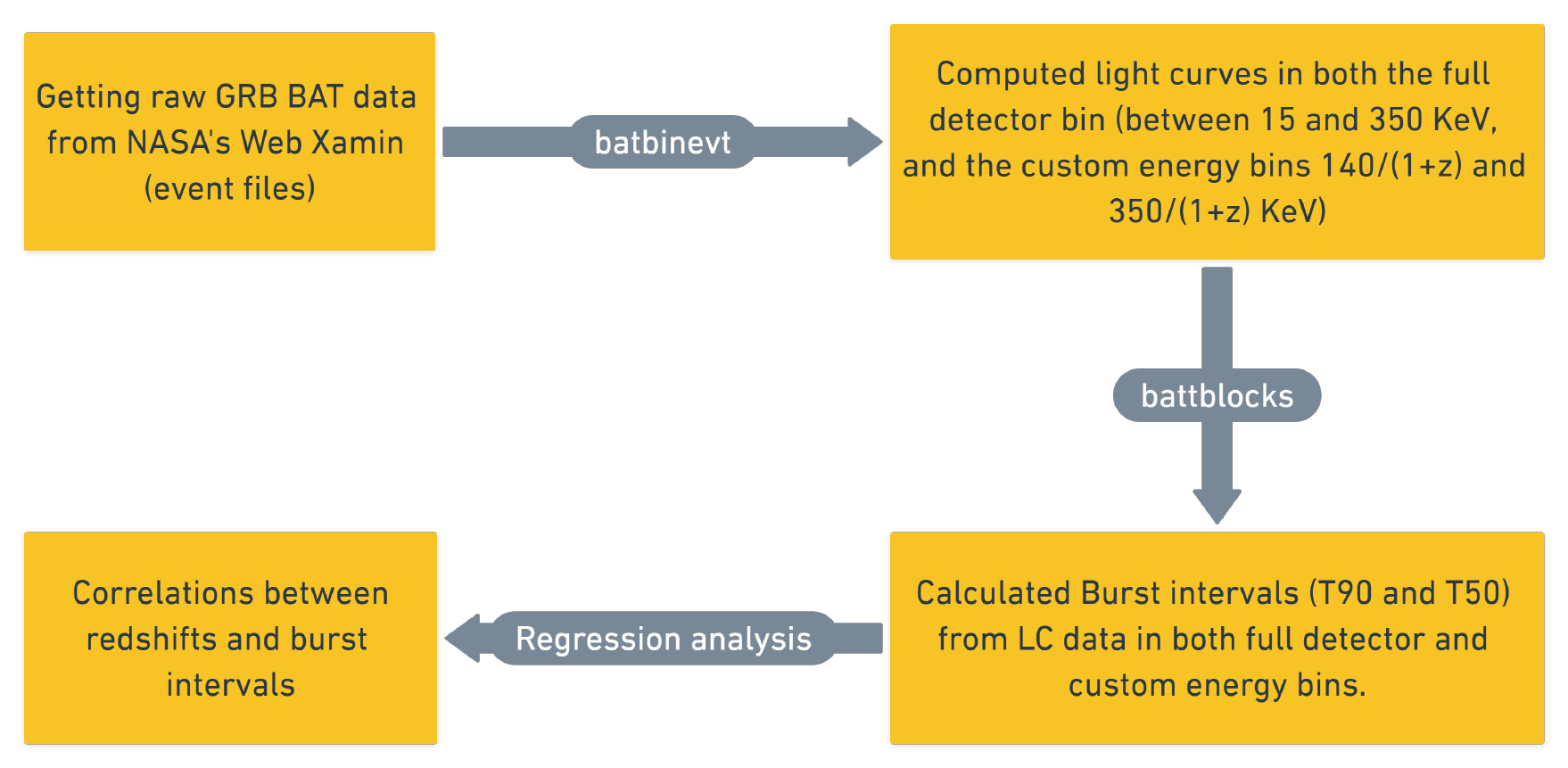}
\caption{A brief data processing pipeline for our analysis of the GRB samples obtained from the latest Swift catalog. The yellow boxes denote the type of products obtained and the black pointers represent the type of command used to generate those products.}
\label{fig:data_processing_pipeline}
\end{figure*}

We now describe the detailed step-by-step procedure for downloading the Swift light curve data and estimating $T_{50}$/$T_{90}$ for every GRB. Fig.~\ref{fig:data_processing_pipeline} demonstrates a brief workflow of this processing pipeline, starting from raw GRB event data to the calculation of the burst intervals.
We have sampled 247 of the 400 GRBs which were eligible for the analysis (containing proper redshift data). The remaining  153 GRBs could not be processed, because most  of the data products (101) for the remaining GRBs did not contain any TDRSS messages, and many (52) did not have data products with enough exposure to construct detailed Bayesian blocks, needed for the estimation of durations.

We use the Burst interval data recorded by the Burst Alert Telescope (BAT) aboard  Swift. The BAT instrument has two basic modes of operation: 1) scan-survey mode and 2) burst mode.
These two modes reflect the two major types of data that BAT produces: hard X-ray survey data and burst positions. Most of BAT’s time is spent in waiting for a burst to occur in its FOV. Finding GRBs within BAT consists of two processes: 1) the detection of the onset of a burst by looking for increases in the event rate across the detector plan, and 2) the formation of an image of the sky using the events detected during the time interval at the beginning of the burst. \cite{markwardt2007Swift}

{\tt Burst Trigger Algorithm}:- The burst trigger algorithm looks for detector count rate excesses that are higher than those predicted from the background and constant sources. The fluctuations in the background and the heterogeneity of the GRB time profiles are the two fundamental barriers in GRB detection. During the 90-minutes the satellite stays in Low Earth Orbit, the detector background rates can change by more than a factor of two. GRBs can last anywhere from a few milliseconds to several minutes, with anywhere from one to several dozen peaks in the emission. As a result, the triggering system must be capable of extrapolating the background and comparing it to the recorded detector count rate over a wide range of timescales and energy bands~\cite{markwardt2007Swift}.

NASA maintains an updated list of all the  BAT observations at the  Swift Archive~\cite{SwiftArchive}. We use {\tt Web Xamin}~\cite{XaminWeb}, which acts as a source for downloading or saving the BAT GRB data for analysis, and {\tt Web Hera}~\cite{HeraWeb} (containing the HEASOFT tools from NASA) to analyze the BAT data. {\tt HEASOFT} tools provided at \url{https://heasarc.gsfc.nasa.gov/docs/tools.html} work on a Linux machine but this analysis has explicitly made use  of the Web Hera interface, owing to its simplicity and minimal computational resources at the user's end. The whole analysis can be classified into the following stages in chronological order:

\begin{itemize}
\item {\bf Querying for and collecting the full GRB data products at the Swift archive through the {\tt Web Xamin} interface}

For this access the {\tt Web Xamin} at \url{https://heasarc.gsfc.nasa.gov/xamin/}. 

We have to focus on the Query Pane for our queries. 

After we have logged in, we can obtain a table of our interest through the {\tt Table Explorer} option. Now there are multiple options to get the data.  We can navigate through the folders in the {\tt Table Explorer} to get the whole catalogue or search a GRB by its name (ex- GRB 050126) or its coordinates in the Target for search. To get the whole catalogue, we would go to {\tt Popular Missions} $\rightarrow$ {\tt Swift} $\rightarrow$ {\tt Swiftbalog} to obtain the data for all the GRBs observed till date. Or if we search for a particular GRB, we can type the name and then click on ‘search on all HEASARC tables’ to run our query. For example, if we want to  search and download the data for GRB 050319, then when we query for the GRB, we can find all the summary   tables related to the GRB. 

In this case too, out of all other tables, we just need the {\tt Swiftbalog} table for getting the BAT GRB data. When we open the {\tt Swiftbalog} table, we get a list of multiple data products. To figure out which event data has data products available, we need to look at the data products; they must contain a folder named as {\tt ‘TDRSS’}. This would most probably be one of the long exposure files (1000-2900 seconds). This is necessary because {\tt batgrbproduct} will only work on those data products that contain this folder. Then, once we have found that particular data product, we can select it and click on ‘Save to Hera’. This will save the data product to Web Hera, where we can further analyse the BAT GRB data.

\item {\bf GRB Analysis in Web HERA to calculate the observer frame energy binned and full detector range light curves}

To run any command, we can use either the Command Window Terminal or the Tool Parameter. For now, we will use Tool Parameter. Type {\tt ‘batgrbproduct’} and click on Get.

We then run {\tt batgrbproduct} keeping all the parameters to their default value. We just have to specify the input folder, where the raw data lies and another folder where the output files will be stored.

 We then go to the results folder and save the file process.log to our local storage or view it elsewhere; we will then find the whole set of operations that {\tt Hera} has performed using {\tt batgrbproduct} command in the process.log file. We then find out the {\tt batbinevt} command (used to prepare custom light curves in user-defined energy bins). A sample {\tt batbinevt} command with its parameters looks like:
\clearpage

\begin{lstlisting}[breaklines=true]

batbinevt infile='./00103780000-results/events/sw00103780000b_all.evt' outfile='./00103780000-results/lc/sw00103780000b_bb_4ms.lc' outtype=LC timedel=0.004 timebinalg=u energybins='15-350' detmask='./00103780000-results/auxil/sw00103780000b_qmap.fits' ecol=ENERGY weighted=YES outunits=RATE clobber=yes

\end{lstlisting}

Notice the parameters {\tt energybins} and {\tt timedel}. {\tt energybins} are used to specify the energy range, and {\tt timedel} is used to specify the $\Delta t$ (the time resolution in seconds). The keyword {\tt LC} denotes a light curve. These light curve files have an extension {\tt .lc}. For short GRBs, we use 16~ms binned light curves, while for long GRBs, we use 64~ms binned light curves (following the same convention as in ~\cite{Kouveliotou93}). This, along with the output name, can be changed to produce the required light curves.

To calculate the light curves for obtaining the raw $T_{50}$ and $T_{90}$ (full detector energy range of 15-350 keV), we just have to copy and paste the command, without changing the energy ranges, copy the command, and run it in Hera. This will create the custom light curves that will be required for raw $T_{90}$ calculation (in the folder light curves).

\item \textbf{Calculation of burst intervals in rest frame and observer frame energy intervals}

For calculation of $T_{90}$ and $T_{50}$ in the rest frame energy interval of 140-350 keV, corresponding to a redshift-dependent  observed energy range, we have to alter the energy bins $E_{1}=140/(1+z)$) and $E_{2}=350/(1+z)$. $E_{1}$ is the lower limit of the observed energy range, and $E_{2}$ is the upper limit.
We make use of the newly created light curves to calculate estimated burst intervals ($T_{90}$ and $T_{50}$) in the different energy ranges using the Bayesian block algorithm \cite{scargle2013bayesian}.

Now using the custom light curve file that we created, we will run the {\tt battblocks} command. The input file would be the custom light curve file. We have to specify any desired name to the {\tt durfile} as well as {\tt output} file. Also, we have to make sure that we don't give an {\tt optional filter}, which implies that the field should be completely empty (without any text). Click on run {\tt battblocks}, and this should give the $T_{90}$ and $T_{50}$ for the GRB as the final result in the Hera command console. 
\end{itemize}

\clearpage
\begin{center}
\begin{longtable*}{|c|c|c|c|c|c|c|c|c|}
    \caption{Durations of 247 Swift GRBs with known redshifts evaluated in different energy bins. $T_{50}$ and $T_{90}$ are evaluated in rest frame energy interval of 140-350 keV and $T_{50,raw}$, $T_{90,raw}$ are evaluated in full detector range of BAT, 15-350 keV. \rthis{We note that we have omitted GRB 130610A and GRB 160410A, as they yielded completely different T90s in the observer frame compared to the SWIFT table. Also we omitted the GRBs 150403A, 180205A and 190106A because their data products could not be found on NASA's Swift database anymore.}}
    \label{Table1:Swift_GRBs}\\
    \hline
    \textbf{GRB Name}  &  \textbf{Obs ID} & \textbf{z} & $\mathbf{T_{90,raw}}$(s) & $\mathbf{T_{50,raw}}$(s) & $\mathbf{E^{*}_1}$(keV) & $\mathbf{E^{*}_2}$(keV) & $\mathbf{T_{90}}$(s) & $\mathbf{T_{50}}$(s)\\
    \hline 

050126	&	103780	&	1.29	&	$	23.248	\pm	2.53	$	&	$	12.432	\pm	2.036	$	&	61.14	&	152.84	&	$	17.808	\pm	2.771	$	&	$	10	\pm	2.815	$	\\
050315	&	111063	&	1.949	&	$	97.248	\pm	14.637	$	&	$	24.976	\pm	1.471	$	&	47.47	&	118.68	&	$	41.744	\pm	10.809	$	&	$	18	\pm	3.065	$	\\
050318	&	111529	&	1.44	&	$	29.056	\pm	0.472	$	&	$	15.344	\pm	10.616	$	&	57.38	&	143.44	&	$	13.776	\pm	1.018	$	&	$	7.216	\pm	0.588	$	\\
050319	&	111622	&	3.24	&	$	153.072	\pm	11.31	$	&	$	124.304	\pm	4.228	$	&	33.02	&	82.55	&	$	139.184	\pm	6.598	$	&	$	122.784	\pm	41.306	$	\\
050401	&	113120	&	2.9	&	$	33.12	\pm	1.037	$	&	$	25.792	\pm	0.588	$	&	35.90	&	89.74	&	$	32.032	\pm	0.77	$	&	$	25.232	\pm	0.497	$	\\
050505	&	117504	&	4.27	&	$	59.53	\pm	7.108	$	&	$	26.992	\pm	6.363	$	&	26.57	&	66.41	&	$	59.928	\pm	7.541	$	&	$	25.408	\pm	7.939	$	\\
050525A	&	130088	&	0.606	&	$	8.832	\pm	0.081	$	&	$	5.232	\pm	0.0226	$	&	87.17	&	217.93	&	$	6.784	\pm	0.136	$	&	$	4.608	\pm	0.086	$	\\
050730	&	148225	&	3.968	&	$	157.584	\pm	14.77	$	&	$	68.352	\pm	10.465	$	&	28.18	&	70.45	&	$	155.904	\pm	25.613	$	&	$	59.472	\pm	21.642	$	\\
050802	&	148646	&	1.52	&	$	18.928	\pm	3.494	$	&	$	8.16	\pm	1.765	$	&	55.56	&	138.89	&	$	11.536	\pm	1.981	$	&	$	6.208	\pm	1.216	$	\\
050803	&	148833	&	0.422	&	$	88.176	\pm	9.982	$	&	$	39.392	\pm	6.757	$	&	98.45	&	246.13	&	$	27.632	\pm	7.789	$	&	$	6.224	\pm	3.507	$	\\
050814	&	150314	&	5.3	&	$	94.208	\pm	14.575	$	&	$	42.256	\pm	7.78	$	&	22.22	&	55.56	&	$	127.936	\pm	40.738	$	&	$	50.544	\pm	14.466	$	\\
050820A	&	151207	&	2.6147	&	$	241.36	\pm	12.277	$	&	$	220.56	\pm	3.117	$	&	38.73	&	96.83	&	$	237.52	\pm	4.766	$	&	$	\rthis{229.298}	\pm	\rthis{22.092}	$	\\
050904	&	153514	&	6.2	&	$	179.5	\pm	10.3	$	&	$	85.4	\pm	7.6	$	&	19.44	&	48.61	&	$	187.744	\pm	14.269	$	&	$	87.216	\pm	8.845	$	\\
050908	&	154112	&	3.35	&	$	19.9	\pm	3.5	$	&	$	6.7	\pm	2	$	&	32.18	&	80.46	&	$	6.032	\pm	0.883	$	&	$	3.088	\pm	0.727	$	\\
050915A	&	155242	&	2.5273	&	$	53.424	\pm	10.044	$	&	$	17.488	\pm	4.585	$	&	39.69	&	99.23	&	$	27.232	\pm	12.728	$	&	$	14.096	\pm	5.21	$	\\
050922C	&	156467	&	2.198	&	$	4.544	\pm	0.442	$	&	$	1.408	\pm	0.09	$	&	43.78	&	109.44	&	$	3.472	\pm	0.821	$	&	$	1.328	\pm	0.124	$	\\
051001	&	157870	&	2.4296	&	$	178.656	\pm	13.787	$	&	$	99.248	\pm	31.747	$	&	40.82	&	102.05	&	$	191.52	\pm	17.696	$	&	$	118.464	\pm	20.638	$	\\
051006	&	158593	&	1.059	&	$	30.56	\pm	4.81	$	&	$	\rthis{8.582}	\pm	\rthis{2.079}	$	&	67.99	&	169.99	&	$	7.456	\pm	1.129	$	&	$	4.448	\pm	0.936	$	\\
051109A	&	163136	&	2.346	&	$	36.8	\pm	4.959	$	&	$	23.2	\pm	15.454	$	&	41.84	&	104.60	&	$	5.264	\pm	0.931	$	&	$	2.48	\pm	0.867	$	\\
051221A	&	173780	&	0.547	&	$	1.392	\pm	0.21	$	&	$	0.544	\pm	0.097	$	&	90.50	&	226.24	&	$	0.224	\pm	0.102	$	&	$	0.08	\pm	0.05	$	\\
060115	&	177408	&	3.53	&	$	144.496	\pm	15.553	$	&	$	92.592	\pm	2.848	$	&	30.91	&	77.26	&	$	173.568	\pm	30.4185	$	&	$	91.792	\pm	4.46	$	\\
060124	&	178750	&	2.297	&	$	13.632	\pm	1.312	$	&	$	7.456	\pm	2.368	$	&	42.46	&	106.16	&	$	8.64	\pm	1.169	$	&	$	5.728	\pm	1.504	$	\\
060206	&	180455	&	4.046	&	$	7.6	\pm	1.919	$	&	$	2.336	\pm	0.195	$	&	27.74	&	69.36	&	$	5.024	\pm	0.653	$	&	$	1.92	\pm	0.215	$	\\
060210	&	180977	&	3.91	&	$	315.596	\pm	90.034	$	&	$	62	\pm	8.17	$	&	28.51	&	71.28	&	$	\rthis{181.432}	\pm	\rthis{13.453}	$	&	$	44.4	\pm	16.55	$	\\
060223A	&	192059	&	4.41	&	$	11.312	\pm	1.087	$	&	$	5.152	\pm	0.653	$	&	25.88	&	64.70	&	$	10.752	\pm	1.472	$	&	$	4.768	\pm	0.715	$	\\
060418	&	205851	&	1.49	&	$	\rthis{80.977}	\pm	\rthis{17.933}	$	&	$	25.408	\pm	0.736	$	&	56.22	&	140.56	&	$	33.296	\pm	3.425	$	&	$	18.704	\pm	2.479	$	\\
060502A	&	208169	&	1.51	&	$	23.712	\pm	8.501	$	&	$	10.752	\pm	0.723	$	&	55.78	&	139.44	&	$	16.464	\pm	2.016	$	&	$	7.536	\pm	0.949	$	\\
060510B	&	209352	&	4.9	&	$	275.984	\pm	20.717	$	&	$	133.28	\pm	14.292	$	&	23.73	&	59.32	&	$	262.096	\pm	12.277	$	&	$	119.104	\pm	18.044	$	\\
060522	&	211117	&	5.11	&	$	71.12	\pm	5.774	$	&	$	42.4	\pm	9.166	$	&	22.91	&	57.28	&	$	71.44	\pm	5.921	$	&	$	34.464	\pm	10.305	$	\\
060526	&	211957	&	3.21	&	$	298.08	\pm	22.499	$	&	$	248.368	\pm	8.458	$	&	33.25	&	83.14	&	$	259.392	\pm	5.641	$	&	$	\rthis{251.125}	\pm	\rthis{75.339}	$	\\
060605	&	213630	&	3.78	&	$	15.392	\pm	1.664	$	&	$	7.6	\pm	1.233	$	&	29.29	&	73.22	&	$	15.984	\pm	3.32	$	&	$	6.704	\pm	2.02	$	\\
060607A	&	213823	&	3.082	&	$	102.32	\pm	20.047	$	&	$	26.528	\pm	2.173	$	&	34.30	&	85.74	&	$	102.928	\pm	13.706	$	&	$	22.56	\pm	3.622	$	\\
060614	&	214805	&	0.127	&	$	109.264	\pm	3.415	$	&	$	43.296	\pm	0.79	$	&	124.22	&	310.56	&	$	18.32	\pm	4.113	$	&	$	11.344	\pm	6.051	$	\\
060707	&	217704	&	3.43	&	$	73.728	\pm	16.668	$	&	$	31.856	\pm	14.128	$	&	31.60	&	79.01	&	$	65.728	\pm	7.743	$	&	$	41.024	\pm	9.893	$	\\
060714	&	219101	&	2.71	&	$	114.912	\pm	7.97	$	&	$	82.368	\pm	3.457	$	&	37.74	&	94.34	&	$	107.008	\pm	6.919	$	&	$	85.408	\pm	7.941	$	\\
060719	&	220020	&	1.532	&	$	66.928	\pm	11.558	$	&	$	34.048	\pm	11.018	$	&	55.29	&	138.23	&	$	53.152	\pm	2.271	$	&	$	\rthis{39.954}	\pm	\rthis{12.996}	$	\\
060814	&	224552	&	0.84	&	$	146.464	\pm	4.768	$	&	$	61.792	\pm	0.701	$	&	76.09	&	190.22	&	$	129.648	\pm	11.505	$	&	$	59.92	\pm	1.527	$	\\
060904B	&	228006	&	0.703	&	$	171.504	\pm	11.59	$	&	$	32.512	\pm	4.59	$	&	82.21	&	205.52	&	$	3.408	\pm	0.543	$	&	$	1.84	\pm	0.498	$	\\
060906	&	228316	&	3.685	&	$	49.568	\pm	4.431	$	&	$	18.768	\pm	2.763	$	&	29.88	&	74.71	&	$	46.944	\pm	13.83	$	&	$	16.672	\pm	3.488	$	\\
060908	&	228581	&	1.8836	&	$	18.928	\pm	1.155	$	&	$	7.312	\pm	0.643	$	&	48.55	&	121.38	&	$	17.968	\pm	1.056	$	&	$	6.64	\pm	0.688	$	\\
060912A	&	229185	&	0.937	&	$	5.056	\pm	0.552	$	&	$	1.76	\pm	0.182	$	&	72.28	&	180.69	&	$	8.512	\pm	2.713	$	&	$	3.36	\pm	1.401	$	\\
060926	&	231231	&	3.208	&	$	9.328	\pm	1.416	$	&	$	3.696	\pm	1.272	$	&	33.27	&	83.17	&	$	21.776	\pm	3.695	$	&	$	\rthis{7.865}	\pm	\rthis{2.454}	$	\\
060927	&	231362	&	5.6	&	$	22.528	\pm	1.277	$	&	$	15.792	\pm	5.217	$	&	21.21	&	53.03	&	$	21.248	\pm	1.395	$	&	$	16.32	\pm	5.891	$	\\
061007	&	232683	&	1.261	&	$	78.32	\pm	2.836	$	&	$	23.136	\pm	0.192	$	&	61.92	&	154.80	&	$	58.432	\pm	0.9	$	&	$	22.08	\pm	0.352	$	\\
061021	&	234905	&	0.3463	&	$	46.448	\pm	5.2	$	&	$	10.24	\pm	1.125	$	&	103.99	&	259.97	&	$	4.928	\pm	0.861	$	&	$	2.128	\pm	0.714	$	\\
061110B	&	238174	&	3.44	&	$	25.504	\pm	1.778	$	&	$	12.48	\pm	1.584	$	&	31.53	&	78.83	&	$	18.128	\pm	1.344	$	&	$	10.704	\pm	1.742	$	\\
061121	&	239899	&	1.314	&	$	35.888	\pm	25.267	$	&	$	7.472	\pm	0.121	$	&	60.50	&	151.25	&	$	\rthis{25.5}	\pm	\rthis{2.7}	$	&	$	7.024	\pm	0.172	$	\\
061217	&	251634	&	0.827	&	$	0.24	\pm	0.067	$	&	$	0.096	\pm	0.057	$	&	76.63	&	191.57	&	$	0.224	\pm	0.045	$	&	$	0.144	\pm	0.071	$	\\
061222A	&	252588	&	2.088	&	$	75.328	\pm	12.018	$	&	$	23.632	\pm	1.586	$	&	45.34	&	113.34	&	$	69.168	\pm	16.343	$	&	$	15.904	\pm	1.215	$	\\
061222B	&	252593	&	3.355	&	$	37.168	\pm	5.947	$	&	$	19.296	\pm	2.544	$	&	32.15	&	80.37	&	$	28.624	\pm	3.671	$	&	$	15.744	\pm	2.787	$	\\
070103	&	254532	&	2.6208	&	$	18.32	\pm	1.14	$	&	$	7.168	\pm	5.625	$	&	38.67	&	96.66	&	$	10.32	\pm	0.563	$	&	$	9.248	\pm	2.432	$	\\
070110	&	255445	&	2.352	&	$	88.608	\pm	14.379	$	&	$	30.8	\pm	5.853	$	&	41.77	&	104.42	&	$	40.464	\pm	3.442	$	&	$	20.656	\pm	4.205	$	\\
070129	&	258408	&	2.3384	&	$	460.976	\pm	63.417	$	&	$	263.744	\pm	30.857	$	&	41.94	&	104.84	&	$	71.136	\pm	5.937	$	&	$	29.136	\pm	22.829	$	\\
070306	&	263361	&	1.497	&	$	259.408	\pm	108.945	$	&	$	25.232	\pm	11.025	$	&	56.07	&	140.17	&	$	128.496	\pm	19.726	$	&	$	\rthis{18.52}	\pm	\rthis{9.651}	$	\\
070318	&	271019	&	0.838	&	$	94	\pm	22.675	$	&	$	33.856	\pm	4.432	$	&	76.17	&	190.42	&	$	22.064	\pm	2.67	$	&	$	10.464	\pm	3.458	$	\\
070506	&	278693	&	2.31	&	$	4.352	\pm	0.614	$	&	$	1.76	\pm	0.593	$	&	42.30	&	105.74	&	$	2.768	\pm	0.531	$	&	$	1.536	\pm	0.582	$	\\
070508	&	278854	&	0.82	&	$	20.736	\pm	0.731	$	&	$	7.568	\pm	0.147	$	&	76.92	&	192.31	&	$	14.768	\pm	1.376	$	&	$	6.592	\pm	0.147	$	\\
070521	&	279935	&	0.553	&	$	38.832	\pm	2.31	$	&	$	14.592	\pm	0.307	$	&	90.15	&	225.37	&	$	31.68	\pm	1.674	$	&	$	7.936	\pm	3.661	$	\\
070529	&	280706	&	2.4996	&	$	109.28	\pm	19.698	$	&	$	55.776	\pm	8.415	$	&	40.00	&	100.01	&	$	76.832	\pm	31.453	$	&	$	52.688	\pm	10.854	$	\\
070611	&	282003	&	2.04	&	$	12.064	\pm	2.426	$	&	$	3.92	\pm	2.031	$	&	46.05	&	115.13	&	$	9.12	\pm	2.737	$	&	$	\rthis{1.292}	\pm	\rthis{0.479}	$	\\
070714B	&	284856	&	0.92	&	$	59.392	\pm	7.645	$	&	$	38.384	\pm	7.273	$	&	72.92	&	182.29	&	$	1.776	\pm	0.279	$	&	$	0.832	\pm	0.249	$	\\
070721B	&	285654	&	3.626	&	$	344.48	\pm	25.742	$	&	$	267.2	\pm	34.467	$	&	30.26	&	75.66	&	$	335.856	\pm	27.91	$	&	$	269.888	\pm	32.864	$	\\
070802	&	286809	&	2.45	&	$	13.984	\pm	2.625	$	&	$	8.912	\pm	2.558	$	&	40.58	&	101.45	&	$	8.992	\pm	2.053	$	&	$	5.728	\pm	2.636	$	\\
070810A	&	287364	&	2.17	&	$	7.488	\pm	1.145	$	&	$	3.376	\pm	0.4	$	&	44.16	&	110.41	&	$	5.056	\pm	0.577	$	&	$	2.832	\pm	0.857	$	\\
071021	&	294974	&	2.452	&	$	236.48	\pm	66.921	$	&	$	60.976	\pm	37.908	$	&	40.56	&	101.39	&	$	62.752	\pm	21.261	$	&	$	\rthis{23.945}	\pm	\rthis{8.677}	$	\\
071031	&	295670	&	2.692	&	$	181.488	\pm	30.035	$	&	$	114.208	\pm	11.346	$	&	37.92	&	94.80	&	$	125.6	\pm	13.378	$	&	$	\rthis{56.174}	\pm	\rthis{16.555}	$	\\
071117	&	296805	&	1.331	&	$	4.176	\pm	0.386	$	&	$	1.52	\pm	0.071	$	&	60.06	&	150.15	&	$	2.48	\pm	0.534	$	&	$	0.928	\pm	0.093	$	\\
080207	&	302728	&	2.0858	&	$	304.064	\pm	8.824	$	&	$	147.344	\pm	19.523	$	&	45.37	&	113.42	&	$	296.496	\pm	9.744	$	&	$	\rthis{11.339}	\pm	\rthis{5.212}	$	\\
080210	&	302888	&	2.641	&	$	45.568	\pm	10.652	$	&	$	11.648	\pm	2.433	$	&	38.45	&	96.13	&	$	20.352	\pm	3.224	$	&	$	8.08	\pm	1.863	$	\\
080310	&	305288	&	2.43	&	$	320.56	\pm	16.622	$	&	$	244.912	\pm	13.281	$	&	40.82	&	102.04	&	$	28.208	\pm	6.805	$	&	$	\rthis{8.742}	\pm	\rthis{2.677}	$	\\
080319B	&	306757	&	0.937	&	$	133.2	\pm	3.638	$	&	$	25.904	\pm	0.124	$	&	72.28	&	180.69	&	$	45.552	\pm	0.385	$	&	$	24.256	\pm	0.249	$	\\
080319C	&	306778	&	1.95	&	$	32.88	\pm	9.962	$	&	$	7.712	\pm	0.492	$	&	47.46	&	118.64	&	$	11.424	\pm	0.985	$	&	$	6.752	\pm	0.374	$	\\
080411	&	309010	&	1.03	&	$	56.288	\pm	0.903	$	&	$	25.025	\pm	0.715	$	&	68.97	&	172.41	&	$	43.84	\pm	2.086	$	&	$	24.208	\pm	0.232	$	\\
080413A	&	309096	&	2.43	&	$	46.352	\pm	0.465	$	&	$	15.344	\pm	0.749	$	&	40.82	&	102.04	&	$	45.104	\pm	1.117	$	&	$	14.624	\pm	0.315	$	\\
080413B	&	309111	&	1.1	&	$	6.064	\pm	1.245	$	&	$	1.68	\pm	0.136	$	&	66.67	&	166.67	&	$	2.32	\pm	0.563	$	&	$	0.864	\pm	0.137	$	\\
080430	&	310613	&	0.76	&	$	16.08	\pm	2.475	$	&	$	5.328	\pm	0.551	$	&	79.55	&	198.86	&	$	9.92	\pm	2.294	$	&	$	5.68	\pm	2.302	$	\\
080516	&	311762	&	3.2	&	$	6.608	\pm	0.753	$	&	$	5.28	\pm	0.329	$	&	33.33	&	83.33	&	$	7.28	\pm	2.727	$	&	$	5.424	\pm	1.028	$	\\
080603B	&	313087	&	2.69	&	$	59.504	\pm	2.252	$	&	$	45.505	\pm	1.124	$	&	37.94	&	94.85	&	$	56.368	\pm	2.226	$	&	$	42.112	\pm	4.018	$	\\
080605	&	313299	&	1.6398	&	$	19.184	\pm	0.924	$	&	$	7.12	\pm	0.113	$	&	53.03	&	132.59	&	$	16.4	\pm	0.972	$	&	$	6.704	\pm	0.147	$	\\
080607	&	313417	&	3.036	&	$	81.888	\pm	2.675	$	&	$	23.184	\pm	1.222	$	&	34.69	&	86.72	&	$	73.488	\pm	10.46	$	&	$	10.336	\pm	2.915	$	\\
080804	&	319016	&	2.2	&	$	28.256	\pm	6.43	$	&	$	10.944	\pm	1.222	$	&	43.75	&	109.38	&	$	26.72	\pm	6.104	$	&	$	9.904	\pm	1.28	$	\\
080810	&	319584	&	3.35	&	$	105.664	\pm	3.074	$	&	$	38.976	\pm	2.557	$	&	32.18	&	80.46	&	$	92.88	\pm	12.074	$	&	$	35.28	\pm	3.155	$	\\
080905B	&	323898	&	2.374	&	$	94.704	\pm	9.295	$	&	$	75.648	\pm	6.082	$	&	41.49	&	103.73	&	$	85.6	\pm	3.644	$	&	$	70.272	\pm	7.92	$	\\
080916A	&	324895	&	0.689	&	$	60.368	\pm	6.296	$	&	$	23.136	\pm	1.002	$	&	82.89	&	207.22	&	$	30.064	\pm	5.38	$	&	$	12.832	\pm	2.922	$	\\
080928	&	326115	&	1.692	&	$	251.152	\pm	61.419	$	&	$	47.888	\pm	67.67	$	&	52.01	&	130.01	&	$	22.256	\pm	3.759	$	&	$	10.384	\pm	1.159	$	\\
081008	&	331093	&	1.967	&	$	166.912	\pm	33.379	$	&	$	111.296	\pm	1.934	$	&	47.19	&	117.96	&	$	167.728	\pm	24.736	$	&	$	106.416	\pm	5.69	$	\\
081029	&	332931	&	3.8479	&	$	102.176	\pm	11.817	$	&	$	55.968	\pm	12.126	$	&	28.88	&	72.20	&	$	218.656	\pm	45.863	$	&	$	99.024	\pm	47.465	$	\\
081222	&	337914	&	2.77	&	$	24.992	\pm	2.563	$	&	$	5.28	\pm	0.2	$	&	37.14	&	92.84	&	$	17.44	\pm	2.853	$	&	$	4.688	\pm	0.264	$	\\
090102	&	338895	&	1.547	&	$	27.44	\pm	2.272	$	&	$	11.36	\pm	1.569	$	&	54.97	&	137.42	&	$	11.968	\pm	1.247	$	&	$	7.696	\pm	0.848	$	\\
090113	&	339852	&	1.7493	&	$	9.04	\pm	0.915	$	&	$	5.872	\pm	0.622	$	&	50.92	&	127.31	&	$	8.208	\pm	0.452	$	&	$	\rthis{2.446}	\pm	\rthis{0.754}	$	\\
090205	&	342121	&	4.67	&	$	10.768	\pm	1.848	$	&	$	5.04	\pm	1.553	$	&	24.69	&	61.73	&	$	8.848	\pm	1.848	$	&	$	4.256	\pm	1.421	$	\\
090418A	&	349510	&	1.608	&	$	56.192	\pm	4.1	$	&	$	31.84	\pm	2.843	$	&	53.68	&	134.20	&	$	53.328	\pm	3.995	$	&	$	32.816	\pm	4.264	$	\\
090423	&	350184	&	8.05	&	$	10.624	\pm	1.008	$	&	$	4.864	\pm	0.544	$	&	15.47	&	38.67	&	$	12.96	\pm	2.808	$	&	$	5.296	\pm	0.581	$	\\
090424	&	350311	&	0.544	&	$	49.744	\pm	2.32	$	&	$	3.376	\pm	0.057	$	&	90.67	&	226.68	&	$	4.16	\pm	0.192	$	&	$	2.896	\pm	0.115	$	\\
090426	&	350479	&	2.609	&	$	1.264	\pm	0.273	$	&	$	0.448	\pm	0.182	$	&	38.79	&	96.98	&	$	0.56	\pm	0.137	$	&	$	0.24	\pm	0.124	$	\\
090510	&	351588	&	0.903	&	$	0.4	\pm	0.113	$	&	$	0.24	\pm	0.05	$	&	73.57	&	183.92	&	$	0.304	\pm	0.066	$	&	$	0.192	\pm	0.086	$	\\
090516A	&	352190	&	4.109	&	$	194.192	\pm	49.188	$	&	$	75.648	\pm	7.839	$	&	27.40	&	68.51	&	$	\rthis{54.234}	\pm	\rthis{13.563}	$	&	$	\rthis{41.13}	\pm	\rthis{8.492}	$	\\
090519	&	352648	&	3.87	&	$	77.184	\pm	14.151	$	&	$	25.456	\pm	14.063	$	&	28.75	&	71.87	&	$	83.632	\pm	15.633	$	&	$	23.808	\pm	16.195	$	\\
090529	&	353540	&	2.625	&	$	79.808	\pm	13.905	$	&	$	37.408	\pm	10.011	$	&	38.62	&	96.55	&	$	62.688	\pm	14.664	$	&	$	33.904	\pm	13.447	$	\\
090530	&	353567	&	1.266	&	$	39.648	\pm	4.278	$	&	$	23.952	\pm	10.948	$	&	61.78	&	154.46	&	$	7.44	\pm	3.193	$	&	$	2.352	\pm	1.746	$	\\
090618	&	355083	&	0.54	&	$	113.536	\pm	0.585	$	&	$	27.952	\pm	0.157	$	&	90.91	&	227.27	&	$	105.904	\pm	1.841	$	&	$	21.456	\pm	0.555	$	\\
090715B	&	357512	&	3	&	$	268.256	\pm	11.4	$	&	$	63.472	\pm	1.191	$	&	35.00	&	87.50	&	$	80.528	\pm	14.67	$	&	$	54.56	\pm	4.011	$	\\
090726	&	358422	&	2.71	&	$	58.064	\pm	12.174	$	&	$	21.392	\pm	9.522	$	&	37.74	&	94.34	&	$	17.488	\pm	4.293	$	&	$	8.672	\pm	3.298	$	\\
090809	&	359530	&	2.737	&	$	6.08	\pm	1.312	$	&	$	2.56	\pm	1.005	$	&	37.46	&	93.66	&	$	6.832	\pm	1.651	$	&	$	3.104	\pm	1.369	$	\\
090812	&	359711	&	2.452	&	$	78.144	\pm	12.498	$	&	$	24.864	\pm	1.296	$	&	40.56	&	101.39	&	$	58.736	\pm	3.411	$	&	$	21.312	\pm	1.234	$	\\
090926B	&	370791	&	1.24	&	$	111.184	\pm	12.297	$	&	$	28.528	\pm	2.178	$	&	62.50	&	156.25	&	$	44	\pm	6.677	$	&	$	17.344	\pm	2.964	$	\\
091018	&	373172	&	0.971	&	$	4.416	\pm	0.602	$	&	$	1.568	\pm	0.086	$	&	71.03	&	177.57	&	$	2.048	\pm	0.283	$	&	$	0.896	\pm	0.237	$	\\
091020	&	373458	&	1.71	&	$	35.84	\pm	3.3943	$	&	$	8.672	\pm	1.006	$	&	51.66	&	129.15	&	$	29.584	\pm	7.922	$	&	$	7.104	\pm	1.129	$	\\
091024	&	373674	&	1.092	&	$	113.696	\pm	14.176	$	&	$	49.584	\pm	3.924	$	&	66.92	&	167.30	&	$	67.36	\pm	5.198	$	&	$	39.072	\pm	2.257	$	\\
091029	&	374210	&	2.752	&	$	34.048	\pm	1.688	$	&	$	13.072	\pm	1.069	$	&	37.31	&	93.28	&	$	35.296	\pm	2.65	$	&	$	15.12	\pm	2.852	$	\\
091109A	&	375246	&	3.076	&	$	21.92	\pm	3.808	$	&	$	10.048	\pm	3.603	$	&	34.35	&	85.87	&	$	17.184	\pm	3.321	$	&	$	8.272	\pm	3.113	$	\\
091127	&	377179	&	0.49	&	$	8.24	\pm	0.69	$	&	$	5.584	\pm	0.534	$	&	93.96	&	234.90	&	$	2.032	\pm	0.486	$	&	$	1.12	\pm	0.252	$	\\
091208B	&	378559	&	1.063	&	$	13.984	\pm	3	$	&	$	\rthis{5.687}	\pm	\rthis{1.071}	$	&	67.86	&	169.66	&	$	0.8	\pm	0.137	$	&	$	0.4	\pm	0.124	$	\\
100219A	&	412982	&	4.7	&	$	23.008	\pm	3.529	$	&	$	11.52	\pm	4.613	$	&	24.56	&	61.40	&	$	26.4	\pm	5.314	$	&	$	14.16	\pm	5.272	$	\\
100413A	&	419404	&	3.9	&	$	194.976	\pm	10.818	$	&	$	104.256	\pm	8.3	$	&	28.57	&	71.43	&	$	190.368	\pm	14.054	$	&	$	102.304	\pm	9.234	$	\\
100424A	&	420367	&	2.465	&	$	90.864	\pm	8.578	$	&	$	40.48	\pm	7.29	$	&	40.40	&	101.01	&	$	51.184	\pm	6.691	$	&	$	27.216	\pm	5.762	$	\\
100615A	&	424733	&	1.398	&	$	39.088	\pm	0.85	$	&	$	23.664	\pm	1.463	$	&	58.38	&	145.95	&	$	39.168	\pm	3.493	$	&	$	26.896	\pm	1.133	$	\\
100621A	&	425151	&	0.542	&	$	65.28	\pm	1.8	$	&	$	25.76	\pm	0.774	$	&	90.79	&	226.98	&	$	35.488	\pm	2.588	$	&	$	16.976	\pm	1.746	$	\\
100728A	&	430151	&	1.567	&	$	211.312	\pm	11.744	$	&	$	72.976	\pm	1.227	$	&	54.54	&	136.35	&	$	168.112	\pm	2.547	$	&	$	63.872	\pm	1.188	$	\\
100728B	&	430172	&	2.106	&	$	11.92	\pm	2.637	$	&	$	4.464	\pm	1.234	$	&	45.07	&	112.69	&	$	10.624	\pm	2.709	$	&	$	4.384	\pm	1.336	$	\\
100816A	&	431764	&	0.8035	&	$	2.56	\pm	0.531	$	&	$	1.008	\pm	0.057	$	&	77.63	&	194.07	&	$	1.792	\pm	0.172	$	&	$	0.896	\pm	0.124	$	\\
100901A	&	433065	&	1.408	&	$	440.672	\pm	24.921	$	&	$	195.76	\pm	21.302	$	&	58.14	&	145.35	&	$	17.664	\pm	2.836	$	&	$	10.976	\pm	3.251	$	\\
100902A	&	433160	&	4.5	&	$	257.584	\pm	18.765	$	&	$	66.432	\pm	8.541	$	&	25.45	&	63.64	&	$	85.136	\pm	8.32	$	&	$	58.896	\pm	6.226	$	\\
100906A	&	433509	&	1.727	&	$	114.304	\pm	1.643	$	&	$	54.576	\pm	8.698	$	&	51.34	&	128.35	&	$	103.184	\pm	2.231	$	&	$	11.936	\pm	1.898	$	\\
101219A	&	440606	&	0.718	&	$	0.832	\pm	0.192	$	&	$	0.416	\pm	0.035	$	&	81.49	&	203.73	&	$	0.528	\pm	0.071	$	&	$	0.304	\pm	0.086	$	\\
110128A	&	443861	&	2.339	&	$	16.592	\pm	3.048	$	&	$	8.24	\pm	2.33	$	&	41.93	&	104.82	&	$	45.408	\pm	14.395	$	&	$	21.248	\pm	13.229	$	\\
110205A	&	444643	&	2.1	&	$	267.504	\pm	16.17	$	&	$	102.72	\pm	7.5	$	&	45.16	&	112.90	&	$	265.856	\pm	33.4	$	&	$	92.128	\pm	11.164	$	\\
110422A	&	451901	&	1.77	&	$	27.552	\pm	1.395	$	&	$	11.344	\pm	0.262	$	&	50.54	&	126.35	&	$	23.792	\pm	0.906	$	&	$	9.904	\pm	0.24	$	\\
110503A	&	452685	&	1.613	&	$	9.824	\pm	3.091	$	&	$	2.912	\pm	0.252	$	&	53.58	&	133.95	&	$	7.232	\pm	1.339	$	&	$	2.032	\pm	0.349	$	\\
110715A	&	457330	&	0.82	&	$	7.696	\pm	1.712	$	&	$	1.584	\pm	0.045	$	&	76.92	&	192.31	&	$	2.976	\pm	0.215	$	&	$	1.328	\pm	0.086	$	\\
110731A	&	458448	&	2.83	&	$	42.096	\pm	12.84	$	&	$	4.88	\pm	0.186	$	&	36.55	&	91.38	&	$	\rthis{31.36}	\pm	\rthis{9.328}	$	&	$	4.72	\pm	0.17	$	\\
110801A	&	458521	&	1.858	&	$	390.512	\pm	9.271	$	&	$	331.728	\pm	7.321	$	&	48.99	&	122.46	&	$	527.44	\pm	45.396	$	&	$	218.144	\pm	69.01	$	\\
110818A	&	500914	&	3.36	&	$	101.664	\pm	20.021	$	&	$	37.792	\pm	6.206	$	&	32.11	&	80.28	&	$	55.904	\pm	6.68	$	&	$	23.456	\pm	3.397	$	\\
111008A	&	505054	&	4.9898	&	$	65.52	\pm	2.73	$	&	$	30.432	\pm	1.666	$	&	23.37	&	58.43	&	$	62.864	\pm	14.042	$	&	$	28.528	\pm	2.202	$	\\
111107A	&	507185	&	2.893	&	$	31.072	\pm	7.974	$	&	$	11.056	\pm	2.946	$	&	35.96	&	89.90	&	$	31.584	\pm	11.236	$	&	$	9.184	\pm	5.284	$	\\
111123A	&	508319	&	3.1516	&	$	219.776	\pm	15.045	$	&	$	81.776	\pm	5.775	$	&	33.72	&	84.30	&	$	176.416	\pm	16.873	$	&	$	80.688	\pm	10.225	$	\\
111228A	&	510649	&	0.714	&	$	100.352	\pm	5.475	$	&	$	46.768	\pm	0.78	$	&	81.68	&	204.20	&	$	48.4	\pm	19.393	$	&	$	\rthis{4.779}	\pm	\rthis{1.291}	$	\\
120118B	&	512003	&	2.943	&	$	22.672	\pm	2.233	$	&	$	9.152	\pm	1.164	$	&	35.51	&	88.76	&	$	23.808	\pm	4.329	$	&	$	9.04	\pm	2.33	$	\\
120119A	&	512035	&	1.728	&	$	76.144	\pm	12.945	$	&	$	18.32	\pm	0.723	$	&	51.32	&	128.30	&	$	252.816	\pm	41.751	$	&	$	24.752	\pm	1.874	$	\\
120326A	&	518626	&	1.798	&	$	72.736	\pm	8.768	$	&	$	5.424	\pm	0.522	$	&	50.04	&	125.09	&	$	12.352	\pm	3.417	$	&	$	3.728	\pm	0.508	$	\\
120327A	&	518731	&	2.813	&	$	65.952	\pm	6.777	$	&	$	23.824	\pm	3.444	$	&	36.72	&	91.79	&	$	83.136	\pm	38.614	$	&	$	23.776	\pm	5.911	$	\\
120404A	&	519380	&	2.876	&	$	38.386	\pm	4.908	$	&	$	17.248	\pm	3.077	$	&	36.12	&	90.30	&	$	40.016	\pm	6.344	$	&	$	17.312	\pm	2.991	$	\\
120712A	&	526351	&	4.15	&	$	16	\pm	2.651	$	&	$	6.96	\pm	0.492	$	&	27.18	&	67.96	&	$	17.424	\pm	3.239	$	&	$	7.024	\pm	0.69	$	\\
120729A	&	529095	&	0.8	&	$	71.44	\pm	16.659	$	&	$	22.224	\pm	2.455	$	&	77.78	&	194.44	&	$	64	\pm	10.44	$	&	$	29.728	\pm	13.899	$	\\
120802A	&	529486	&	3.796	&	$	15.2	\pm	1.712	$	&	$	6.32	\pm	0.568	$	&	29.19	&	72.98	&	$	15.696	\pm	2.553	$	&	$	6.672	\pm	0.829	$	\\
120805A	&	530031	&	3.1	&	$	30.448	\pm	4.98	$	&	$	15.168	\pm	7.564	$	&	34.15	&	85.37	&	$	36.544	\pm	5.751	$	&	$	19.856	\pm	6.959	$	\\
120811C	&	530689	&	2.671	&	$	24.224	\pm	3.175	$	&	$	7.008	\pm	0.489	$	&	38.14	&	95.34	&	$	23.04	\pm	4.555	$	&	$	6.672	\pm	0.659	$	\\
120815A	&	531003	&	2.358	&	$	6.944	\pm	1.022	$	&	$	3.472	\pm	0.861	$	&	41.69	&	104.23	&	$	2.72	\pm	0.758	$	&	$	1.44	\pm	0.967	$	\\
120909A	&	533060	&	3.93	&	$	105.504	\pm	7.571	$	&	$	45.088	\pm	2.249	$	&	28.40	&	70.99	&	$	61.456	\pm	61.113	$	&	$	28.176	\pm	6.739	$	\\
120922A	&	534394	&	3.1	&	$	175.808	\pm	30.454	$	&	$	94.608	\pm	4.037	$	&	34.15	&	85.37	&	$	160.464	\pm	38.491	$	&	$	99.152	\pm	12.865	$	\\
121024A	&	536580	&	2.298	&	$	11.344	\pm	1.18	$	&	$	6.208	\pm	1.266	$	&	42.45	&	106.12	&	$	11.472	\pm	1.319	$	&	$	6.608	\pm	1.611	$	\\
121027A	&	536831	&	1.773	&	$	68.432	\pm	9.656	$	&	$	32.192	\pm	7.014	$	&	50.49	&	126.22	&	$	61.184	\pm	9.355	$	&	$	\rthis{17.454}	\pm	\rthis{4.37}	$	\\
121128A	&	539866	&	2.2	&	$	23.056	\pm	1.843	$	&	$	7.264	\pm	0.436	$	&	43.75	&	109.38	&	$	18.192	\pm	1.497	$	&	$	5.488	\pm	0.43	$	\\
121201A	&	540178	&	3.385	&	$	27.712	\pm	1.812	$	&	$	18.672	\pm	2.977	$	&	31.93	&	79.82	&	$	41.024	\pm	7.971	$	&	$	22.032	\pm	4.686	$	\\
130131B	&	547420	&	2.539	&	$	4.352	\pm	0.3	$	&	$	2.688	\pm	0.846	$	&	39.56	&	98.90	&	$	4.4	\pm	0.409	$	&	$	2.576	\pm	0.763	$	\\
130427A	&	554620	&	0.34	&	$	316.272	\pm	7.63	$	&	$	32.064	\pm	1.296	$	&	104.48	&	261.19	&	$	128.336	\pm	4.859	$	&	$	4.016	\pm	0.079	$	\\
130514A	&	555821	&	3.6	&	$	215.488	\pm	17.628	$	&	$	104.56	\pm	2.813	$	&	30.43	&	76.09	&	$	174.784	\pm	19.402	$	&	$	97.344	\pm	5.01	$	\\
130603B	&	557310	&	0.356	&	$	0.192	\pm	0.032	$	&	$	0.048	\pm	0.016	$	&	103.24	&	258.11	&	$	0.064	\pm	0.035	$	&	$	0.032	\pm	0.016	$	\\
130604A	&	557354	&	1.06	&	$	37.792	\pm	4.627	$	&	$	17.632	\pm	2.579	$	&	67.96	&	169.90	&	$	28.016	\pm	4.249	$	&	$	15.088	\pm	4.185	$	\\
130606A	&	557589	&	5.913	&	$	163.088	\pm	2.983	$	&	$	71.696	\pm	5.921	$	&	20.25	&	50.63	&	$	163.344	\pm	2.805	$	&	$	\rthis{76.936}	\pm	\rthis{14.425}	$	\\
130701A	&	559482	&	1.155	&	$	\rthis{30.498}	\pm	\rthis{7.721}	$	&	$	2.528	\pm	0.204	$	&	64.97	&	162.41	&	$	3.632	\pm	0.315	$	&	$	1.6	\pm	0.315	$	\\
130831A	&	568849	&	0.4791	&	$	32.512	\pm	1.689	$	&	$	8.032	\pm	0.726	$	&	94.65	&	236.63	&	$	21.968	\pm	6.399	$	&	$	5.408	\pm	2.68	$	\\
130925A	&	571830	&	0.347	&	$	456.752	\pm	42.832	$	&	$	130.096	\pm	2.671	$	&	103.93	&	259.84	&	$	123.136	\pm	44.636	$	&	$	54.112	\pm	27.817	$	\\
131004A	&	573190	&	0.717	&	$	1.12	\pm	0.172	$	&	$	0.432	\pm	0.065	$	&	81.54	&	203.84	&	$	0.832	\pm	0.222	$	&	$	0.368	\pm	0.244	$	\\
131030A	&	576238	&	1.293	&	$	41.808	\pm	4.054	$	&	$	8.496	\pm	0.227	$	&	61.06	&	152.64	&	$	19.696	\pm	1.704	$	&	$	7.328	\pm	0.222	$	\\
131117A	&	577968	&	4.18	&	$	6.032	\pm	0.644	$	&	$	3.808	\pm	0.713	$	&	27.03	&	67.57	&	$	5.792	\pm	0.679	$	&	$	3.6	\pm	0.734	$	\\
131227A	&	582184	&	5.3	&	$	17.744	\pm	1.323	$	&	$	10.304	\pm	1.681	$	&	22.22	&	55.56	&	$	18.096	\pm	1.625	$	&	$	10.224	\pm	1.625	$	\\
140114A	&	583861	&	3	&	$	143.376	\pm	12.658	$	&	$	53.664	\pm	3.288	$	&	35.00	&	87.50	&	$	103.328	\pm	14.59	$	&	$	37.888	\pm	7.521	$	\\
140206A	&	585834	&	2.73	&	$	96.096	\pm	18.87	$	&	$	11.264	\pm	19.12	$	&	37.53	&	93.83	&	$	83.072	\pm	9.18	$	&	$	\rthis{10.179}	\pm	\rthis{3.544}	$	\\
140213A	&	586569	&	1.2076	&	$	59.92	\pm	2.691	$	&	$	6.528	\pm	0.272	$	&	63.42	&	158.54	&	$	17.312	\pm	3.8	$	&	$	5.088	\pm	0.227	$	\\
140304A	&	590206	&	5.283	&	$	15.424	\pm	1.793	$	&	$	9.232	\pm	0.502	$	&	22.28	&	55.71	&	$	15.632	\pm	2.044	$	&	$	9.264	\pm	0.851	$	\\
140419A	&	596426	&	3.956	&	$	98.816	\pm	11.03	$	&	$	41.072	\pm	1.356	$	&	28.25	&	70.62	&	$	78.624	\pm	4.454	$	&	$	38.464	\pm	1.402	$	\\
140423A	&	596901	&	3.26	&	$	96.656	\pm	4.8	$	&	$	47.136	\pm	2.198	$	&	32.86	&	82.16	&	$	115.936	\pm	16.846	$	&	$	52	\pm	4.125	$	\\
140506A	&	598284	&	0.889	&	$	113.232	\pm	10.643	$	&	$	58.576	\pm	3.623	$	&	74.11	&	185.28	&	$	60.784	\pm	57.44	$	&	$	\rthis{26.564}	\pm	\rthis{11.931}	$	\\
140512A	&	598819	&	0.725	&	$	156.4	\pm	4.609	$	&	$	90.08	\pm	25.051	$	&	81.16	&	202.90	&	$	149.616	\pm	13.693	$	&	$	73.52	\pm	43.51	$	\\
140515A	&	599037	&	6.32	&	$	23.456	\pm	2.048	$	&	$	9.776	\pm	5.341	$	&	19.13	&	47.81	&	$	23.136	\pm	1.958	$	&	$	10.048	\pm	5.955	$	\\
140518A	&	599287	&	4.707	&	$	60.624	\pm	2.409	$	&	$	48.24	\pm	3.093	$	&	24.53	&	61.33	&	$	60.992	\pm	2.595	$	&	$	49.312	\pm	2.671	$	\\
140614A	&	601646	&	4.233	&	$	68.64	\pm	9.568	$	&	$	32.544	\pm	9.56	$	&	26.75	&	66.88	&	$	164.752	\pm	57.526	$	&	$	54.592	\pm	43.213	$	\\
140703A	&	603243	&	3.14	&	$	79.008	\pm	8.665	$	&	$	34.96	\pm	7.704	$	&	33.82	&	84.54	&	$	76.144	\pm	11.648	$	&	$	29.904	\pm	9.801	$	\\
140907A	&	611933	&	1.21	&	$	43.888	\pm	7.536	$	&	$	13.616	\pm	0.784	$	&	63.35	&	158.37	&	$	25.648	\pm	3.993	$	&	$	10.672	\pm	0.917	$	\\
141004A	&	614390	&	0.57	&	$	3.904	\pm	1.091	$	&	$	1.12	\pm	0.178	$	&	89.17	&	222.93	&	$	0.416	\pm	0.08	$	&	$	0.224	\pm	0.09	$	\\
141026A	&	616502	&	3.35	&	$	148.16	\pm	14.323	$	&	$	83.856	\pm	11.565	$	&	32.18	&	80.46	&	$	154.704	\pm	19.59	$	&	$	88.384	\pm	34.148	$	\\
141109A	&	618024	&	2.993	&	$	151.312	\pm	13.295	$	&	$	49.168	\pm	6.437	$	&	35.06	&	87.65	&	$	142.96	\pm	31.975	$	&	$	47.28	\pm	8.818	$	\\
141220A	&	621915	&	1.3195	&	$	7.504	\pm	0.489	$	&	$	3.072	\pm	0.92	$	&	60.36	&	150.89	&	$	7.424	\pm	0.77	$	&	$	2.928	\pm	0.602	$	\\
141221A	&	622006	&	1.452	&	$	36.928	\pm	4.064	$	&	$	14.688	\pm	3.075	$	&	57.10	&	142.74	&	$	36.432	\pm	12.491	$	&	$	6.496	\pm	4.708	$	\\
141225A	&	622476	&	0.915	&	$	40.336	\pm	7.33	$	&	$	16.848	\pm	3.373	$	&	73.11	&	182.77	&	$	17.68	\pm	3.238	$	&	$	8.96	\pm	3.206	$	\\
150206A	&	630019	&	2.087	&	$	92.72	\pm	27.123	$	&	$	18.288	\pm	0.656	$	&	45.35	&	113.38	&	$	78.128	\pm	15.265	$	&	$	17.856	\pm	0.444	$	\\
150301B	&	633180	&	1.5169	&	$	13.04	\pm	1.465	$	&	$	4.64	\pm	0.443	$	&	55.62	&	139.06	&	$	15.424	\pm	7.267	$	&	$	4.592	\pm	1.098	$	\\
150314A	&	634795	&	1.758	&	$	14.79	\pm	2.681	$	&	$	5.344	\pm	0.151	$	&	50.76	&	126.90	&	$	10.8	\pm	0.349	$	&	$	4.336	\pm	0.187	$	\\
150323A	&	635887	&	0.593	&	$	149.12	\pm	6.815	$	&	$	12.928	\pm	1.264	$	&	87.88	&	219.71	&	$	13.84	\pm	1.687	$	&	$	7.808	\pm	1.165	$	\\
150423A	&	638808	&	1.394	&	$	0.224	\pm	0.035	$	&	$	0.128	\pm	0.035	$	&	58.48	&	146.20	&	$	0.08	\pm	0.022	$	&	$	0.048	\pm	0.016	$	\\
150727A	&	650530	&	0.313	&	$	86.128	\pm	12.748	$	&	$	30.192	\pm	3.624	$	&	106.63	&	266.57	&	$	23.904	\pm	4.388	$	&	$	14.976	\pm	6.15	$	\\
150910A	&	655097	&	1.359	&	$	126.72	\pm	27.241	$	&	$	50.64	\pm	8.607	$	&	59.35	&	148.37	&	$	32.576	\pm	10.868	$	&	$	18.64	\pm	11.31	$	\\
151021A	&	660671	&	2.33	&	$	114.56	\pm	4.125	$	&	$	36.736	\pm	1.245	$	&	42.04	&	105.11	&	$	104.672	\pm	7.647	$	&	$	32.832	\pm	1.545	$	\\
151027A	&	661775	&	0.81	&	$	128.88	\pm	5.924	$	&	$	96.352	\pm	8.243	$	&	77.35	&	193.37	&	$	124.048	\pm	17.66	$	&	$	109.04	\pm	16.476	$	\\
151027B	&	661869	&	4.063	&	$	23.424	\pm	4.197	$	&	$	11.472	\pm	2.494	$	&	27.65	&	69.13	&	$	18.256	\pm	4.521	$	&	$	10.8	\pm	3.158	$	\\
151111A	&	663074	&	3.5	&	$	76.88	\pm	12.341	$	&	$	30.448	\pm	3.4	$	&	31.11	&	77.78	&	$	38.272	\pm	4.518	$	&	$	18.064	\pm	2.36	$	\\
151112A	&	663179	&	4.1	&	$	13.424	\pm	1.125	$	&	$	7.936	\pm	1.431	$	&	27.45	&	68.63	&	$	20.4	\pm	5.227	$	&	$	10.048	\pm	2.301	$	\\
151215A	&	667392	&	2.59	&	$	3.072	\pm	0.443	$	&	$	1.968	\pm	0.323	$	&	39.00	&	97.49	&	$	1.2	\pm	0.229	$	&	$	0.56	\pm	0.359	$	\\
160121A	&	671231	&	1.96	&	$	10.848	\pm	1.67	$	&	$	4.88	\pm	0.823	$	&	47.30	&	118.24	&	$	7.44	\pm	1.022	$	&	$	4	\pm	0.871	$	\\
160131A	&	672236	&	0.97	&	$	216.096	\pm	21.979	$	&	$	54.64	\pm	2.462	$	&	71.07	&	177.66	&	$	272.544	\pm	51.957	$	&	$	50.096	\pm	6.88	$	\\
160203A	&	672525	&	3.52	&	$	19.92	\pm	3.116	$	&	$	9.408	\pm	3.083	$	&	30.97	&	77.43	&	$	14.672	\pm	3.941	$	&	$	7.424	\pm	2.75	$	\\
160227A	&	676423	&	2.38	&	$	315.856	\pm	70.295	$	&	$	186.464	\pm	13.206	$	&	41.42	&	103.55	&	$	235.664	\pm	13.489	$	&	$	160.352	\pm	44.885	$	\\
160327A	&	680655	&	4.99	&	$	34.64	\pm	8.411	$	&	$	9.952	\pm	1.282	$	&	23.37	&	58.43	&	$	31.712	\pm	8.388	$	&	$	8.624	\pm	1.773	$	\\
160804A	&	707231	&	0.736	&	$	153.168	\pm	16.881	$	&	$	50.784	\pm	2.622	$	&	80.65	&	201.61	&	$	46.528	\pm	6.453	$	&	$	18.56	\pm	4.806	$	\\
161014A	&	717500	&	2.823	&	$	25.52	\pm	3.939	$	&	$	12.592	\pm	3.887	$	&	36.62	&	91.55	&	$	17.984	\pm	1.808	$	&	$	8.192	\pm	3.969	$	\\
161017A	&	718023	&	2.01	&	$	217.728	\pm	6.527	$	&	$	126.64	\pm	5.22	$	&	46.51	&	116.28	&	$	215.44	\pm	13.341	$	&	$	120.416	\pm	5.937	$	\\
161117A	&	722604	&	1.549	&	$	126.672	\pm	1.23	$	&	$	74.816	\pm	0.523	$	&	54.92	&	137.31	&	$	121.104	\pm	1.126	$	&	$	81.056	\pm	7.988	$	\\
161129A	&	724438	&	0.645	&	$	35.376	\pm	2.128	$	&	$	14.656	\pm	2.572	$	&	85.10	&	212.77	&	$	32.512	\pm	5.204	$	&	$	14.48	\pm	5.299	$	\\
170113A	&	732526	&	1.968	&	$	20.592	\pm	4.477	$	&	$	10.608	\pm	0.649	$	&	47.17	&	117.92	&	$	18.608	\pm	4.749	$	&	$	10.368	\pm	2.902	$	\\
170202A	&	736407	&	3.65	&	$	48.416	\pm	11.521	$	&	$	14.864	\pm	1.262	$	&	30.1	&	75.27	&	$	43.312	\pm	15.926	$	&	$	14.48	\pm	2.135	$	\\
170405A	&	745797	&	3.51	&	$	144.48	\pm	49.554	$	&	$	44.96	\pm	2.055	$	&	31.04	&	77.61	&	$	\rthis{68.352}	\pm	\rthis{9.956}	$	&	$	39.104	\pm	6.281	$	\\
170531B	&	755354	&	2.366	&	$	172.384	\pm	9.642	$	&	$	134.208	\pm	10.509	$	&	41.59	&	103.98	&	$	167.472	\pm	9.085	$	&	$	138.032	\pm	18.292	$	\\
170604A	&	755867	&	1.329	&	$	26.48	\pm	2.808	$	&	$	9.008	\pm	1.087	$	&	60.11	&	150.28	&	$	4.928	\pm	0.754	$	&	$	2.608	\pm	0.758	$	\\
170607A	&	756284	&	0.557	&	$	269.808	\pm	43.921	$	&	$	215.632	\pm	1.009	$	&	89.92	&	224.79	&	$	11.936	\pm	2.638	$	&	$	5.68	\pm	1.424	$	\\
170705A	&	760064	&	2.01	&	$	224	\pm	15.031	$	&	$	107.04	\pm	70.704	$	&	46.51	&	116.28	&	$	210.048	\pm	30.829	$	&	$	\rthis{26.594}	\pm	\rthis{7.351}	$	\\
170714A	&	762535	&	0.793	&	$	397.984	\pm	39.658	$	&	$	240.352	\pm	48.829	$	&	78.08	&	195.20	&	$	0.32	\pm	0.024	$	&	$	0.16	\pm	0.024	$	\\
170903A	&	770528	&	0.886	&	$	29.264	\pm	3.162	$	&	$	10.912	\pm	5.073	$	&	74.23	&	185.58	&	$	27.904	\pm	5.51	$	&	$	13.264	\pm	10.102	$	\\
171020A	&	780845	&	1.87	&	$	26.496	\pm	2.718	$	&	$	14.704	\pm	2.894	$	&	48.78	&	121.95	&	$	16.512	\pm	2.588	$	&	$	8.08	\pm	1.924	$	\\
171222A	&	799669	&	2.409	&	$	174.112	\pm	22.922	$	&	$	99.808	\pm	19.292	$	&	41.06	&	102.67	&	$	167.232	\pm	32.646	$	&	$	95.344	\pm	42.602	$	\\
180115A	&	805318	&	2.487	&	$	41.504	\pm	3.066	$	&	$	30.832	\pm	3.026	$	&	47.35	&	100.37	&	$	46.064	\pm	5.001	$	&	$	34.432	\pm	10.67	$	\\
180314A	&	814129	&	1.445	&	$	24.368	\pm	1.367	$	&	$	9.968	\pm	0.4305	$	&	57.26	&	143.15	&	$	20.768	\pm	0.958	$	&	$	9.328	\pm	0.801	$	\\
180325A	&	817564	&	2.25	&	$	90.8	\pm	3.385	$	&	$	4.976	\pm	1.159	$	&	43.07	&	107.69	&	$	455.968	\pm	62.01	$	&	$	124.672	\pm	42.425	$	\\
180329B	&	819490	&	1.998	&	$	224.848	\pm	35.111	$	&	$	142.768	\pm	3.528	$	&	46.70	&	116.74	&	$	16.448	\pm	5.562	$	&	$	6.864	\pm	1.3	$	\\
180510B	&	831816	&	1.305	&	$	133.152	\pm	55.954	$	&	$	28	\pm	7.388	$	&	60.74	&	151.84	&	$	19.056	\pm	5.933	$	&	$	8.08	\pm	3.97	$	\\
180620B	&	843211	&	1.1175	&	$	229.968	\pm	21.917	$	&	$	45.12	\pm	9.44	$	&	66.12	&	165.29	&	$	144.208	\pm	70.882	$	&	$	25.28	\pm	11.174	$	\\
180624A	&	844192	&	2.855	&	$	465.216	\pm	36.332	$	&	$	325.552	\pm	9.305	$	&	36.32	&	90.79	&	$	461.728	\pm	59.093	$	&	$	322.544	\pm	18.492	$	\\
180720B	&	848890	&	0.654	&	$	130.864	\pm	10.726	$	&	$	23.488	\pm	1.649	$	&	84.64	&	211.61	&	$	49.12	\pm	10.04	$	&	$	14.304	\pm	2.846	$	\\
180728A	&	850471	&	0.117	&	$	8.688	\pm	0.321	$	&	$	2.464	\pm	0.035	$	&	125.33	&	313.34	&	$	5.28	\pm	2.197	$	&	$	1.824	\pm	0.23	$	\\
181010A	&	866434	&	1.39	&	$	12.704	\pm	1.047	$	&	$	7.872	\pm	0.723	$	&	58.57	&	146.44	&	$	10.256	\pm	1.486	$	&	$	7.36	\pm	1.809	$	\\
181020A	&	867987	&	2.938	&	$	243.568	\pm	13.439	$	&	$	17.296	\pm	6.28	$	&	35.55	&	88.88	&	$	229.328	\pm	16.505	$	&	$	11.008	\pm	2.856	$	\\
181110A	&	871316	&	1.505	&	$	134.768	\pm	11.635	$	&	$	46.912	\pm	1.436	$	&	55.89	&	139.72	&	$	87.616	\pm	16.402	$	&	$	30.8	\pm	4.379	$	\\
190114A	&	930285	&	3.3765	&	$	70.416	\pm	5.15	$	&	$	30.992	\pm	1.143	$	&	31.99	&	79.97	&	$	70.08	\pm	7.642	$	&	$	31.488	\pm	1.681	$	\\
190324A	&	927839	&	1.1715	&	$	299.984	\pm	81.712	$	&	$	32.144	\pm	27.954	$	&	64.47	&	161.18	&	$	7.168	\pm	1.957	$	&	$	3.168	\pm	0.25	$	\\
190719C	&	915381	&	2.469	&	$	186.208	\pm	9.309	$	&	$	99.456	\pm	6.052	$	&	40.36	&	100.89	&	$	177.856	\pm	6.448	$	&	$	94.272	\pm	23.013	$	\\
191004B	&	894718	&	3.503	&	$	18.816	\pm	7.008	$	&	$	5.712	\pm	0.279	$	&	31.09	&	77.73	&	$	\rthis{15.626}	\pm	\rthis{3.17}	$	&	$	5.232	\pm	0.357	$	\\
191011A	&	883832	&	1.722	&	$	372.32	\pm	13.76	$	&	$	55.136	\pm	1.552	$	&	51.43	&	128.58	&	$	193.312	\pm	13.36	$	&	$	17.648	\pm	0.426	$	\\
200205B	&	954520	&	1.465	&	$	116.56	\pm	2.732	$	&	$	53.056	\pm	2.838	$	&	56.8	&	141.99	&	$	114.304	\pm	5.166	$	&	$	40.016	\pm	13.073	$	\\
200829A	&	993768	&	1.25	&	$	64.416	\pm	17.16	$	&	$	2.848	\pm	0.057	$	&	62.22	&	155.56	&	$	6.352	\pm	0.283	$	&	$	2.192	\pm	0.057	$	\\
201020A	&	1000926	&	2.903	&	$	14.17	\pm	2.945	$	&	$	\rthis{3.728}	\pm	\rthis{1.859}	$	&	35.87	&	89.67	&	$	7.216	\pm	0.893	$	&	$	3.264	\pm	0.708	$	\\
201021C	&	1001130	&	1.07	&	$	21.584	\pm	3.301	$	&	$	8.528	\pm	2.132	$	&	67.63	&	169.08	&	$	9.136	\pm	2.172	$	&	$	4.16	\pm	2.515	$	\\
201104B	&	1004168	&	1.954	&	$	10.448	\pm	2.883	$	&	$	5.552	\pm	0.417	$	&	47.39	&	118.48	&	$	8.56	\pm	0.23	$	&	$	5.104	\pm	0.486	$	\\

\hline 
\end{longtable*}
\end{center}

\bibliography{references}

\providecommand{\noopsort}[1]{}\providecommand{\singleletter}[1]{#1}%
\begin{thebibliography}{54}
\expandafter\ifx\csname natexlab\endcsname\relax\def\natexlab#1{#1}\fi
\expandafter\ifx\csname bibnamefont\endcsname\relax
  \def\bibnamefont#1{#1}\fi
\expandafter\ifx\csname bibfnamefont\endcsname\relax
  \def\bibfnamefont#1{#1}\fi
\expandafter\ifx\csname citenamefont\endcsname\relax
  \def\citenamefont#1{#1}\fi
\expandafter\ifx\csname url\endcsname\relax
  \def\url#1{\texttt{#1}}\fi
\expandafter\ifx\csname urlprefix\endcsname\relax\def\urlprefix{URL }\fi
\providecommand{\bibinfo}[2]{#2}
\providecommand{\eprint}[2][]{\url{#2}}

\bibitem[{\citenamefont{{Zhang} et~al.}(2013)\citenamefont{{Zhang}, {Fan},
  {Shao}, and {Wei}}}]{zhang2013cosmological}
\bibinfo{author}{\bibfnamefont{F.-W.} \bibnamefont{{Zhang}}},
  \bibinfo{author}{\bibfnamefont{Y.-Z.} \bibnamefont{{Fan}}},
  \bibinfo{author}{\bibfnamefont{L.}~\bibnamefont{{Shao}}}, \bibnamefont{and}
  \bibinfo{author}{\bibfnamefont{D.-M.} \bibnamefont{{Wei}}},
  \bibinfo{journal}{\apjl} \textbf{\bibinfo{volume}{778}}, \bibinfo{eid}{L11}
  (\bibinfo{year}{2013}), \eprint{1309.5612}.

\bibitem[{\citenamefont{{Kumar} and {Zhang}}(2015)}]{Kumar}
\bibinfo{author}{\bibfnamefont{P.}~\bibnamefont{{Kumar}}} \bibnamefont{and}
  \bibinfo{author}{\bibfnamefont{B.}~\bibnamefont{{Zhang}}},
  \bibinfo{journal}{\physrep} \textbf{\bibinfo{volume}{561}},
  \bibinfo{pages}{1} (\bibinfo{year}{2015}), \eprint{1410.0679}.

\bibitem[{\citenamefont{{Kouveliotou} et~al.}(1993)\citenamefont{{Kouveliotou},
  {Meegan}, {Fishman}, {Bhat}, {Briggs}, {Koshut}, {Paciesas}, and
  {Pendleton}}}]{Kouveliotou93}
\bibinfo{author}{\bibfnamefont{C.}~\bibnamefont{{Kouveliotou}}},
  \bibinfo{author}{\bibfnamefont{C.~A.} \bibnamefont{{Meegan}}},
  \bibinfo{author}{\bibfnamefont{G.~J.} \bibnamefont{{Fishman}}},
  \bibinfo{author}{\bibfnamefont{N.~P.} \bibnamefont{{Bhat}}},
  \bibinfo{author}{\bibfnamefont{M.~S.} \bibnamefont{{Briggs}}},
  \bibinfo{author}{\bibfnamefont{T.~M.} \bibnamefont{{Koshut}}},
  \bibinfo{author}{\bibfnamefont{W.~S.} \bibnamefont{{Paciesas}}},
  \bibnamefont{and} \bibinfo{author}{\bibfnamefont{G.~N.}
  \bibnamefont{{Pendleton}}}, \bibinfo{journal}{\apjl}
  \textbf{\bibinfo{volume}{413}}, \bibinfo{pages}{L101} (\bibinfo{year}{1993}).

\bibitem[{\citenamefont{{Levan} et~al.}(2016)\citenamefont{{Levan}, {Crowther},
  {de Grijs}, {Langer}, {Xu}, and {Yoon}}}]{Levan}
\bibinfo{author}{\bibfnamefont{A.}~\bibnamefont{{Levan}}},
  \bibinfo{author}{\bibfnamefont{P.}~\bibnamefont{{Crowther}}},
  \bibinfo{author}{\bibfnamefont{R.}~\bibnamefont{{de Grijs}}},
  \bibinfo{author}{\bibfnamefont{N.}~\bibnamefont{{Langer}}},
  \bibinfo{author}{\bibfnamefont{D.}~\bibnamefont{{Xu}}}, \bibnamefont{and}
  \bibinfo{author}{\bibfnamefont{S.-C.} \bibnamefont{{Yoon}}},
  \bibinfo{journal}{\ssr} \textbf{\bibinfo{volume}{202}}, \bibinfo{pages}{33}
  (\bibinfo{year}{2016}), \eprint{1611.03091}.

\bibitem[{\citenamefont{{Piran}}(2004)}]{Piran}
\bibinfo{author}{\bibfnamefont{T.}~\bibnamefont{{Piran}}},
  \bibinfo{journal}{Reviews of Modern Physics} \textbf{\bibinfo{volume}{76}},
  \bibinfo{pages}{1143} (\bibinfo{year}{2004}), \eprint{astro-ph/0405503}.

\bibitem[{\citenamefont{{Nakar}}(2007)}]{Nakar}
\bibinfo{author}{\bibfnamefont{E.}~\bibnamefont{{Nakar}}},
  \bibinfo{journal}{\physrep} \textbf{\bibinfo{volume}{442}},
  \bibinfo{pages}{166} (\bibinfo{year}{2007}), \eprint{astro-ph/0701748}.

\bibitem[{\citenamefont{{Berger}}(2014)}]{Berger}
\bibinfo{author}{\bibfnamefont{E.}~\bibnamefont{{Berger}}},
  \bibinfo{journal}{\araa} \textbf{\bibinfo{volume}{52}}, \bibinfo{pages}{43}
  (\bibinfo{year}{2014}), \eprint{1311.2603}.

\bibitem[{\citenamefont{{Horv{\'a}th}}(1998)}]{Horvath98}
\bibinfo{author}{\bibfnamefont{I.}~\bibnamefont{{Horv{\'a}th}}},
  \bibinfo{journal}{\apj} \textbf{\bibinfo{volume}{508}}, \bibinfo{pages}{757}
  (\bibinfo{year}{1998}), \eprint{astro-ph/9803077}.

\bibitem[{\citenamefont{{Horv{\'a}th}}(2002)}]{Horvath02}
\bibinfo{author}{\bibfnamefont{I.}~\bibnamefont{{Horv{\'a}th}}},
  \bibinfo{journal}{\aap} \textbf{\bibinfo{volume}{392}}, \bibinfo{pages}{791}
  (\bibinfo{year}{2002}), \eprint{astro-ph/0205004}.

\bibitem[{\citenamefont{{Horv{\'a}th} et~al.}(2006)\citenamefont{{Horv{\'a}th},
  {Bal{\'a}zs}, {Bagoly}, {Ryde}, and {M{\'e}sz{\'a}ros}}}]{Horvath06}
\bibinfo{author}{\bibfnamefont{I.}~\bibnamefont{{Horv{\'a}th}}},
  \bibinfo{author}{\bibfnamefont{L.~G.} \bibnamefont{{Bal{\'a}zs}}},
  \bibinfo{author}{\bibfnamefont{Z.}~\bibnamefont{{Bagoly}}},
  \bibinfo{author}{\bibfnamefont{F.}~\bibnamefont{{Ryde}}}, \bibnamefont{and}
  \bibinfo{author}{\bibfnamefont{A.}~\bibnamefont{{M{\'e}sz{\'a}ros}}},
  \bibinfo{journal}{\aap} \textbf{\bibinfo{volume}{447}}, \bibinfo{pages}{23}
  (\bibinfo{year}{2006}), \eprint{astro-ph/0509909}.

\bibitem[{\citenamefont{{Horv{\'a}th} et~al.}(2008)\citenamefont{{Horv{\'a}th},
  {Bal{\'a}zs}, {Bagoly}, and {Veres}}}]{Horvath08}
\bibinfo{author}{\bibfnamefont{I.}~\bibnamefont{{Horv{\'a}th}}},
  \bibinfo{author}{\bibfnamefont{L.~G.} \bibnamefont{{Bal{\'a}zs}}},
  \bibinfo{author}{\bibfnamefont{Z.}~\bibnamefont{{Bagoly}}}, \bibnamefont{and}
  \bibinfo{author}{\bibfnamefont{P.}~\bibnamefont{{Veres}}},
  \bibinfo{journal}{\aap} \textbf{\bibinfo{volume}{489}}, \bibinfo{pages}{L1}
  (\bibinfo{year}{2008}), \eprint{0808.1067}.

\bibitem[{\citenamefont{{Zhang} et~al.}(2009)\citenamefont{{Zhang}, {Zhang},
  {Virgili}, {Liang}, {Kann}, {Wu}, {Proga}, {Lv}, {Toma}, {M{\'e}sz{\'a}ros}
  et~al.}}]{Zhang09}
\bibinfo{author}{\bibfnamefont{B.}~\bibnamefont{{Zhang}}},
  \bibinfo{author}{\bibfnamefont{B.-B.} \bibnamefont{{Zhang}}},
  \bibinfo{author}{\bibfnamefont{F.~J.} \bibnamefont{{Virgili}}},
  \bibinfo{author}{\bibfnamefont{E.-W.} \bibnamefont{{Liang}}},
  \bibinfo{author}{\bibfnamefont{D.~A.} \bibnamefont{{Kann}}},
  \bibinfo{author}{\bibfnamefont{X.-F.} \bibnamefont{{Wu}}},
  \bibinfo{author}{\bibfnamefont{D.}~\bibnamefont{{Proga}}},
  \bibinfo{author}{\bibfnamefont{H.-J.} \bibnamefont{{Lv}}},
  \bibinfo{author}{\bibfnamefont{K.}~\bibnamefont{{Toma}}},
  \bibinfo{author}{\bibfnamefont{P.}~\bibnamefont{{M{\'e}sz{\'a}ros}}},
  \bibnamefont{et~al.}, \bibinfo{journal}{\apj} \textbf{\bibinfo{volume}{703}},
  \bibinfo{pages}{1696} (\bibinfo{year}{2009}), \eprint{0902.2419}.

\bibitem[{\citenamefont{{Horv{\'a}th} et~al.}(2010)\citenamefont{{Horv{\'a}th},
  {Bagoly}, {Bal{\'a}zs}, {de Ugarte Postigo}, {Veres}, and
  {M{\'e}sz{\'a}ros}}}]{Horvath10}
\bibinfo{author}{\bibfnamefont{I.}~\bibnamefont{{Horv{\'a}th}}},
  \bibinfo{author}{\bibfnamefont{Z.}~\bibnamefont{{Bagoly}}},
  \bibinfo{author}{\bibfnamefont{L.~G.} \bibnamefont{{Bal{\'a}zs}}},
  \bibinfo{author}{\bibfnamefont{A.}~\bibnamefont{{de Ugarte Postigo}}},
  \bibinfo{author}{\bibfnamefont{P.}~\bibnamefont{{Veres}}}, \bibnamefont{and}
  \bibinfo{author}{\bibfnamefont{A.}~\bibnamefont{{M{\'e}sz{\'a}ros}}},
  \bibinfo{journal}{\apj} \textbf{\bibinfo{volume}{713}}, \bibinfo{pages}{552}
  (\bibinfo{year}{2010}), \eprint{1003.0632}.

\bibitem[{\citenamefont{{Bromberg} et~al.}(2013)\citenamefont{{Bromberg},
  {Nakar}, {Piran}, and {Sari}}}]{Bromberg}
\bibinfo{author}{\bibfnamefont{O.}~\bibnamefont{{Bromberg}}},
  \bibinfo{author}{\bibfnamefont{E.}~\bibnamefont{{Nakar}}},
  \bibinfo{author}{\bibfnamefont{T.}~\bibnamefont{{Piran}}}, \bibnamefont{and}
  \bibinfo{author}{\bibfnamefont{R.}~\bibnamefont{{Sari}}},
  \bibinfo{journal}{\apj} \textbf{\bibinfo{volume}{764}}, \bibinfo{eid}{179}
  (\bibinfo{year}{2013}), \eprint{1210.0068}.

\bibitem[{\citenamefont{{Kulkarni} and {Desai}}(2017)}]{Kulkarni}
\bibinfo{author}{\bibfnamefont{S.}~\bibnamefont{{Kulkarni}}} \bibnamefont{and}
  \bibinfo{author}{\bibfnamefont{S.}~\bibnamefont{{Desai}}},
  \bibinfo{journal}{\apss} \textbf{\bibinfo{volume}{362}}, \bibinfo{eid}{70}
  (\bibinfo{year}{2017}), \eprint{1612.08235}.

\bibitem[{\citenamefont{{Tarnopolski}}(2019)}]{Tarno19}
\bibinfo{author}{\bibfnamefont{M.}~\bibnamefont{{Tarnopolski}}},
  \bibinfo{journal}{\apj} \textbf{\bibinfo{volume}{887}}, \bibinfo{eid}{97}
  (\bibinfo{year}{2019}), \eprint{1910.08968}.

\bibitem[{\citenamefont{{Horv{\'a}th} et~al.}(2018)\citenamefont{{Horv{\'a}th},
  {T{\'o}th}, {Hakkila}, {T{\'o}th}, {Bal{\'a}zs}, {R{\'a}cz}, {Pint{\'e}r},
  and {Bagoly}}}]{Horvath18}
\bibinfo{author}{\bibfnamefont{I.}~\bibnamefont{{Horv{\'a}th}}},
  \bibinfo{author}{\bibfnamefont{B.~G.} \bibnamefont{{T{\'o}th}}},
  \bibinfo{author}{\bibfnamefont{J.}~\bibnamefont{{Hakkila}}},
  \bibinfo{author}{\bibfnamefont{L.~V.} \bibnamefont{{T{\'o}th}}},
  \bibinfo{author}{\bibfnamefont{L.~G.} \bibnamefont{{Bal{\'a}zs}}},
  \bibinfo{author}{\bibfnamefont{I.~I.} \bibnamefont{{R{\'a}cz}}},
  \bibinfo{author}{\bibfnamefont{S.}~\bibnamefont{{Pint{\'e}r}}},
  \bibnamefont{and} \bibinfo{author}{\bibfnamefont{Z.}~\bibnamefont{{Bagoly}}},
  \bibinfo{journal}{\apss} \textbf{\bibinfo{volume}{363}}, \bibinfo{eid}{53}
  (\bibinfo{year}{2018}), \eprint{1710.11509}.

\bibitem[{\citenamefont{Abbott et~al.}(2017)}]{LIGO}
\bibinfo{author}{\bibfnamefont{B.~P.} \bibnamefont{Abbott}}
  \bibnamefont{et~al.} (\bibinfo{collaboration}{LIGO Scientific, Virgo,
  Fermi-GBM, INTEGRAL}), \bibinfo{journal}{Astrophys. J. Lett.}
  \textbf{\bibinfo{volume}{848}}, \bibinfo{pages}{L13} (\bibinfo{year}{2017}),
  \eprint{1710.05834}.

\bibitem[{\citenamefont{{Boran} et~al.}(2018)\citenamefont{{Boran}, {Desai},
  {Kahya}, and {Woodard}}}]{Woodard}
\bibinfo{author}{\bibfnamefont{S.}~\bibnamefont{{Boran}}},
  \bibinfo{author}{\bibfnamefont{S.}~\bibnamefont{{Desai}}},
  \bibinfo{author}{\bibfnamefont{E.~O.} \bibnamefont{{Kahya}}},
  \bibnamefont{and} \bibinfo{author}{\bibfnamefont{R.~P.}
  \bibnamefont{{Woodard}}}, \bibinfo{journal}{\prd}
  \textbf{\bibinfo{volume}{97}}, \bibinfo{eid}{041501} (\bibinfo{year}{2018}),
  \eprint{1710.06168}.

\bibitem[{\citenamefont{{Klebesadel} et~al.}(1973)\citenamefont{{Klebesadel},
  {Strong}, and {Olson}}}]{firstgrb}
\bibinfo{author}{\bibfnamefont{R.~W.} \bibnamefont{{Klebesadel}}},
  \bibinfo{author}{\bibfnamefont{I.~B.} \bibnamefont{{Strong}}},
  \bibnamefont{and} \bibinfo{author}{\bibfnamefont{R.~A.}
  \bibnamefont{{Olson}}}, \bibinfo{journal}{\apjl}
  \textbf{\bibinfo{volume}{182}}, \bibinfo{pages}{L85} (\bibinfo{year}{1973}).

\bibitem[{\citenamefont{{Paczynski}}(1995)}]{Paczynski}
\bibinfo{author}{\bibfnamefont{B.}~\bibnamefont{{Paczynski}}},
  \bibinfo{journal}{\pasp} \textbf{\bibinfo{volume}{107}},
  \bibinfo{pages}{1167} (\bibinfo{year}{1995}), \eprint{astro-ph/9505096}.

\bibitem[{\citenamefont{{Lamb}}(1995)}]{Lamb}
\bibinfo{author}{\bibfnamefont{D.~Q.} \bibnamefont{{Lamb}}},
  \bibinfo{journal}{\pasp} \textbf{\bibinfo{volume}{107}},
  \bibinfo{pages}{1152} (\bibinfo{year}{1995}).

\bibitem[{\citenamefont{{van Paradijs} et~al.}(1997)\citenamefont{{van
  Paradijs}, {Groot}, {Galama}, {Kouveliotou}, {Strom}, {Telting}, {Rutten},
  {Fishman}, {Meegan}, {Pettini} et~al.}}]{vanPara}
\bibinfo{author}{\bibfnamefont{J.}~\bibnamefont{{van Paradijs}}},
  \bibinfo{author}{\bibfnamefont{P.~J.} \bibnamefont{{Groot}}},
  \bibinfo{author}{\bibfnamefont{T.}~\bibnamefont{{Galama}}},
  \bibinfo{author}{\bibfnamefont{C.}~\bibnamefont{{Kouveliotou}}},
  \bibinfo{author}{\bibfnamefont{R.~G.} \bibnamefont{{Strom}}},
  \bibinfo{author}{\bibfnamefont{J.}~\bibnamefont{{Telting}}},
  \bibinfo{author}{\bibfnamefont{R.~G.~M.} \bibnamefont{{Rutten}}},
  \bibinfo{author}{\bibfnamefont{G.~J.} \bibnamefont{{Fishman}}},
  \bibinfo{author}{\bibfnamefont{C.~A.} \bibnamefont{{Meegan}}},
  \bibinfo{author}{\bibfnamefont{M.}~\bibnamefont{{Pettini}}},
  \bibnamefont{et~al.}, \bibinfo{journal}{\nat} \textbf{\bibinfo{volume}{386}},
  \bibinfo{pages}{686} (\bibinfo{year}{1997}).

\bibitem[{\citenamefont{{Gehrels} et~al.}(2004)\citenamefont{{Gehrels},
  {Chincarini}, {Giommi}, {Mason}, {Nousek}, {Wells}, {White}, {Barthelmy},
  {Burrows}, {Cominsky} et~al.}}]{SWIFT}
\bibinfo{author}{\bibfnamefont{N.}~\bibnamefont{{Gehrels}}},
  \bibinfo{author}{\bibfnamefont{G.}~\bibnamefont{{Chincarini}}},
  \bibinfo{author}{\bibfnamefont{P.}~\bibnamefont{{Giommi}}},
  \bibinfo{author}{\bibfnamefont{K.~O.} \bibnamefont{{Mason}}},
  \bibinfo{author}{\bibfnamefont{J.~A.} \bibnamefont{{Nousek}}},
  \bibinfo{author}{\bibfnamefont{A.~A.} \bibnamefont{{Wells}}},
  \bibinfo{author}{\bibfnamefont{N.~E.} \bibnamefont{{White}}},
  \bibinfo{author}{\bibfnamefont{S.~D.} \bibnamefont{{Barthelmy}}},
  \bibinfo{author}{\bibfnamefont{D.~N.} \bibnamefont{{Burrows}}},
  \bibinfo{author}{\bibfnamefont{L.~R.} \bibnamefont{{Cominsky}}},
  \bibnamefont{et~al.}, \bibinfo{journal}{\apj} \textbf{\bibinfo{volume}{611}},
  \bibinfo{pages}{1005} (\bibinfo{year}{2004}), \eprint{astro-ph/0405233}.

\bibitem[{\citenamefont{{Piran}}(1992)}]{Piran92}
\bibinfo{author}{\bibfnamefont{T.}~\bibnamefont{{Piran}}},
  \bibinfo{journal}{\apjl} \textbf{\bibinfo{volume}{389}}, \bibinfo{pages}{L45}
  (\bibinfo{year}{1992}).

\bibitem[{\citenamefont{{Paczynski}}(1992)}]{Paczynski92}
\bibinfo{author}{\bibfnamefont{B.}~\bibnamefont{{Paczynski}}},
  \bibinfo{journal}{\nat} \textbf{\bibinfo{volume}{355}}, \bibinfo{pages}{521}
  (\bibinfo{year}{1992}).

\bibitem[{\citenamefont{{Goldhaber} et~al.}(2001)\citenamefont{{Goldhaber},
  {Groom}, {Kim}, {Aldering}, {Astier}, {Conley}, {Deustua}, {Ellis}, {Fabbro},
  {Fruchter} et~al.}}]{Goldhaber}
\bibinfo{author}{\bibfnamefont{G.}~\bibnamefont{{Goldhaber}}},
  \bibinfo{author}{\bibfnamefont{D.~E.} \bibnamefont{{Groom}}},
  \bibinfo{author}{\bibfnamefont{A.}~\bibnamefont{{Kim}}},
  \bibinfo{author}{\bibfnamefont{G.}~\bibnamefont{{Aldering}}},
  \bibinfo{author}{\bibfnamefont{P.}~\bibnamefont{{Astier}}},
  \bibinfo{author}{\bibfnamefont{A.}~\bibnamefont{{Conley}}},
  \bibinfo{author}{\bibfnamefont{S.~E.} \bibnamefont{{Deustua}}},
  \bibinfo{author}{\bibfnamefont{R.}~\bibnamefont{{Ellis}}},
  \bibinfo{author}{\bibfnamefont{S.}~\bibnamefont{{Fabbro}}},
  \bibinfo{author}{\bibfnamefont{A.~S.} \bibnamefont{{Fruchter}}},
  \bibnamefont{et~al.}, \bibinfo{journal}{\apj} \textbf{\bibinfo{volume}{558}},
  \bibinfo{pages}{359} (\bibinfo{year}{2001}), \eprint{astro-ph/0104382}.

\bibitem[{\citenamefont{{Hawkins}}(2010)}]{Hawkins}
\bibinfo{author}{\bibfnamefont{M.~R.~S.} \bibnamefont{{Hawkins}}},
  \bibinfo{journal}{\mnras} \textbf{\bibinfo{volume}{405}},
  \bibinfo{pages}{1940} (\bibinfo{year}{2010}), \eprint{1004.1824}.

\bibitem[{\citenamefont{{Norris} et~al.}(1994)\citenamefont{{Norris},
  {Nemiroff}, {Scargle}, {Kouveliotou}, {Fishman}, {Meegan}, {Paciesas}, and
  {Bonnell}}}]{Norris}
\bibinfo{author}{\bibfnamefont{J.~P.} \bibnamefont{{Norris}}},
  \bibinfo{author}{\bibfnamefont{R.~J.} \bibnamefont{{Nemiroff}}},
  \bibinfo{author}{\bibfnamefont{J.~D.} \bibnamefont{{Scargle}}},
  \bibinfo{author}{\bibfnamefont{C.}~\bibnamefont{{Kouveliotou}}},
  \bibinfo{author}{\bibfnamefont{G.~J.} \bibnamefont{{Fishman}}},
  \bibinfo{author}{\bibfnamefont{C.~A.} \bibnamefont{{Meegan}}},
  \bibinfo{author}{\bibfnamefont{W.~S.} \bibnamefont{{Paciesas}}},
  \bibnamefont{and} \bibinfo{author}{\bibfnamefont{J.~T.}
  \bibnamefont{{Bonnell}}}, \bibinfo{journal}{\apj}
  \textbf{\bibinfo{volume}{424}}, \bibinfo{pages}{540} (\bibinfo{year}{1994}),
  \eprint{astro-ph/9312049}.

\bibitem[{\citenamefont{{Che} et~al.}(1997)\citenamefont{{Che}, {Yang}, {Wu},
  and {Li}}}]{Che97}
\bibinfo{author}{\bibfnamefont{H.}~\bibnamefont{{Che}}},
  \bibinfo{author}{\bibfnamefont{Y.}~\bibnamefont{{Yang}}},
  \bibinfo{author}{\bibfnamefont{M.}~\bibnamefont{{Wu}}}, \bibnamefont{and}
  \bibinfo{author}{\bibfnamefont{Q.~B.} \bibnamefont{{Li}}},
  \bibinfo{journal}{\apjl} \textbf{\bibinfo{volume}{483}}, \bibinfo{pages}{L25}
  (\bibinfo{year}{1997}).

\bibitem[{\citenamefont{{Deng} and {Schaefer}}(1998)}]{Deng98}
\bibinfo{author}{\bibfnamefont{M.}~\bibnamefont{{Deng}}} \bibnamefont{and}
  \bibinfo{author}{\bibfnamefont{B.~E.} \bibnamefont{{Schaefer}}},
  \bibinfo{journal}{\apjl} \textbf{\bibinfo{volume}{502}},
  \bibinfo{pages}{L109} (\bibinfo{year}{1998}).

\bibitem[{\citenamefont{{Lee} et~al.}(2000)\citenamefont{{Lee}, {Bloom}, and
  {Petrosian}}}]{Lee2000}
\bibinfo{author}{\bibfnamefont{A.}~\bibnamefont{{Lee}}},
  \bibinfo{author}{\bibfnamefont{E.~D.} \bibnamefont{{Bloom}}},
  \bibnamefont{and}
  \bibinfo{author}{\bibfnamefont{V.}~\bibnamefont{{Petrosian}}},
  \bibinfo{journal}{\apjs} \textbf{\bibinfo{volume}{131}}, \bibinfo{pages}{21}
  (\bibinfo{year}{2000}), \eprint{astro-ph/0002218}.

\bibitem[{\citenamefont{{Chang}}(2001)}]{Chang01}
\bibinfo{author}{\bibfnamefont{H.-Y.} \bibnamefont{{Chang}}},
  \bibinfo{journal}{\apjl} \textbf{\bibinfo{volume}{557}}, \bibinfo{pages}{L85}
  (\bibinfo{year}{2001}), \eprint{astro-ph/0106220}.

\bibitem[{\citenamefont{{Sakamoto} et~al.}(2011)\citenamefont{{Sakamoto},
  {Barthelmy}, {Baumgartner}, {Cummings}, {Fenimore}, {Gehrels}, {Krimm},
  {Markwardt}, {Palmer}, {Parsons} et~al.}}]{Sakamoto}
\bibinfo{author}{\bibfnamefont{T.}~\bibnamefont{{Sakamoto}}},
  \bibinfo{author}{\bibfnamefont{S.~D.} \bibnamefont{{Barthelmy}}},
  \bibinfo{author}{\bibfnamefont{W.~H.} \bibnamefont{{Baumgartner}}},
  \bibinfo{author}{\bibfnamefont{J.~R.} \bibnamefont{{Cummings}}},
  \bibinfo{author}{\bibfnamefont{E.~E.} \bibnamefont{{Fenimore}}},
  \bibinfo{author}{\bibfnamefont{N.}~\bibnamefont{{Gehrels}}},
  \bibinfo{author}{\bibfnamefont{H.~A.} \bibnamefont{{Krimm}}},
  \bibinfo{author}{\bibfnamefont{C.~B.} \bibnamefont{{Markwardt}}},
  \bibinfo{author}{\bibfnamefont{D.~M.} \bibnamefont{{Palmer}}},
  \bibinfo{author}{\bibfnamefont{A.~M.} \bibnamefont{{Parsons}}},
  \bibnamefont{et~al.}, \bibinfo{journal}{\apjs}
  \textbf{\bibinfo{volume}{195}}, \bibinfo{eid}{2} (\bibinfo{year}{2011}),
  \eprint{1104.4689}.

\bibitem[{\citenamefont{{Kocevski} and {Petrosian}}(2013)}]{Petrosian13}
\bibinfo{author}{\bibfnamefont{D.}~\bibnamefont{{Kocevski}}} \bibnamefont{and}
  \bibinfo{author}{\bibfnamefont{V.}~\bibnamefont{{Petrosian}}},
  \bibinfo{journal}{\apj} \textbf{\bibinfo{volume}{765}}, \bibinfo{eid}{116}
  (\bibinfo{year}{2013}).

\bibitem[{\citenamefont{{Gruber} et~al.}(2011)\citenamefont{{Gruber},
  {Greiner}, {von Kienlin}, {Rau}, {Briggs}, {Connaughton}, {Goldstein}, {van
  der Horst}, {Nardini}, {Bhat} et~al.}}]{Gruber}
\bibinfo{author}{\bibfnamefont{D.}~\bibnamefont{{Gruber}}},
  \bibinfo{author}{\bibfnamefont{J.}~\bibnamefont{{Greiner}}},
  \bibinfo{author}{\bibfnamefont{A.}~\bibnamefont{{von Kienlin}}},
  \bibinfo{author}{\bibfnamefont{A.}~\bibnamefont{{Rau}}},
  \bibinfo{author}{\bibfnamefont{M.~S.} \bibnamefont{{Briggs}}},
  \bibinfo{author}{\bibfnamefont{V.}~\bibnamefont{{Connaughton}}},
  \bibinfo{author}{\bibfnamefont{A.}~\bibnamefont{{Goldstein}}},
  \bibinfo{author}{\bibfnamefont{A.~J.} \bibnamefont{{van der Horst}}},
  \bibinfo{author}{\bibfnamefont{M.}~\bibnamefont{{Nardini}}},
  \bibinfo{author}{\bibfnamefont{P.~N.} \bibnamefont{{Bhat}}},
  \bibnamefont{et~al.}, \bibinfo{journal}{\aap} \textbf{\bibinfo{volume}{531}},
  \bibinfo{eid}{A20} (\bibinfo{year}{2011}), \eprint{1104.5495}.

\bibitem[{\citenamefont{{Lien} et~al.}(2016)\citenamefont{{Lien}, {Sakamoto},
  {Barthelmy}, {Baumgartner}, {Cannizzo}, {Chen}, {Collins}, {Cummings},
  {Gehrels}, {Krimm} et~al.}}]{Lien}
\bibinfo{author}{\bibfnamefont{A.}~\bibnamefont{{Lien}}},
  \bibinfo{author}{\bibfnamefont{T.}~\bibnamefont{{Sakamoto}}},
  \bibinfo{author}{\bibfnamefont{S.~D.} \bibnamefont{{Barthelmy}}},
  \bibinfo{author}{\bibfnamefont{W.~H.} \bibnamefont{{Baumgartner}}},
  \bibinfo{author}{\bibfnamefont{J.~K.} \bibnamefont{{Cannizzo}}},
  \bibinfo{author}{\bibfnamefont{K.}~\bibnamefont{{Chen}}},
  \bibinfo{author}{\bibfnamefont{N.~R.} \bibnamefont{{Collins}}},
  \bibinfo{author}{\bibfnamefont{J.~R.} \bibnamefont{{Cummings}}},
  \bibinfo{author}{\bibfnamefont{N.}~\bibnamefont{{Gehrels}}},
  \bibinfo{author}{\bibfnamefont{H.~A.} \bibnamefont{{Krimm}}},
  \bibnamefont{et~al.}, \bibinfo{journal}{\apj} \textbf{\bibinfo{volume}{829}},
  \bibinfo{eid}{7} (\bibinfo{year}{2016}), \eprint{1606.01956}.

\bibitem[{\citenamefont{{Crawford}}(2018)}]{Crawford}
\bibinfo{author}{\bibfnamefont{D.~F.} \bibnamefont{{Crawford}}},
  \bibinfo{journal}{arXiv e-prints} \bibinfo{eid}{arXiv:1804.10274}
  (\bibinfo{year}{2018}), \eprint{1804.10274}.

\bibitem[{\citenamefont{{Tarnopolski}}(2020)}]{Tarno}
\bibinfo{author}{\bibfnamefont{M.}~\bibnamefont{{Tarnopolski}}},
  \bibinfo{journal}{\apj} \textbf{\bibinfo{volume}{897}}, \bibinfo{eid}{77}
  (\bibinfo{year}{2020}), \eprint{2004.13623}.

\bibitem[{\citenamefont{{Golkhou} and {Butler}}(2014)}]{Butler}
\bibinfo{author}{\bibfnamefont{V.~Z.} \bibnamefont{{Golkhou}}}
  \bibnamefont{and} \bibinfo{author}{\bibfnamefont{N.~R.}
  \bibnamefont{{Butler}}}, \bibinfo{journal}{\apj}
  \textbf{\bibinfo{volume}{787}}, \bibinfo{eid}{90} (\bibinfo{year}{2014}),
  \eprint{1403.4254}.

\bibitem[{\citenamefont{{Gopika} and {Desai}}(2020)}]{Gopika_2020}
\bibinfo{author}{\bibfnamefont{K.}~\bibnamefont{{Gopika}}} \bibnamefont{and}
  \bibinfo{author}{\bibfnamefont{S.}~\bibnamefont{{Desai}}},
  \bibinfo{journal}{Physics of the Dark Universe}
  \textbf{\bibinfo{volume}{30}}, \bibinfo{eid}{100707} (\bibinfo{year}{2020}),
  \eprint{2006.12320}.

\bibitem[{\citenamefont{{Pradyumna} et~al.}(2021)\citenamefont{{Pradyumna},
  {Gupta}, {Seeram}, and {Desai}}}]{Pradyumna}
\bibinfo{author}{\bibfnamefont{S.}~\bibnamefont{{Pradyumna}}},
  \bibinfo{author}{\bibfnamefont{S.}~\bibnamefont{{Gupta}}},
  \bibinfo{author}{\bibfnamefont{S.}~\bibnamefont{{Seeram}}}, \bibnamefont{and}
  \bibinfo{author}{\bibfnamefont{S.}~\bibnamefont{{Desai}}},
  \bibinfo{journal}{Physics of the Dark Universe}
  \textbf{\bibinfo{volume}{31}}, \bibinfo{eid}{100765} (\bibinfo{year}{2021}),
  \eprint{2011.06421}.

\bibitem[{\citenamefont{{Bora} and {Desai}}(2021)}]{Bora_feb}
\bibinfo{author}{\bibfnamefont{K.}~\bibnamefont{{Bora}}} \bibnamefont{and}
  \bibinfo{author}{\bibfnamefont{S.}~\bibnamefont{{Desai}}},
  \bibinfo{journal}{\jcap} \textbf{\bibinfo{volume}{2021}}, \bibinfo{eid}{012}
  (\bibinfo{year}{2021}), \eprint{2008.10541}.

\bibitem[{\citenamefont{{Tian} et~al.}(2020)\citenamefont{{Tian}, {Umetsu},
  {Ko}, {Donahue}, and {Chiu}}}]{Tian}
\bibinfo{author}{\bibfnamefont{Y.}~\bibnamefont{{Tian}}},
  \bibinfo{author}{\bibfnamefont{K.}~\bibnamefont{{Umetsu}}},
  \bibinfo{author}{\bibfnamefont{C.-M.} \bibnamefont{{Ko}}},
  \bibinfo{author}{\bibfnamefont{M.}~\bibnamefont{{Donahue}}},
  \bibnamefont{and} \bibinfo{author}{\bibfnamefont{I.~N.}
  \bibnamefont{{Chiu}}}, \bibinfo{journal}{\apj}
  \textbf{\bibinfo{volume}{896}}, \bibinfo{eid}{70} (\bibinfo{year}{2020}),
  \eprint{2001.08340}.

\bibitem[{\citenamefont{Foreman-Mackey
  et~al.}(2013)\citenamefont{Foreman-Mackey, Hogg, Lang, and
  Goodman}}]{foreman2013emcee}
\bibinfo{author}{\bibfnamefont{D.}~\bibnamefont{Foreman-Mackey}},
  \bibinfo{author}{\bibfnamefont{D.~W.} \bibnamefont{Hogg}},
  \bibinfo{author}{\bibfnamefont{D.}~\bibnamefont{Lang}}, \bibnamefont{and}
  \bibinfo{author}{\bibfnamefont{J.}~\bibnamefont{Goodman}},
  \bibinfo{journal}{Publications of the Astronomical Society of the Pacific}
  \textbf{\bibinfo{volume}{125}}, \bibinfo{pages}{306} (\bibinfo{year}{2013}).

\bibitem[{\citenamefont{{P{\'e}langeon}
  et~al.}(2008)\citenamefont{{P{\'e}langeon}, {Atteia}, {Nakagawa}, {Hurley},
  {Yoshida}, {Vanderspek}, {Suzuki}, {Kawai}, {Pizzichini}, {Bo{\"e}r}
  et~al.}}]{Pelangeon}
\bibinfo{author}{\bibfnamefont{A.}~\bibnamefont{{P{\'e}langeon}}},
  \bibinfo{author}{\bibfnamefont{J.~L.} \bibnamefont{{Atteia}}},
  \bibinfo{author}{\bibfnamefont{Y.~E.} \bibnamefont{{Nakagawa}}},
  \bibinfo{author}{\bibfnamefont{K.}~\bibnamefont{{Hurley}}},
  \bibinfo{author}{\bibfnamefont{A.}~\bibnamefont{{Yoshida}}},
  \bibinfo{author}{\bibfnamefont{R.}~\bibnamefont{{Vanderspek}}},
  \bibinfo{author}{\bibfnamefont{M.}~\bibnamefont{{Suzuki}}},
  \bibinfo{author}{\bibfnamefont{N.}~\bibnamefont{{Kawai}}},
  \bibinfo{author}{\bibfnamefont{G.}~\bibnamefont{{Pizzichini}}},
  \bibinfo{author}{\bibfnamefont{M.}~\bibnamefont{{Bo{\"e}r}}},
  \bibnamefont{et~al.}, \bibinfo{journal}{\aap} \textbf{\bibinfo{volume}{491}},
  \bibinfo{pages}{157} (\bibinfo{year}{2008}), \eprint{0811.3304}.

\bibitem[{\citenamefont{{Barthelmy} et~al.}(2005)\citenamefont{{Barthelmy},
  {Barbier}, {Cummings}, {Fenimore}, {Gehrels}, {Hullinger}, {Krimm},
  {Markwardt}, {Palmer}, {Parsons} et~al.}}]{Barthelmy}
\bibinfo{author}{\bibfnamefont{S.~D.} \bibnamefont{{Barthelmy}}},
  \bibinfo{author}{\bibfnamefont{L.~M.} \bibnamefont{{Barbier}}},
  \bibinfo{author}{\bibfnamefont{J.~R.} \bibnamefont{{Cummings}}},
  \bibinfo{author}{\bibfnamefont{E.~E.} \bibnamefont{{Fenimore}}},
  \bibinfo{author}{\bibfnamefont{N.}~\bibnamefont{{Gehrels}}},
  \bibinfo{author}{\bibfnamefont{D.}~\bibnamefont{{Hullinger}}},
  \bibinfo{author}{\bibfnamefont{H.~A.} \bibnamefont{{Krimm}}},
  \bibinfo{author}{\bibfnamefont{C.~B.} \bibnamefont{{Markwardt}}},
  \bibinfo{author}{\bibfnamefont{D.~M.} \bibnamefont{{Palmer}}},
  \bibinfo{author}{\bibfnamefont{A.}~\bibnamefont{{Parsons}}},
  \bibnamefont{et~al.}, \bibinfo{journal}{\ssr} \textbf{\bibinfo{volume}{120}},
  \bibinfo{pages}{143} (\bibinfo{year}{2005}), \eprint{astro-ph/0507410}.

\bibitem[{\citenamefont{{Littlejohns} et~al.}(2013)\citenamefont{{Littlejohns},
  {Tanvir}, {Willingale}, {Evans}, {O'Brien}, and {Levan}}}]{Littlejohns}
\bibinfo{author}{\bibfnamefont{O.~M.} \bibnamefont{{Littlejohns}}},
  \bibinfo{author}{\bibfnamefont{N.~R.} \bibnamefont{{Tanvir}}},
  \bibinfo{author}{\bibfnamefont{R.}~\bibnamefont{{Willingale}}},
  \bibinfo{author}{\bibfnamefont{P.~A.} \bibnamefont{{Evans}}},
  \bibinfo{author}{\bibfnamefont{P.~T.} \bibnamefont{{O'Brien}}},
  \bibnamefont{and} \bibinfo{author}{\bibfnamefont{A.~J.}
  \bibnamefont{{Levan}}}, \bibinfo{journal}{\mnras}
  \textbf{\bibinfo{volume}{436}}, \bibinfo{pages}{3640} (\bibinfo{year}{2013}),
  \eprint{1309.7045}.

\bibitem[{\citenamefont{{von Kienlin} et~al.}(2020)\citenamefont{{von Kienlin},
  {Meegan}, {Paciesas}, {Bhat}, {Bissaldi}, {Briggs}, {Burns}, {Cleveland},
  {Gibby}, {Giles} et~al.}}]{von}
\bibinfo{author}{\bibfnamefont{A.}~\bibnamefont{{von Kienlin}}},
  \bibinfo{author}{\bibfnamefont{C.~A.} \bibnamefont{{Meegan}}},
  \bibinfo{author}{\bibfnamefont{W.~S.} \bibnamefont{{Paciesas}}},
  \bibinfo{author}{\bibfnamefont{P.~N.} \bibnamefont{{Bhat}}},
  \bibinfo{author}{\bibfnamefont{E.}~\bibnamefont{{Bissaldi}}},
  \bibinfo{author}{\bibfnamefont{M.~S.} \bibnamefont{{Briggs}}},
  \bibinfo{author}{\bibfnamefont{E.}~\bibnamefont{{Burns}}},
  \bibinfo{author}{\bibfnamefont{W.~H.} \bibnamefont{{Cleveland}}},
  \bibinfo{author}{\bibfnamefont{M.~H.} \bibnamefont{{Gibby}}},
  \bibinfo{author}{\bibfnamefont{M.~M.} \bibnamefont{{Giles}}},
  \bibnamefont{et~al.}, \bibinfo{journal}{\apj} \textbf{\bibinfo{volume}{893}},
  \bibinfo{eid}{46} (\bibinfo{year}{2020}), \eprint{2002.11460}.

\bibitem[{\citenamefont{Markwardt et~al.}(2007)\citenamefont{Markwardt,
  Barthelmy, Cummings, Hullinger, Krimm, and Parsons}}]{markwardt2007Swift}
\bibinfo{author}{\bibfnamefont{C.}~\bibnamefont{Markwardt}},
  \bibinfo{author}{\bibfnamefont{S.}~\bibnamefont{Barthelmy}},
  \bibinfo{author}{\bibfnamefont{J.}~\bibnamefont{Cummings}},
  \bibinfo{author}{\bibfnamefont{D.}~\bibnamefont{Hullinger}},
  \bibinfo{author}{\bibfnamefont{H.}~\bibnamefont{Krimm}}, \bibnamefont{and}
  \bibinfo{author}{\bibfnamefont{A.}~\bibnamefont{Parsons}},
  \bibinfo{journal}{NASA/GSFC, Greenbelt, MD} \textbf{\bibinfo{volume}{6}}
  (\bibinfo{year}{2007}).

\bibitem[{\citenamefont{Swift~Team}(2021)}]{SwiftArchive}
\bibinfo{author}{\bibfnamefont{N.}~\bibnamefont{Swift~Team}},
  \emph{\bibinfo{title}{Swift archive}},
  \bibinfo{howpublished}{\url{https://swift.gsfc.nasa.gov/archive/}}
  (\bibinfo{year}{2021}).

\bibitem[{\citenamefont{NASA}(2021{\natexlab{a}})}]{XaminWeb}
\bibinfo{author}{\bibnamefont{NASA}}, \emph{\bibinfo{title}{Xamin web
  interface}},
  \bibinfo{howpublished}{\url{https://heasarc.gsfc.nasa.gov/xamin/}}
  (\bibinfo{year}{2021}{\natexlab{a}}).

\bibitem[{\citenamefont{NASA}(2021{\natexlab{b}})}]{HeraWeb}
\bibinfo{author}{\bibnamefont{NASA}}, \emph{\bibinfo{title}{Web hera}},
  \bibinfo{howpublished}{\url{https://hera.gsfc.nasa.gov/hera/}}
  (\bibinfo{year}{2021}{\natexlab{b}}).

\bibitem[{\citenamefont{{Scargle} et~al.}(2013)\citenamefont{{Scargle},
  {Norris}, {Jackson}, and {Chiang}}}]{scargle2013bayesian}
\bibinfo{author}{\bibfnamefont{J.~D.} \bibnamefont{{Scargle}}},
  \bibinfo{author}{\bibfnamefont{J.~P.} \bibnamefont{{Norris}}},
  \bibinfo{author}{\bibfnamefont{B.}~\bibnamefont{{Jackson}}},
  \bibnamefont{and} \bibinfo{author}{\bibfnamefont{J.}~\bibnamefont{{Chiang}}},
  \bibinfo{journal}{arXiv e-prints} \bibinfo{eid}{arXiv:1304.2818}
  (\bibinfo{year}{2013}), \eprint{1304.2818}.

\end{thebibliography}

\end{document}